\documentclass[a4paper,11pt]{article}
\pdfoutput=1 

\usepackage{jcappub} 
\makeatletter\def\@fpheader{\relax}\makeatother 

\usepackage{float}
\usepackage[section]{placeins}
\usepackage{amssymb,amsmath,latexsym,bm,amsfonts}
\usepackage{graphicx,color,xcolor,longtable}
\usepackage{indentfirst}
\usepackage{subfigure}
\usepackage{breqn}

\newcommand{\MatBK}[3]{\ensuremath{\left\langle #1 \middle| #2 \middle| #3 \right\rangle}}

\newcommand{\be}{\begin{equation}}
\newcommand{\ee}{\end{equation}}
\newcommand{\bpm}{\begin{pmatrix}}
\newcommand{\epm}{\end{pmatrix}}

\newcommand{\PBK}[1]{\ensuremath{\begin{pmatrix}#1\end{pmatrix}}}

\newcommand{\EV}[1]{\langle #1 \rangle}
\newcommand{\beqn}{\begin{eqnarray}}
\newcommand{\eeqn}{\end{eqnarray}}

\DeclareMathOperator{\sgn}{sgn}
\DeclareMathOperator{\Real}{Re}

\DeclareMathOperator{\Imag}{Im}

\DeclareMathOperator{\Ci}{Ci}
\DeclareMathOperator{\Si}{Si}

\title{On the initial condition of inflationary fluctuations}

\author{Hongliang Jiang}
\author[*]{, Yi Wang}
\author{and Siyi Zhou}

\emailAdd{phyw@ust.hk}

\affiliation{Department of Physics, The Hong Kong University of Science and Technology,
Clear Water Bay, Kowloon, Hong Kong, P.R.China}

\abstract{
  It is usually assumed that the inflationary fluctuations start from the Bunch-Davies (BD) vacuum and the $i\varepsilon$ prescription is used when interactions are calculated. We show that those assumptions can be verified explicitly by calculating the loop corrections to the inflationary two-point and three-point correlation functions. Those loop corrections can be resumed to exponential factors, which suppress non-BD coefficients and behave as the $i\varepsilon$ factor for the case of the BD initial condition. A new technique of loop chain diagram resummation is developed for this purpose. For the non-BD initial conditions which is setup at finite time and has not fully decayed, explicit correction to the two-point and three-point correlation functions are calculated. Especially, non-Gaussianity in the folded limit is regularized due to the interactions.
}

\begin{document}

\maketitle

\section{Introduction and Summary} \label{sec:int-sum}

Inflation is the leading paradigm of the early universe cosmology. The fluctuations generated during inflation provide seeds for the cosmic microwave background (CMB) and the large scale structure (LSS) formation \cite{Mukhanov:1990me} . The standard calculation of those primordial fluctuations follows from the quantum theoretical in-in formalism following two assumptions, namely the standard vacuum initial condition (known as the Bunch-Davies vacuum \cite{Bunch:1978yq}, or BD vacuum for short) and the $i\varepsilon$ prescription. Those assumptions are inherited from the flat space quantum field theory, but have to be reconsidered in cosmology.

\begin{itemize}
  \item The BD vacuum initial condition. This is the simplest choice of initial state in the simplest models of inflation, because inflation is an attractor solution. However, it has been debated for long because of the following issues:
  \begin{itemize}
    \item Beyond the attractor stage of inflation. Scale dependent features during inflation can override the BD vacuum initial conditions . For example, inflation may be just enough and the start of observable inflation may be close to the absolute start of inflation \cite{Chen:2013tna}. As another example, there may be features on the inflationary potential such that the inflationary fluctuations after the features are in an excited state before horizon crossing \cite{Chen:2010bka}.
    \item Beyond the simplest theory of fluctuations. In cosmology it is convenient to follow the time evolution of a comoving perturbation mode. The comoving mode originates from scales much smaller than the inflationary Hubble scale, and its physical wavelength expands with the cosmological expansion. During the expansion of its physical wavelength, the dynamics of the mode may be governed by different effective field theories, or no effective field theory at all when its physical wavelength is shorter than the Planck scale (the trans-Planckian problem \cite{Martin:2000xs}).
    \item Beyond the leading order calculation of gravitational fluctuations. Practically, the BD vacuum is selected as the lowest energy state. However, when gravitational fluctuations are concerned, energy is a gauge dependent quantity. Different gauge can have different definition of time, and thus different definition of energy. This is similar to the case that in the Minkowski vacuum, accelerating observer sees Unruh radiation  \cite{PhysRevD.14.870}, which appears to be no longer the lowest energy state. In the literature in each gauge people choose the lowest energy state as the physical ``vacuum'' state. This cannot be right. Only the vacuum state of one gauge should be physical and the vacuum in other gauges should be the gauge transformation of the same physical vacuum.
  \end{itemize}

  \item The $i\varepsilon$ prescription. This prescription is not relevant in the tree level power spectrum calculation, but become important for the non-trivial in-in calculation for higher point correlation functions or loop diagrams. In flat space in-out formalism, the $i\varepsilon$ prescription is proposed to project the physical interacting vacuum onto the vacuum of the free theory, because only the vacuum of the free theory is operationally defined by the free quantum fields (or the interacting picture fields) and can be practically used in the perturbative calculation. One can relate the free vacuum $|0\rangle$ and the interacting vacuum $|\Omega\rangle$ by
  \begin{align}
    \label{eq:free-int-vac}
    e^{-iHT} |0\rangle
    = e^{-iE_0T} \langle\Omega|0\rangle |\Omega\rangle
    + \sum_{n>0} e^{-iE_nT} \langle n|0\rangle |n\rangle~,
  \end{align}
  where $T$ is the duration of the interaction, $H$ is the full Hamiltonian, $E_0$ is the energy of the ground state defined by $E_0 \equiv \langle\Omega| H |\Omega\rangle$, and $E_n \equiv \langle n| H |n\rangle$ for non-perturbative states $|n\rangle$ with higher energies. One can then send $T$ to $\infty$ by $T\rightarrow \infty (1-i\varepsilon)$. Then all but the first term in the RHS of \eqref{eq:free-int-vac} vanishes, and we obtain a relation between $|0\rangle$
  and $|\Omega\rangle$. The following assumptions are involved in this prescription:
  \begin{itemize}
    \item One can adiabatically turn off the interactions. This assumption works fine in flat space calculation of the S-matrix because we are preparing the initial states in the far past with large spatial separation. Following cluster decomposition~\cite{PhysRev.132.2788}, or any explicit law of forces, the states can indeed be considered to be non-interacting. Actually, under some mild  assumptions, the validness of relating interacting vacuum to the free vacuum  in this way can be rigorously proved in quantum field theory, known as Gell-Mann and Low theorem \cite{PhysRev.84.350}.    However, in cosmology, we are interested in considering the time evolution of the initial vacuum state. The state is initially of sub-Hubble size and all (virtual) particles stay close to each other. Thus we are no longer sure about the validity of turning off interactions in the calculation of cosmological perturbations.
    \item There exists enough time duration $T$ for the $i\varepsilon$ prescription. This assumption is again tricky in cosmology, because this statement is again coordinate dependent. For inflation, one can use conformal time or proper time. When the conformal time is used, one indeed have nearly infinite (though still not really infinite because inflation cannot be eternal to the past) amount of conformal time in the past. However, when using proper time, the amount of time duration gets shortened exponentially. One can indeed argue that before horizon crossing, the conformal time is more relevant. But explicit calculation is needed to verify the argument. Even we use the conformal time, a mathematically infinitesimal $i\epsilon$ does not work because of the finiteness of conformal time, even if the duration is exponentially long.
  \end{itemize}

\end{itemize}

In this work, we aim to provide a systematic method towards resolving the above puzzles. This is an extension of our previous work \cite{Jiang:2015hfa}. We show that interaction is the key to the vacuum and the $i \epsilon$ problems.

Interaction exists in the early universe. The theory of gravity is nonlinear. The gravitational nonlinearity provides a lower bound on the interaction of perturbations during inflation. In terms of the non-Gaussianity estimator $f_{\text{NL}}$, the minimal gravitational nonlinearity corresponds to $f_{\text{NL}}\sim \mathcal{O}(0.01)$. Large non-Gaussianities are predicted in some inflation models and the current observational bound is $f_{\text{NL}}$ of order 10 or 100, depending on the shapes of non-Gaussianity.

For this purpose, in our previous work, we calculate the one loop correction of the two point function with non-BD initial conditions. We have shown that, with the help of interactions, the non-BD initial conditions dissipates exponentially fast towards large scales. The one loop correction of the non-BD coefficients can be classified into two types, namely the correction to the amplitude and phase of the non-BD coefficient. The correction to the amplitude of non-BD coefficients corresponds to the contributions close to the folded limit of the interaction vertex. This amplitude correction is negative and can be resumed onto the exponent by dynamical RG method \cite{Boyanovsky:2003ui,Boyanovsky:2004ph,Burgess:2009bs}. As a result, for sub-horizon fluctuations, we have
\begin{align}
  \label{eq:previous-result}
  c_{\bm{k}}^{\mathrm{eff}} = c_{\bm{k}}
  \exp \left [ -\Gamma (\tau - \tau_0) \right ]~,
  \quad
  \Gamma \sim f_{\text{NL}}^2 P_\zeta k^5 \tau_0^4~,
\end{align}
where $c_{\bm{k}}$ is the absolute value of the tree level non-BD coefficient, $c_{\bm{k}}^{\mathrm{eff}}$ is that with dynamical-RG-resumed one loop corrections, and $\tau_0$ is the initial time where the non-BD initial condition is setup. For $f_{\text{NL}}\sim \mathcal{O}(1)$, the characteristic scale on the exponent is between $k\tau_0 \sim 4$, indicating that non-BD initial  conditions which are setup at sub-Horizon scales as deep as 4 e-folds start to decay exponentially. For larger non-Gaussianities, the decay of non-BD initial conditions become significantly faster. As a result, smaller non-Gaussianities, which seem not great for the purpose of probing interactions during inflation, have the advantage of better preserving the initial state of inflation.

In this work, we solidify the previous calculation by an explicit loop calculation, fixing the previously undetermined order one coefficient. For $(\partial_t\zeta)^3$ interaction, the result is
\begin{align}  \label{eq:result-2pt}
  c_{\bm k}^{\text{eff}}(\tau) =
   c_{\bm k}  \exp\Big(- \frac{19683\pi}{20000} P_\zeta  f_{\text{NL}}^2 k^5(\tau_{ }^5-\tau_0^5 )\Big) ~.
\end{align}
The dynamical RG resummation method which has been used in our previous work is also checked explicitly using a direct resummation of one particle reducible multi-loop diagrams. We show that the two results agree up to a two-loop contribution, which is under control when proper scale of renormalization is chosen.

There are model dependent and model independent components in \eqref{eq:result-2pt}. The numerical factor is of course model dependent. The 5th power in $\tau$ is also model dependent. If the interaction were marginal (i.e. dimension 4 after canonically normalize $\zeta$), then one expects linear dependence in $\tau$, because the total amount of interaction should be proportional to the length of interaction time. Here, the operator under our consideration has dimension 6. Thus for each interaction vertex there arises two additional powers of $\tau$ due to UV sensitivity. As a result the exponent scales as $\tau^5$. For inflation with standard kinetic term and Einstein gravity, the interactions have dimension 5 and we should expect the exponent scaling as $\tau^3$. The dependence on $P_\zeta$, $f_{\text{NL}}$ and the exponential structure of the decay, on the other hand, should be model independent. Also, the interaction scales linearly in $\tau-\tau_0$ when $\tau-\tau_0$ is small. This is model independent from the physical interpretation of a decay rate.

Technically, it is interesting to note that, in the sub-horizon limit, the reducible multi-loop diagrams (as a chain of one loop diagrams) dominate over the irreducible ones. The reason is as follows. We hope to pick up the highest power of $|k\tau_0|$ in the calculation. The highest power comes from the diagrams where the largest number of vertices can freely take values from $\tau\sim \tau_0$ to $|k\tau|\sim 1$, which is a large range. In the reducible diagrams, the vertices group into freely moving pairs, each pair represent a loop and the relative time difference is constrained by the uncertainty principle. However, for diagrams which contain irreducible multi-loop parts, more vertices are constrained by the uncertainty principle and thus do not show up at leading power of $|k\tau_0|$. This further assures the validity of the dynamical RG method.

\begin{figure}[h]
 \centering
  \includegraphics[width=0.6\textwidth]{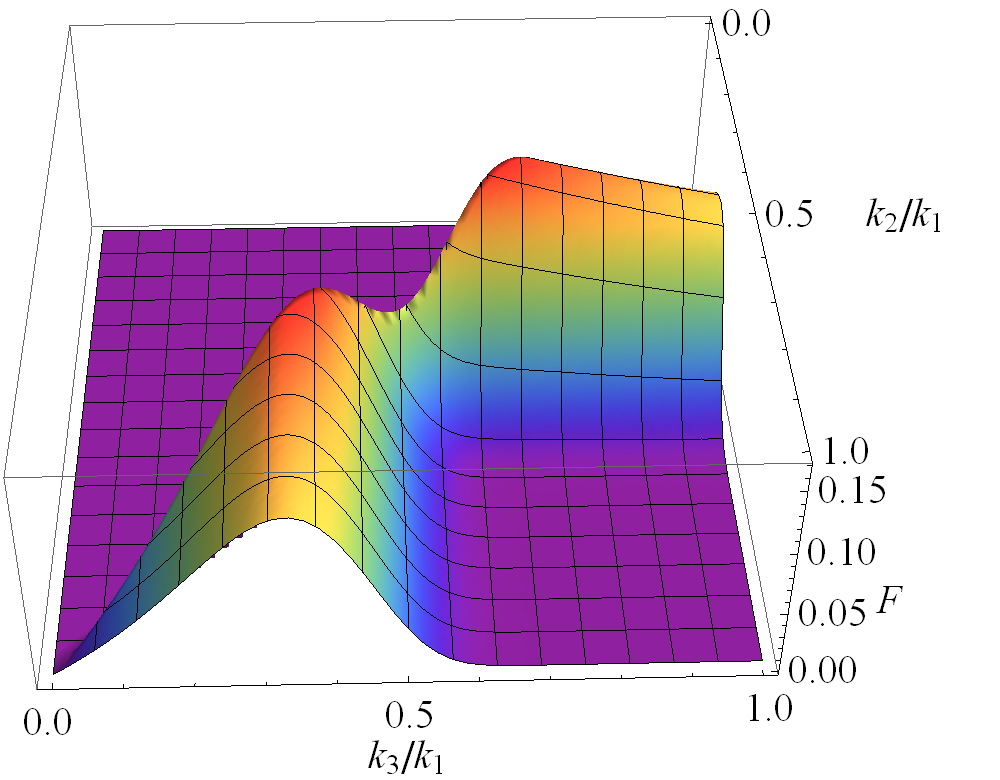}
  \caption{A typical plot for non-BD non-Gaussianity shape after including loop corrections with sharp initial time cut-off.}
 \label{NonBDTypical}
 \end{figure}

We then study the one loop correction of the three point correlation function. In the case of the three point function, the three external legs can carry different momenta and thus the dynamical RG method becomes no longer accurate. We can nevertheless still calculate the multi-loop reducible diagrams and sum them up explicitly. The result corresponds to adding a decaying factor to   the propagator:
\begin{align}
  \label{eq:decay-propagator}
G(\tau_a,\tau_b)
  \quad \rightarrow \quad
G(\tau_a,\tau_b)\exp\Big(-\# k^5(\tau_b^5-\tau_a^5)\Big)
\end{align}
 With the help of the resumed propagator, the folded limit of non-Gaussianity no longer diverges. The folded contribution of non-Gaussianity vanishes if taking $\tau_0 \rightarrow -\infty$. Once a finite initial time $\tau_0$ is given, explicit loop-corrected shapes of non-Gaussianities can be obtained.  For example, if the non-BD modes are set up at relatively early time, the non-Gaussianity may show some nontrivial shape   like Figure~\ref{NonBDTypical}. One of the underlying reason is that the large $k$ modes decay faster, while small $k$ modes decay relatively slowly and thus leave more prominent non-BD initial information on the observations.

Also, it is known that once given an initial time $\tau_0$, one can no longer use the $i\varepsilon$ terms to suppress the boundary terms in the UV. As a result, the tree level result depends on $\tau_0$ strongly and oscillations are present if the cutoff is sharp. Such dependence presents also for the BD initial condition. We show here that those $\tau_0$ dependent terms also decay with a similar exponent. In other words, the interactions practically serve as the $i\varepsilon$, and indeed pick up the physical initial state.

The rest of the paper is organized as follows: In Section \ref{Model}, we write down a simple interaction model and review the basic formalism. In Section \ref{Two-point function}, we calculate the two point correlation function. After recapitulating the one-loop folded limit cut-off result as given in our previous work, we come up with the new technique: loop chain diagram resummation. The dynamical RG method is also used to double check the result. In Section \ref{Three point function}, we calculate the loop corrected three point function. The decay of both the non-BD terms and the non-$i\varepsilon$ suppressed terms are manifest.

\section{Our Model}\label{Model}

We start from  general single field inflation with $\mathcal{L}=P(\phi,X)$~\cite{Chen:2006nt, Chen:2009bc}. The second and third order action up to the first order in slow parameter $\epsilon$ can be derived as
\beqn
S_2&=&\int dt d^3 x\; \Big[ a^3 \frac{\epsilon}{c_s^2}\dot{\zeta}^2-a \epsilon (\partial \zeta)^2 \Big]~,\\
S_3&=&\int dt d^3 x\;  a^3\Big\{ -\Big[\Sigma\Big(1-\frac{1}{c_s^2}\Big)  +2\lambda   \Big]\frac{\dot{\zeta}^3}{H^3}
-\frac{3\epsilon}{c_s^4}(1-c_s^2)\zeta\dot{\zeta}^2 +\frac{1}{a^2c_s^2}(1-c_s^2)\zeta(\partial \zeta)^2    \Big\}~,
\eeqn
where the dot ``\; $\dot{}$ \;'' denotes the derivative with respect to time $t$ and the prime ``\; $\rq{}$ \;'' denotes the derivative with respect to conformal time $\tau$. Also we set the reduced Planck mass $M_p=1$.

The relevant  quantities of this model are
\be
\Sigma=\frac{H^2 \epsilon}{c_s^2},\quad P_\zeta=\frac{H^2}{8\pi^2 c_s\epsilon}~.
\ee

From non-interacting $S_2$, we can quantize the field  $\zeta$:
\be
\zeta_{\bm k }^I(\tau)=u_{\bm k}(\tau)a_{\bm k}+u_{\bm k}^*(\tau)a_{-\bm k}^\dagger~,
\ee
with superscript ``$I$'' for interacting picture. The mode function is given by
\be
u_{\bm k }(\tau)=\frac{H}{ \sqrt{4 \epsilon c_s k^3}}\Big[ C_+(\bm k )(1+i k c_s \tau)e^{-ik c_s\tau}
       +C_-(\bm k ) (1-i k c_s \tau)e^{ik c_s\tau}  \Big]~.
\ee

As our motivation is to see the effects of interactions, we can use the sub-horizon limit approximation $|c_s k\tau| \gg 1$. The reason is that after horizon crossing, the modes are nearly frozen and can not evolve anymore. So, interactions can play no role in the super-horizon case.

In such limit, the mode function and its derivative have the following approximate behaviors
\be
u_{\bm k}\propto kc_s\tau e^{\pm i k c_s\tau},  \quad
\dot{u}_{\bm k}\propto \frac{1}{a}k^2c_s^2\tau e^{\pm i k c_s\tau}  ~.
\ee

For $\zeta$ field, the leading order time dependence is similar and note that the real space derivative corresponds to momentum multiplication in momentum space $\partial \zeta \leftrightarrow k \zeta_k$. So, we have the following relation
\be
\frac{1}{c_s} \dot{\zeta}\sim \frac{1}{a}\partial \zeta \sim \frac{1}{a} k^2 c_s \tau \sim  \frac{k}{a}\zeta~.
\ee

In the 3rd order action $S_3$ for interaction, the ratios of different terms are
\beqn
\frac{\text{2nd term}}{\text{3rd term}}&=&-3\frac{(\dot{\zeta}/c_s)^2}{(\partial \zeta/a)^2}\sim -3~,  \\
\frac{\text{1st term}}{\text{2nd term}}&=&\frac{1}{3H}\Big[  \frac{2\lambda c_s^4}{\epsilon(1-c_s^2)H^2} -1\Big] \frac{\dot{\zeta}}{\zeta}
\sim -\frac{1}{3 }\Big[  \frac{2\lambda c_s^4}{\epsilon(1-c_s^2)H^2} -1\Big] c_s k\tau ~.
\eeqn
We can easily see that in the sub-horizon limit, usually the first term is much larger than other two terms. So, one can just consider the first term and discard other two. This is because the highest dimensional operator is the most sensitive to the UV physics. Furthermore, for simplicity, we set sound speed $c_s$  to be 1.

Based on these arguments, we can consider a simple model of inflation described by
\be
S=\int dt d^3 x\; \Big[\epsilon a^3\dot{\zeta}^2-\epsilon a(\partial_i \zeta)^2
                          -2a^3\frac{\lambda}{H^3}\dot{\zeta}^3      \Big]
=S_2+\int d\tau d^3 x\; \Big[ -2a \frac{\lambda}{H^3}\zeta\rq{}^3   \Big]~,
\ee

The Hamiltonian for interaction is
\be
H_I(\tau)=-\int d^3 x\Big[     -2a \frac{\lambda}{H^3}\zeta\rq{}^3    \Big]
=\int d^3 x\Big[     -2 \frac{\lambda}{H^4}\frac{1}{\tau}\zeta\rq{}^3   \Big]~,
\ee
where the scale factor $a(\tau)\approx-\frac{1}{H\tau}$ for quasi de-Sitter space during inflaton.

In this simplified model, the mode function is given by
\be
u_{\bm  k}(\tau)=\frac{H}{2 \sqrt{\epsilon k^3}}\Big[C_+(\bm k)(1+i k \tau)e^{-i k \tau}+C_-(\bm k)(1-ik\tau)e^{i k \tau}\Big] ~,
\ee
where the coefficients $C_+,C_-$ are subject to the following constraint required by the consitency of quantization
\be\label{consistency_quantization}
|C_+|^2-|C_-|^2=1~.
\ee
In the usual case, the requirement of the vacuum state as a minimal energy state or the matching of de-Sitter space-time in the sub-horizon limit with Minkowski space-time will give rise to another condition $C_-=0$. This is the so called Bunch-Davies vacuum~\cite{Bunch:1978yq}. But, here we consider small $C_-$ , corresponding to non-Bunch-Davies case. To the first order, we have
$ C_+(\bm k)   \approx 1, C_-(\bm k)  \approx c_{\bm k} e^{i\theta_{\bm k} }$.
So the mode function and its derivative are given by
\beqn
u_{\bm k}(\tau)&\approx&\frac{H}{2 \sqrt{\epsilon  }} k^{-3/2}\Big[ (1+i k \tau)e^{-i k \tau}+ c_{\bm k} e^{i\theta_{\bm k} }(1-ik\tau)e^{i k \tau}\Big]\\
&\approx&
 \frac{H}{2 \sqrt{\epsilon  }} ik^{-1/2}\tau\Big[e^{-i k \tau}- c_{\bm k} e^{i\theta_{\bm k} }e^{i k \tau}\Big]     \qquad\qquad    (-k\tau \gg 1) ~, \\
u_{\bm k}\rq{}(\tau)&\approx& \frac{H}{2 \sqrt{\epsilon }} k^{1/2}\tau\Big( e^{-i k \tau}+ c_{\bm k} e^{i\theta_{\bm k} } e^{i k \tau}\Big) ~.
\eeqn

In order to make the story simple and clear in some sense, in the following calculations, we assume that the mode functions do not depend on the directions of momentum. Namely, we require $u_{\bm k}=u_k, \theta_{\bm k}=\theta_k$. The calculations and conclusions are expected to be more general independent of these assumptions except possible complications.

\section{Two-point function}\label{Two-point function}

\subsection{General consideration: tree level and one-loop level}
The interaction Hamiltonian in interaction picture is
\be
H_I(\tau)=\int d^3 x\Big[     -2 \frac{\lambda}{H^4}\frac{1}{\tau}{{\zeta^I}\rq{}}^3 \Big]
=\int \prod_{j=1}^3 \frac{d^3 \bm p_j}{(2\pi)^3}   \Big[-2 \frac{\lambda}{H^4}\frac{1}{\tau}
{\zeta_{\bm p_1}^I}\rq{}(\tau) {\zeta_{\bm p_2}^I}\rq{}(\tau){\zeta_{\bm p_3}^I}\rq{}(\tau)  \Big]
(2\pi)^3\delta^3\Big(\sum_{j=1}^3 \bm p_j \Big) ~.
\ee

The two-point correlation function can be calculated by using  the in-in formalism (see Appendix~\ref{in-in_formalism}):
\beqn
\EV{\zeta_{\bm k_1}(\tau)\zeta_{\bm k_2}(\tau)}
&=&\MatBK{0}{\zeta^I_{\bm k_1}(\tau)\zeta^I_{\bm k_2}(\tau)}{0}
+2\Imag \int_{\tau_0}^{\tau} d\tau_1 \MatBK{0}{\zeta^I_{\bm k_1}(\tau)\zeta^I_{\bm k_2}(\tau) H_I(\tau_1)}{0}\nonumber\\
&&+\int_{\tau_0}^{\tau} d\tau_1 \int_{\tau_0}^{\tau}d\tau_2 \MatBK{0}{H_I(\tau_1)\zeta^I_{\bm k_1}(\tau)\zeta^I_{\bm k_2}(\tau)H_I(\tau_2)}{0}\nonumber\\
&&-2 \Real \int_{\tau_0}^{\tau} d\tau_1 \int_{\tau_0}^{\tau_1}d\tau_2 \MatBK{0}{\zeta^I_{\bm k_1}(\tau)\zeta^I_{\bm k_2}(\tau)H_I(\tau_1)H_I(\tau_2)}{0}+\cdots
\eeqn

The zeroth order of two-point correlation function is given by (note our notation $\bm k=\bm k_1$):
\beqn\label{2pt_tree}
\MatBK{0}{\zeta^I_{\bm k_1}(\tau)\zeta^I_{\bm k_2}(\tau)}{0}&=&(2\pi)^3\delta^3(\bm k_1+\bm k_2) u_{\bm k }(\tau)u_{\bm k }^*(\tau)
 \nonumber\\
&\approx & (2\pi)^3\delta^3(\bm k_1+\bm k_2)   \frac{H^2}{4 \epsilon  }
k^{-1} \tau^2 \Big[1-2  c_{\bm k}\cos(2k\tau+\theta_{\bm k})   \Big]~,
\eeqn
where we consider the sub-horizon limit $-k\tau\gg1$ and only keep terms up to the first order in $ c_{\bm k}$.

The first order correction of two-point correlation function vanishes due to odd number of operators  or imbalance of creation and annihilation creators.

Next, we consider the second order loop corrections. There are two types of corrections:  non-BD mode in the loop and non-BD mode  in the external line. When the non-BD modes are in the external line, the physical meaning is very obvious if we cut the loop. This process can be thought as the decay of non-BD mode  in the external line into two BD modes in the loop. Furthermore, in order to match with tree level result as  will  be elaborated later \footnote{The physical meaning of matching the tree level result is that, the contributions coming from $c_{\bm p}$ and $c_{\bm q}$ correspond to processes where two long modes fuse into a short mode. When the short mode is far away from vacuum and the long mode is nearer to the vacuum (considering more time of decay), this is unlikely to happen. However, there is an important exception: Near thermal equilibrium, the detailed balance makes sure that the decay of the short mode is indeed balanced by the fusion of the long mode. Our calculation thus does not apply for such near equilibrium cases. An approach of Boltzmann equation would help and we hope to explore this possibility in the future.}, we need to have something like $c_{\bm k}$ which also implies a non-BD mode in the external leg. Thus, for simplicity, we can just consider this case by setting $c_{\bm p},c_{\bm q}=0$.

We are interested in the sub-horizon limit which means that $| k\tau |\gg 1$. Usually, this doesn't imply $ |(p+q-k) \tau |\gg 1$ in the folded limit. But, for simplicity, let's first consider the \emph{unfolded} case. In such a case, when evaluating the above equations, we only keep those terms which have highest power in $\tau$ and zeroth and first order in $ c_{\bm k}$. Then we can use the following integration formula $\int \tau^n e^{iQ\tau} d\tau  \approx\frac{\tau^n}{iQ}e^{iQ\tau}+\cdots$.

The second order symmetric part is:
\beqn
&&\int_{\tau_0}^{\tau} d\tau_1 \int_{\tau_0}^{\tau}d\tau_2 \MatBK{0}{H_I(\tau_1)\zeta^I_{\bm k_1}(\tau)\zeta^I_{\bm k_2}(\tau)H_I(\tau_2)}{0}\nonumber \\
&=&\Big(-2 \frac{\lambda}{H^4}\Big)^2\int_{\tau_0}^{\tau} \frac{d\tau_1}{\tau_1} \int_{\tau_0}^{\tau}\frac{d\tau_2}{\tau_2} \int  \prod_i \frac{d^3 \bm p_i}{(2\pi)^3} (2\pi)^3\delta^3(\sum_i \bm p_i)
 \int \prod_i \frac{d^3 \bm q_i}{(2\pi)^3} (2\pi)^3 \delta^3(\sum_i \bm q_i)
\times\nonumber \\ &&\Big[
\EV{\zeta^I_{\bm p_1}(\tau_1)\zeta^I_{\bm k_1}(\tau)}_0\EV{\zeta^I_{\bm p_2}(\tau_1)\zeta^I_{\bm q_1}(\tau)}_0
\EV{\zeta^I_{\bm p_3}(\tau_1)\zeta^I_{\bm q_2}(\tau)}_0\EV{\zeta^I_{\bm k_2}(\tau_1)\zeta^I_{\bm q_3}(\tau)}_0
+\text{different contractions} \Big]
 \nonumber \\
  &=&\Big(-2 \frac{\lambda}{H^4}\Big)^2\Big[ 3\times3\times2\times 2\Big]
(2\pi)^3\delta^3(\bm k_1+\bm k_2)u_{\bm k}^*(\tau )    u_{\bm k}(\tau)   \int \frac{d^3 q}{(2\pi)^3}
 kpq\Big(\frac{H}{2 \sqrt{\epsilon }}\Big)^6  f_S  ~,
\eeqn
where
\beqn
f_S&=&
\Big(\frac{H}{2 \sqrt{\epsilon }}\Big)^{-6}  \frac{1}{kpq}
\int_{\tau_0}^{\tau} \frac{d\tau_1}{\tau_1} \int_{\tau_0}^{\tau}\frac{d\tau_2}{\tau_2}
u_{\bm k}\rq{}(\tau_1)     u_{\bm p}\rq{}(\tau_1)     u_{\bm q}\rq{}(\tau_1)
u_{\bm k}\rq{}^*(\tau_2) u_{\bm p}\rq{}^*(\tau_2)    u_{\bm q}\rq{}^*(\tau_2) \nonumber\\
&=&\frac{\tau ^4}{(k+p+q)^2}+\frac{\tau
   _0^4}{(k+p+q)^2}-\frac{2 \tau _0^2 \tau ^2 \cos[ (k+p+q)(\tau-\tau_0)]}{(k+p+q)^2}
 +  \nonumber\\
&&\Bigg[   2 c_{\bm k} \Big(\frac{  \tau ^4   \cos  (\theta _{\bm k}+2 k
   \tau  )+ \tau _0^4    \cos  (\theta_{\bm k}+2 k \tau _0 )
   -2\tau _0^2   \tau ^2 \cos [\theta_{\bm k}+k
  (\tau +\tau _0 )] \cos [ (\tau -\tau _0 )   (p+q) ]}{(-k+p+q) (k+p+q)}  \Big)
     \nonumber\\  && \quad +2 \text{ permutations of }\bm k,\bm p,\bm q\Bigg] ~,
\eeqn
where we have defined $\bm k=\bm k_1, \bm p=\bm k+\bm q$.

The second order asymmetric part is:
\beqn
&&-2 \Real \int_{\tau_0}^{\tau} d\tau_1 \int_{\tau_0}^{\tau_1}d\tau_2 \MatBK{0}{\zeta^I_{\bm k_1}(\tau)\zeta^I_{\bm k_2}(\tau)H_I(\tau_1)H_I(\tau_2)}{0}    \nonumber\\
 &=&\Big(-2 \frac{\lambda}{H^4}\Big)^2\Big[ 3\times3\times2\times 2\Big]
(2\pi)^3\delta^3(\bm k_1+\bm k_2)u_{\bm k}^*(\tau )    u_{\bm k}(\tau)   \int \frac{d^3 q}{(2\pi)^3}
 kpq\Big(\frac{H}{2 \sqrt{\epsilon }}\Big)^6  f_A~,  
\eeqn
where
\beqn\label{unfolded_asym}
f_A/(-2)&=&\Real \Bigg[\Big(\frac{H}{2 \sqrt{\epsilon }}\Big)^{-6} \frac{1}{kpq}\frac{u_{\bm k}(\tau)}{u_{\bm k}^*(\tau)}
\int_{\tau_0}^{\tau}\frac{d\tau_1}{\tau_1 } \int_{\tau_0}^{\tau_1}\frac{d\tau_2}{\tau_2}\;
  u_{\bm k}\rq{}^*(\tau_1)    u_{\bm p}\rq{}(\tau_1)     u_{\bm q}\rq{}(\tau_1)
u_{\bm k}\rq{}^*(\tau_2) u_{\bm p}\rq{}^*(\tau_2)    u_{\bm q}\rq{}^*(\tau_2) \Bigg]   \nonumber\\
&=&\frac{\tau _0^4 \cos (2 k\tau -2k\tau_0)}{2 k(k-p-q)}+\frac{\tau ^4}{2 k (k+p+q)}
        -\frac{\tau _0^2 \tau ^2 \cos[(k+p+q)(\tau -\tau _0)] }{(k-p-q) (k+p+q)} \nonumber\\
&&+\Bigg[ \frac{2 (\tau ^5-  \tau   _0^5)    (p+q) c_{\bm k}
 \sin \left(\theta_{\bm k}+2 k \tau \right)}{5 (p+q-k) (k+p+q)}
 +\mathcal{O}(\tau^4, \tau_0^4, \tau^2\tau_0^2)\Bigg]~.
\eeqn
where we use the approximation $\frac{u_{\bm k}(\tau)}{u_{\bm k}^*(\tau)}=-e^{-2ik\tau}[1-2i  c_{\bm k} \sin(2k\tau+\theta_{\bm k} )]$.
It should be noted that we have $\tau^5$ terms now. This is because the exponential parts of mode functions  cancel and the power of $\tau$ increases after integration.
So, the final  result is
\beqn\label{2pt_all}
\EV{\zeta_{\bm k_1}(\tau)\zeta_{\bm k_2}(\tau)}&=&(2\pi)^3\delta^3(\bm k_1+\bm k_2)u_{\bm k}^*(\tau )    u_{\bm k}(\tau)  \Big[1+\frac{9\lambda^2}{4 H^2  \epsilon^3}\int \frac{d^3 q}{(2\pi)^3}   kpq(f_S+f_A) \Big] \nonumber\\
&\approx&(2\pi)^3\delta^3(\bm k_1+\bm k_2)\Big(\frac{H}{2 \sqrt{\epsilon }} \Big)^2
k^{-1} \tau^2   \nonumber\\ && \times
  \Big[1-2  c_{\bm k}\cos(2k\tau+\theta_{\bm k})
+\frac{9\lambda^2}{4 H^2  \epsilon^3}\int \frac{d^3 q}{(2\pi)^3}   kpq(f_S+f_A) +\cdots  \Big]~.
\eeqn
Because we are considering the sub-horizon limit,  $\tau^5$ terms dominate.  Note that in Eq.~(\ref{unfolded_asym}), the oscillation is sine function form while the tree level result in Eq.~(\ref{2pt_tree}) is cosine function form. This means that the unfolded part can not modify the amplitude of effective non-BD coefficient at leading order.

\subsection{Folded limit momentum cut-off}

In the previous calculations, we focus on the unfolded case and thus have the approximations $\int \tau^n e^{iQ\tau} d\tau  \approx\frac{\tau^n}{iQ}e^{iQ\tau}+\cdots$ for $|Q\tau| \gg 1$. While in the folded limit, $|Q\tau|\ll 1 $ with $Q=p+q-k$, so the  appropriate way is to expand $ e^{i Q\tau}=1+iQ\tau+\frac{i^2}{2}Q^2\tau^2+\cdots$. Thus, the integral can only be approximated as $\int d\tau \tau^n e^{iQ\tau}=\int d\tau \tau^n(1+iQ\tau+\frac{i^2}{2}Q^2\tau^2+\cdots)\approx \frac{\tau^{ n+1 }}{n+1}+\cdots$. It should be noted that  the leading term of the integral at small $Q$ case is real now.

With this in mind, we need to reexamine the  asymmetric part:
\beqn\label{loop_high}
f_A/(-2)&=&\Real \Bigg[\Big(\frac{H}{2 \sqrt{\epsilon }}\Big)^{-6} \frac{1}{kpq}\frac{u_{\bm k}(\tau)}{u_{\bm k}^*(\tau)}
\int_{\tau_0}^{\tau}\frac{d\tau_1}{\tau_1 } \int_{\tau_0}^{\tau_1}\frac{d\tau_2}{\tau_2}\;
  u_{\bm k}\rq{}^*(\tau_1)    u_{\bm p}\rq{}(\tau_1)     u_{\bm q}\rq{}(\tau_1)
  u_{\bm k}\rq{}^*(\tau_2) u_{\bm p}\rq{}^*(\tau_2)    u_{\bm q}\rq{}^*(\tau_2) \Bigg] \nonumber\\
&=&\Real \Bigg[ -e^{-2ik\tau}
\int_{\tau_0}^{\tau} d\tau_1  \int_{\tau_0}^{\tau_1} d\tau_2 \;
 \tau_1^2\tau_2^2 \Big(  c_{\bm k}e^{-i\theta_{\bm k}} e^{i( p+q+k)(\tau_2-\tau_1)}+ c_{\bm k}e^{-i\theta_{\bm k}} e^{i( p+q-k)(\tau_2-\tau_1)}+\cdots  \Big) \Bigg]~,
 \nonumber\\
  \eeqn
 where the ellipsis denotes the relatively irrelevant terms.

Now that the tree level result has different function form with unfolded loop correction, we may expect the dominant loop correction comes from the folded limit. So, we can Taylor expand the integrand and perform time integration, yielding
 \be
 f_A=-2\Real \Bigg[ -e^{-2ik\tau}
\int_{\tau_0}^{\tau} d\tau_1  \int_{\tau_0}^{\tau_1} d\tau_2 \;
 \tau_1^2\tau_2^2\;   c_{\bm k}e^{-i\theta_{\bm k}}\cdot  1   \Bigg]=- c_{\bm k} \cos(2k\tau+\theta_{\bm k})\frac{2(\tau^3-\tau_0^3)^2}{18}~.
\ee
As we expect, the tree level function form emerges which would be the dominant loop contributions. After that, loop momentum integration can be performed as follows by choosing a momentum cut-off near the folded limit $(p+q-k)\le \Lambda$,
\beqn
\int  \frac{d^3 q}{(2\pi)^3}   kpq f_A
&\sim&\int_{0\le(p+q-k)\le \Lambda}   \frac{d^3 q}{(2\pi)^3} kpq  c_{\bm k}\cos(2k\tau+\theta_{\bm k})  \frac{2}{18}(\tau^3-\tau_0^3)^2   \nonumber\\
 &=&\int_1^{1+\Lambda/k}d\mu \int_{-1}^1 d\nu  \frac{\pi k^3}{4} (\mu^2-\nu^2) \frac{1}{(2\pi)^3}
 \frac{\mu^2-\nu^2}{4}   k^3    c_{\bm k}\cos(2k\tau+\theta_{\bm k})  \frac{2}{18}(\tau^3-\tau_0^3)^2  \nonumber\\
&=&  c_{\bm k}\cos(2k\tau+\theta_{\bm k}) (\tau^3-\tau_0^3)^2 \frac{k^5 \Lambda}{1080\pi^2} ~.
\eeqn
When evaluating the above momentum integral, we use the elliptical coordinate system (see Appendix~\ref{elliptical_coordinate}).
While for the unfolded part, we can use Eq.~\eqref{unfolded_asym}, so
$
\int_{ p+q-k> \Lambda} \frac{d^3 q}{(2\pi)^3}   kpq f_A\propto  c_{\bm k}\sin(2k\tau+\theta_{\bm k})
$. Naively, the coefficient is infinity due to non-bounded momentum integration.  But physically,  if we take the renormalization counter term into considerations, the coefficient should be finite. Nevertheless, its dependence on non-BD coefficients are different from the folded limit one and tree level result.

Note that in order for the expansion to be valid, we require $|(p+q-k)(\tau_2-\tau_1) |\ll1 $  which leads to the condition $\Lambda(\tau-\tau_0)\lesssim 1$. A reasonable choice of cut-off is $\Lambda\approx 1/(\tau-\tau_0)$ in spite of an order one discrepancy. Collecting all the facts, we get final two-point correlation function under one-loop correlation,
\beqn\label{cut_off_result}
\EV{\zeta_{\bm k_1}(\tau)\zeta_{\bm k_2}(\tau) } &=& (2\pi)^3\delta^3(\bm k_1+\bm k_2)
 \frac{H^2}{4  \epsilon }   k^{-1}\tau^2   \Big[1 -2  c_{\bm k}\cos(2k\tau+\theta_{\bm k})
  \nonumber \\&&\times
  \Big(1-   \frac{ \lambda^2k^5}{480\pi^2 H^2  \epsilon^3} (\tau^2+\tau\tau_0+\tau_0^2)^2 (\tau-\tau_0) \Big) +\cdots\Big]~.
\eeqn
where the ellipsis includes the loop corrections of BD modes and higher order corrections to non-BD modes. It is very interesting to see that the loop corrections to the non-BD coefficients are negative, implying the decay of non-BD modes.

\subsection{Rigorous treatment of momentum integral}

From previous calculations, in the sub-horizon limit, by power counting, the $\tau^5$ terms are the dominant one. From Eq.~(\ref{2pt_all}), after loop correction,
 \be
-2  c_{\bm k}  \cos(2k\tau+\theta_{\bm k})\rightarrow-2  c_{\bm k} \cos(2k\tau+\theta_{\bm k})+\frac{9\lambda^2}{4H^2 \epsilon^3} \int\frac{d^3 q}{(2\pi)^3}   kpq(f_S+f_A) ~.
\ee
In our sub-horizon limit approximations, $f_S+f_A $ actually is given by Eq.~(\ref{loop_high}).
We can regard the right hand side as non-BD contribution with effective parameters running with time, so
\beqn
-2\Big( c_{\bm k}^{\text{eff}}\cos(2k\tau+\theta_{\bm k}^{\text{eff}})- c_{\bm k}\cos(2k\tau+\theta_{\bm k})\Big)
 &=& 2\frac{9\lambda^2}{4H^2 \epsilon^3}  c_{\bm k}\Big(\cos(2k\tau+\theta_{\bm k})\Real I-\sin(2k\tau+\theta_{\bm k}) \Imag  I \Big)~,
 \nonumber\\
\eeqn

The effective one can be written as
\be -2 c_{\bm k}^{\text{eff}} \cos(2k\tau+\theta_{\bm k}^{\text{eff}})
=-2\Big( c_{\bm k} \cos(2k\tau+\theta_{\bm k})+\delta  c_{\bm k} \cos(2k\tau+\theta_{\bm k})- c_{\bm k} \sin(2k\tau+\theta_{\bm k})\delta \theta_{\bm k}\Big)+\cdots ~,
\ee
where $\delta  c_{\bm k}= c_{\bm k}^{\text{eff}}- c_{\bm k},\delta \theta_{\bm k}=\theta_{\bm k}^{\text{eff}}-\theta_{\bm k}$ are very small when $\tau$ and $\tau_0$ are very close. By matching the form, we can get
\be\label{RG_eq}
 c_{\bm k}^{\text{eff}}- c_{\bm k}=\delta  c_{\bm k}=-  c_{\bm k}\frac{9\lambda^2}{4H^2 \epsilon^3}  \Real I,  \qquad
 c_{\bm k} \delta \theta_{\bm k}=c_{\bm k} \frac{9\lambda^2}{4H^2 \epsilon^3}  \Imag I ~.
\ee

The integral $I$ can be simplified by using elliptical coordinate system (see Appendix~\ref{elliptical_coordinate})
\begin{eqnarray*}
I&=&  \int \frac{d^3 q}{(2\pi)^3}\; k p q
\int_{\tau_0}^{\tau} d\tau_1   \int_{\tau_0}^{\tau_1}\ d\tau_2\;
 \tau_1^2\tau_2^2  \; \Big(  e^{i( p+q-k)(\tau_2-\tau_1)}+  e^{i( p+q+k)(\tau_2-\tau_1)}\Big) \\
   &=&\frac{k^6}{128\pi^2}\int_1^\infty d\mu \int_{-1}^1 d\nu
   (\mu^2-\nu^2)^2   \int_{\tau_0}^{\tau} d\tau_1   \int_{\tau_0}^{\tau_1}\ d\tau_2\;
 \tau_1^2\tau_2^2  \;
  \Big( e^{i( \mu-1)k(\tau_2-\tau_1)}+ e^{i(\mu+1)k(\tau_2-\tau_1)}\Big)\\
  &=&\frac{k^6}{128\pi^2} (I_-+I_+) ~.
\end{eqnarray*}

Let's focus on the  integral $I_-$ first. We define $z=\mu-1$ and  find that
\beqn
 S(z)&=&\int_{-1}^1 d\nu   (\mu^2-\nu^2)^2   \int_{\tau_0}^{\tau} d\tau_1   \int_{\tau_0}^{\tau_1}\ d\tau_2\;
 \tau_1^2\tau_2^2  \;   e^{i( \mu-1)k(\tau_2-\tau_1)}
 \nonumber \\  &=&
 \text{polynomial terms of } z+\frac{1}{z}\Big[\Big(-\frac{16 i \tau ^5}{75 k}+\frac{8 \tau ^4}{3 k^2}+\frac{64 i \tau ^3}{9 k^3}+\frac{16 i \tau _0^5}{75 k}+\frac{8 \tau _0^4}{3 k^2}-\frac{64 i \tau _0^3}{9 k^3}\Big)
 \nonumber \\&&
 +e^{-i k z (\tau -\tau _0)} \Big(-\frac{16 \tau _0^2 \tau ^2}{3 k^2}-\frac{64 i \tau _0 \tau ^2}{3 k^3}+\frac{16 \tau ^2}{k^4}+\frac{64 i \tau _0^2 \tau }{3 k^3}-\frac{32 \tau _0 \tau }{k^4}-\frac{8 i \tau }{k^5}
 +\frac{16 \tau _0^2}{k^4}+\frac{8 i \tau _0}{k^5}\Big)  \Big]
  \nonumber \\&& +
 \text{fractional terms like  } \{\frac{1}{z^2},\frac{e^{-i k z (\tau -\tau _0)}}{z^2},\frac{1}{z^3}, \cdots ,\frac{1}{z^6},\frac{e^{-i k z (\tau -\tau _0)}}{z^6}\}     ~.
\eeqn

Then, we need to integrate over $\mu$ or $z$.
From the expression for $S(z)$, we can write it as the following general form
\be
S(z)=\sum_{n=1}^6 \frac{a_n e^{-ikz(\tau-\tau_0)}-b_n}{z^n}+\text{polynomial of }z ~.
\ee
The polynomial part of $S(z)$ implies the power divergence if we integrate over $z$. They can be discarded if we choose to believe that these terms can be canceled by the  the local counter terms.  So, we only need to care the fractional part $S_{f}$. In the UV region and sub-horizon limit, intuitively, the most relevant contributions come from those terms with lowest power in $1/z$ and highest power in $\tau$, i.e. $ \frac{16 i(\tau_0^5- \tau ^5)}{75 kz}$ for the current problem. But, this navie choice is a little problematic due to the divergence near $z\sim 0$. Fortunately, we can simplify this problem by introducing a new basis functions $T_n(z)$ (see Appendix~\ref{basis_function})
\be
S_f(z)=\sum_{n=1}^6  \frac{a_n e^{ izu }-b_n}{z^n}=\sum_{n=1}^6 A_n T_n(z) \;,\quad\text{with} \qquad u=-k(\tau-\tau_0)~.
\ee

We can easily solve the above equation to obtain $A_n$. In particular, $A_1$ is given by
\be
A_1= -\frac{64 i \tau ^3}{9 k^3}+\frac{64 i \tau _0^3}{9 k^3}-\frac{8 \tau ^4}{3 k^2}-\frac{8 \tau _0^4}{3 k^2}+\frac{16 i \tau ^5}{75 k}-\frac{16 i \tau _0^5}{75 k}~.
\ee
Thus we have $A_1=b_1$ as  emphasized in Appendix~\ref{basis_function}.
So,
\beqn
  I_-&=& \int_1^\infty d\mu \int_{-1}^1 d\nu
   (\mu^2-\nu^2)^2   \int_{\tau_0}^{\tau} d\tau_1   \int_{\tau_0}^{\tau_1}\ d\tau_2\;
 \tau_1^2\tau_2^2  \;   e^{i( \mu-1)k(\tau_2-\tau_1)} \\
 &\rightarrow&\int_0^\infty dz S_f(z)= \sum_{n=1}^6   A_n \int_0^\infty dz   T_n(z) ~.
\eeqn
As mentioned  in Appendix~\ref{basis_function},  $T_1$ is not integrable and thus, we need to choose a momentum cut off on $z$,
\be
\sum_{n=2}^6   A_n \int_0^\infty dz   T_n(z) = \frac{136 i \tau ^5}{375 k}-\frac{16 i \tau _0 \tau ^4}{75 k}-\frac{8 i \tau _0^2 \tau ^3}{75 k}+\frac{8 i \tau _0^3 \tau ^2}{75 k}+\frac{16 i \tau _0^4 \tau }{75 k}-\frac{136 i \tau _0^5}{375 k}+\mathcal{O}(\tau^4,\tau_0^4,\cdots)~,
\ee
\be\label{one_loop_all}
A_1 \int_0^\Lambda dz   T_1(z) =
A_1\Big(\Ci(k \Lambda  (\tau - \tau_0))-i \Si(k \Lambda  (\tau - \tau_0))-\log (k \Lambda  (\tau - \tau_0))-\gamma_E\Big)  ~.
\ee
In the sub-horizon limit, we only need to keep the highest power term in $\tau$, i.e. $\tau^5$ terms. Then, $I_-$ can be approximated as
\beqn\label{one_loop}
  I_-&=&\frac{8 i (\tau -\tau _0) (17 \tau ^4+7 \tau _0 \tau ^3+2 \tau _0^2 \tau ^2+7 \tau _0^3 \tau +17 \tau _0^4)}{375 k}
\nonumber \\&&
+ \frac{16 i (\tau^5-\tau _0^5)}{75 k}\Big(\Ci(k \Lambda  (\tau - \tau_0))
    -i \Si(k \Lambda  (\tau - \tau_0))-\log (k \Lambda  (\tau - \tau_0))-\gamma_E\Big) ~.
\eeqn
It's interesting to note that
\be\label{one_loop_real}
 \Real I_-= \frac{16 (\tau^5-\tau _0^5)}{75 k}  \Si(k \Lambda  (\tau - \tau_0))\rightarrow \frac{8\pi (\tau^5-\tau _0^5)}{75 k} \quad  \text{when }\quad k \Lambda  (\tau - \tau_0) \rightarrow+\infty~.
\ee

Similarly, we can get
\beqn
\Real I_+&=& \frac{16 (\tau^5-\tau _0^5)}{75 k}\Big(\Si(k \Lambda  (\tau - \tau_0)) - \Si(2 k (\tau - \tau_0)) \Big)
\nonumber\\
&\xrightarrow{k \Lambda  (\tau - \tau_0) \rightarrow+\infty}&
\frac{8\pi (\tau^5-\tau _0^5)}{75 k}\Big(1 -\frac{2}{\pi}\Si(2 k (\tau - \tau_0)) \Big) ~.
\eeqn
When $k(\tau-\tau_0)\gtrsim 1 , \Real I_+\sim 0$.

\be
I=\frac{k^6}{128\pi^2} (I_-+I_+)
  \rightarrow  \Real I=\frac{k^6}{128\pi^2} \frac{8\pi (\tau^5-\tau _0^5)}{75 k}=\frac{1}{1200\pi } k^5(\tau^5-\tau _0^5)~,
\ee
so
\be
 c_{\bm k}^{\text{eff}}= c_{\bm k}- c_{\bm k} \frac{9\lambda^2}{4H^2 \epsilon^3}\Real I
= c_{\bm k}\Big(1- \frac{3\lambda^2}{1600\pi H^2 \epsilon^3} k^5(\tau^5-\tau _0^5)\Big)~.
\ee
Compared with the folded limit cut-off result~(\ref{cut_off_result}), they only differ by a decay factor, i.e. $3(\tau^4+\tau^3\tau_0+\tau^2\tau_0^2+\tau^1\tau_0^3+\tau_0^4 )/1600$ and $( \tau^4+2\tau^3\tau_0+3\tau^2\tau_0^2+2\tau^1\tau_0^3+\tau_0^4 )/(480\pi)$,
which is of order $\mathcal{O}(1)$. Another important observation is that $\tau^5$ terms appear  in unfolded case, folded limit cut-off result and present rigorous treatment with universal power. So, our strategy can be like this: analyze the unfolded case first, extract the highest power terms and then transform back to the folded limit. Very amazingly, the transformations can be simplified by just changing $1/z$ to $T_1(z)$ with exactly the same coefficients (remember $b_1=A_1$).

The one loop correction is small as long as the initial time and final time are close enough. Once the time difference becomes large, the one loop perturbation is not valid any more. We need to try to cure the secular growth with time, either through the dynamical renormalization group method or by turning to higher order loop analysis.

\subsection{Dynamical Renormalization Group method}
The one loop corrections of effective parameters are

\be
\delta  c_{\bm k}=-  c_{\bm k}\frac{9\lambda^2}{4H^2 \epsilon^3}  \Real I,  \qquad
  \delta \theta_{\bm k}= \frac{9\lambda^2}{4H^2 \epsilon^3}  \Imag I  ~.
\ee
Once we come to realize that $ c_{\bm k}$ should be $ c_{\bm k}^{\text{eff}}$ and running, the $ c_{\bm k}$  above should be replaced with
$ c_{\bm k}^{\text{eff}}(\tau)$ . This yields
\be
\frac{ c_{\bm k}^{\text{eff}}(\tau)- c_{\bm k}}{ c_{\bm k}^{\text{eff}}(\tau)}=-\frac{9\lambda^2}{4H^2 \epsilon^3}  \Real I  ~,
\ee
with initial condition  $ c_{\bm k}^{\text{eff}}(\tau_0)= c_{\bm k}$. It is easy to get
\be
 c_{\bm k}^{\text{eff}}(\tau)= c_{\bm k}\exp\Big( - \frac{3\lambda^2}{1600\pi H^2 \epsilon^3} k^5(\tau^5-\tau _0^5)\Big)  ~.
\ee

The idea essentially is the dynamical renormalization group method~\cite{Boyanovsky:2003ui,Boyanovsky:2004ph,Burgess:2009bs}. The effects of early time modes to later time modes through loop corrections can be viewed as the modifications of effective parameters, yielding the running effective parameters with time. For DRG method, the physical picture is very clear and enlightened. In the next section, we are going to provide another way which is more  rigorous in mathematics.

\subsection{Multi-loop analysis and loop chain diagram resummation}

In the following, we are going to consider the higher loop corrections to the non-BD coefficients. In principle, there are infinite ways to draw the corresponding Feynman diagrams, nested or non-nested. As we see before, all the modes which run in the loop are BD modes. Non-BD modes in the loop will not affect the effective non-BD coefficients $ c_{\bm k}$ and thus are not considered.
Furthermore, we only consider the  non-nested loop-chain diagram which consists of loops connected in series. Due to the time sequence of interacting vertices, there are still lots of diagrams which have the loop-chain topology but differ in the time ordering. Thanks to the sub-horizon limit, we only need to keep the highest power term in $\tau$. In such a limit,   we are going to show that only those V-shaped diagrams dominate.

\begin{figure}
\subfigure[BD loop ]{\includegraphics[width=0.20\textwidth]{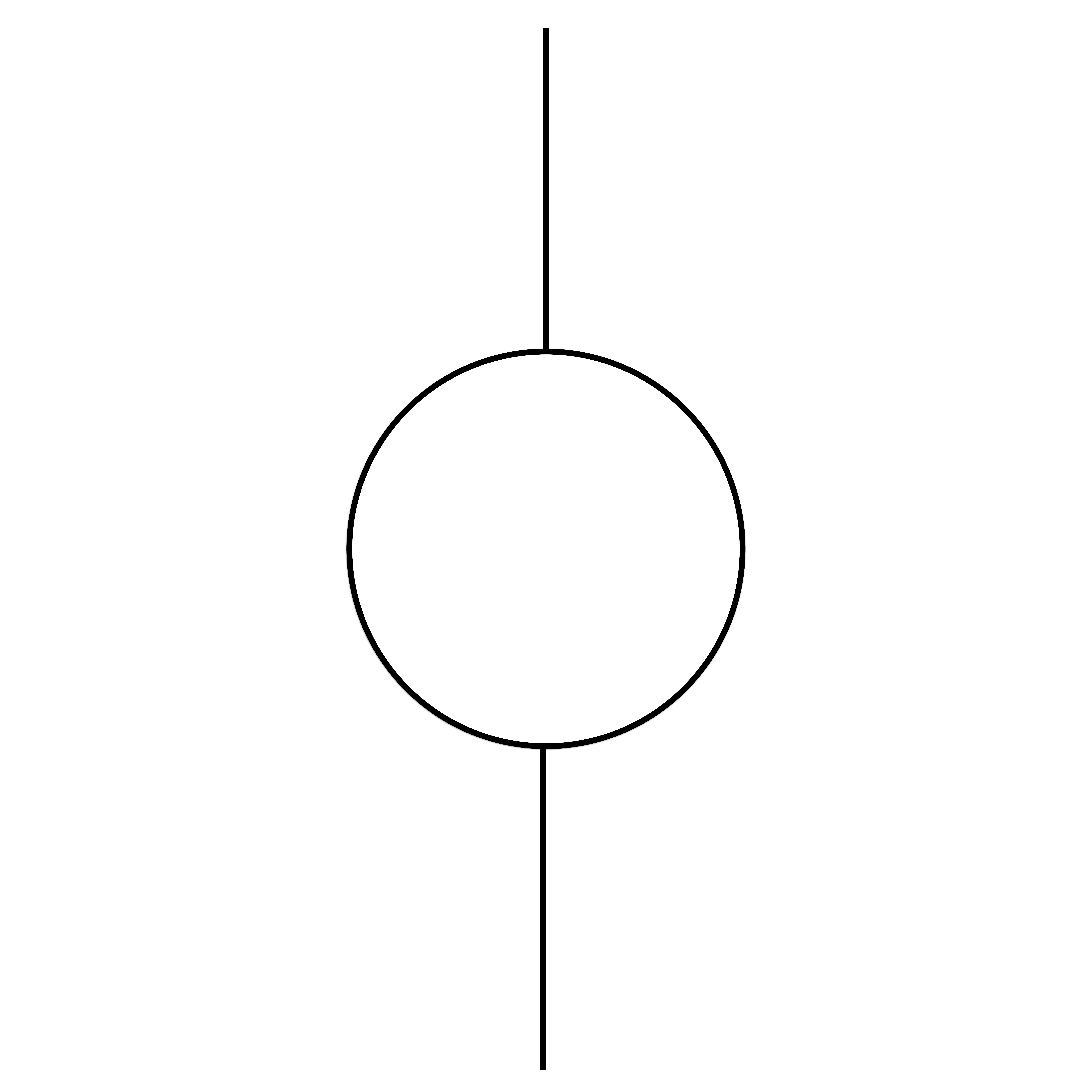} \label{loop_component_a}}
\quad
\subfigure[Non-BD loop ]{\includegraphics[width=0.20\textwidth]{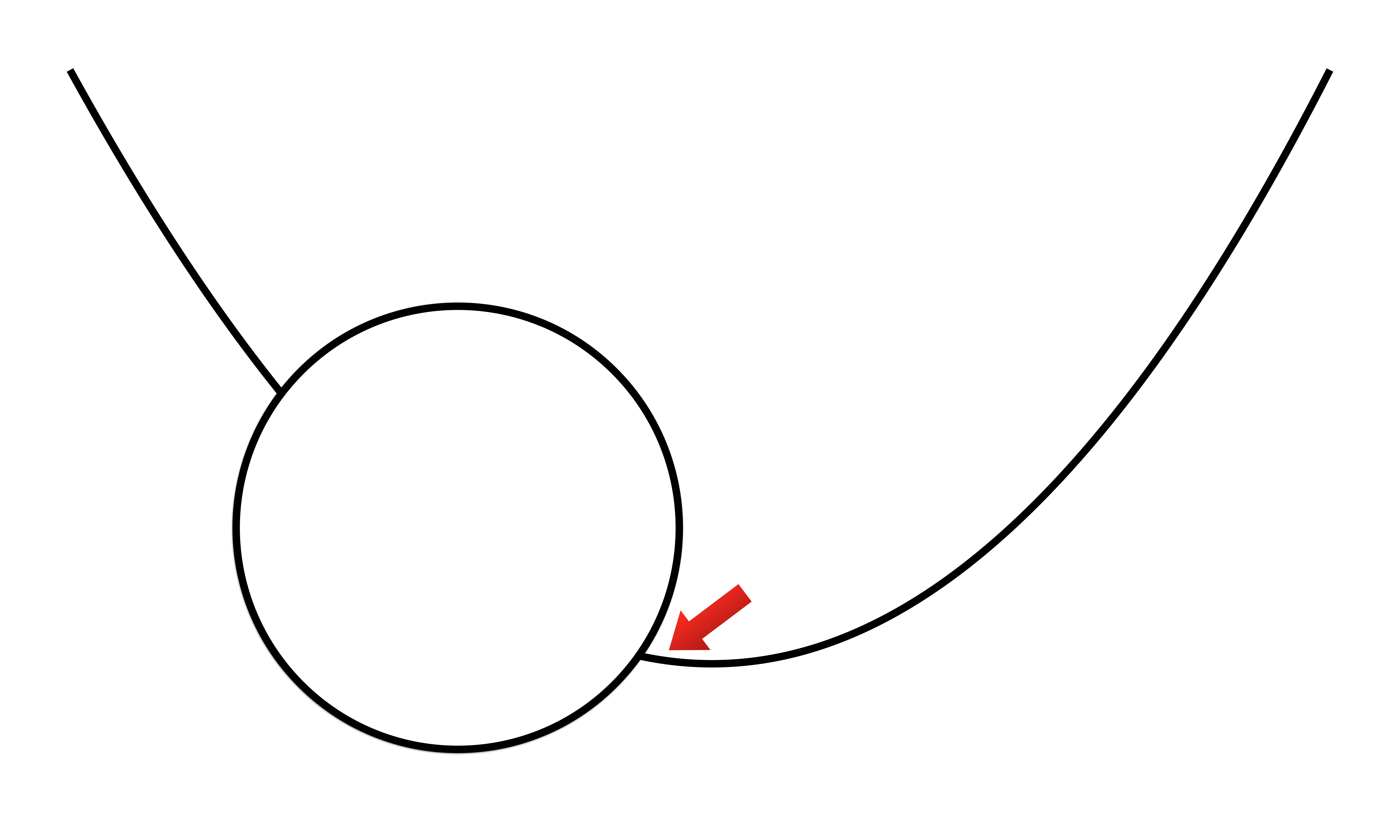} \label{loop_component_b}}
\quad
\subfigure[Non-BD loop ]{\includegraphics[width=0.20\textwidth]{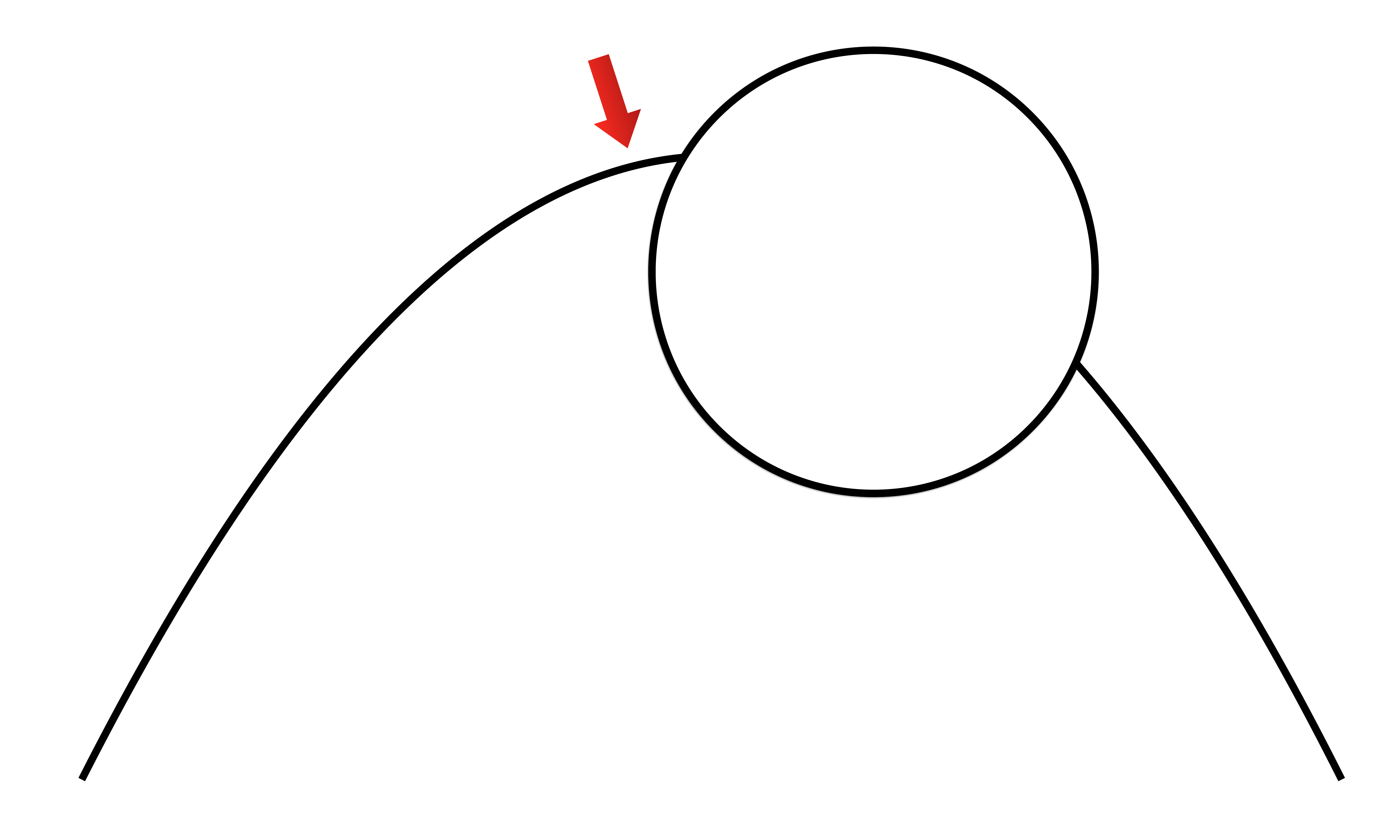} \label{loop_component_c}}
\quad
\subfigure[V-shaped loop chain  diagram]{\includegraphics[width=0.25\textwidth]{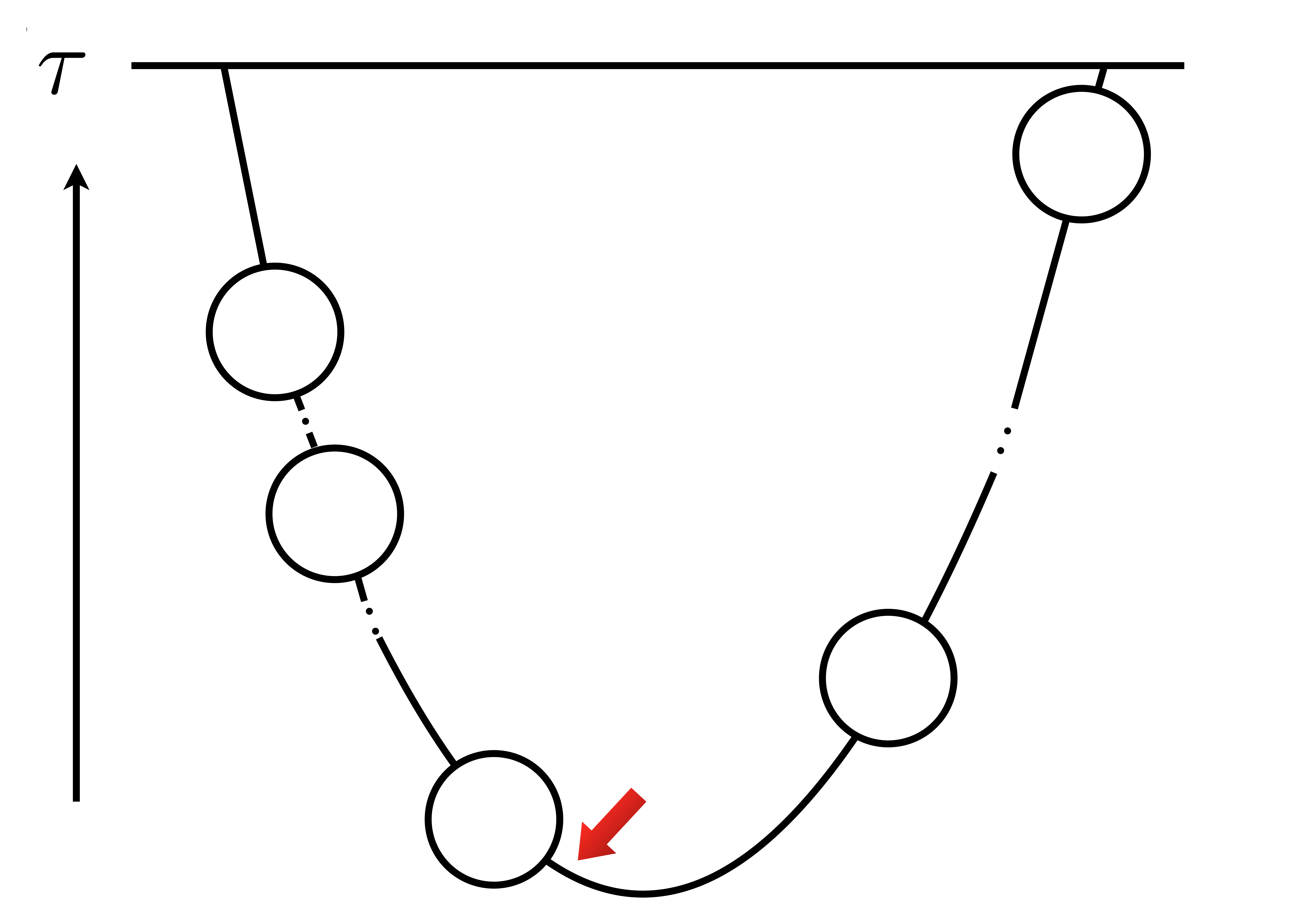} \label{Vshape_loop}}
\caption{Basic building bricks of two-point function Feynman diagrams. All the modes running in the loop are BD modes. The red arrow indicates possible non-BD modes which can yield $\tau^5$ terms. Non-BD loop means that one of the two modes which connect loop and other components of the whole diagram is non-BD mode (We only consider the first order in $ c_{\bm k}$), while BD loop means that all the six modes which run in the loop and connect loop with other components are BD modes. For any loop, the possible time ordering has three possibilities (a), (b) and (c). If we want to get $\tau^5$ term, we require (a) to be BD loop and (b),(c) to be non-BD loop with non-BD modes indicated by the red arrow. After some considerations, only the V-shaped diagrams are dominant as shown in (d).  }
\label{loop_component}
\end{figure}

Let us analyze the basic component of the diagram first. For each loop, there are three possibilities of time sequence as shown in Figure~\ref{loop_component}.

In  Figure~\ref{loop_component_a}, suppose $\tau_U>\tau_1>\tau_2>\tau_L$ ($\tau_U,\tau_L$ are upper and lower limit of integration and $\tau_1,\tau_2$ are time of two interacting vertices of the loop) and  note that loop modes are BD-modes, we have the  following integral
\beqn
I_a &\propto& \int \frac{d^3 q}{(2\pi)^3} k p q  \int_{\tau_L}^{\tau_U}\frac{d\tau_1 }{\tau_1} \int_{\tau_L}^{\tau_1}  \frac{d\tau_2 }{\tau_2}
u_{\bm p}\rq{}(\tau_1)u_{\bm p}\rq{}^*(\tau_2) u_{\bm q}\rq{}(\tau_1)u_{\bm q}\rq{}^*(\tau_2) u_{\bm k}\rq{}^*(\tau_1)u_{\bm k}\rq{}(\tau_2)
\nonumber\\
 &\propto&  \int \frac{d^3 q}{(2\pi)^3} k p q  \int_{\tau_L}^{\tau_U} d\tau_1   \int_{\tau_L}^{\tau_1} d\tau_2 \tau_1^2 \tau_2^2
e^{i(p+q)(\tau_2-\tau_1)}\Big(e^{ik\tau_1}+ c_{\bm k} e^{-i\theta_{\bm k}}e^{-ik\tau_1}\Big)\Big(e^{-ik\tau_2}+ c_{\bm k} e^{ i\theta_{\bm k}}e^{ik\tau_2}\Big)
\nonumber\\
 &\propto&  \int \frac{d^3 q}{(2\pi)^3} k p q  \int_{\tau_L}^{\tau_U} d\tau_1   \int_{\tau_L}^{\tau_1} d\tau_2 \tau_1^2 \tau_2^2
 \Big(e^{i(p+q-k)(\tau_2-\tau_1)}+ c_{\bm k} e^{-i\theta_{\bm k}}e^{i(p+q)(\tau_2-\tau_1)}e^{-ik(\tau_1+\tau_2)}+\cdots \Big)
\nonumber\\
 &\propto&  \int \frac{d^3 q}{(2\pi)^3} k p q  \Big(-i\frac{\tau_U^5-\tau_L^5}{5(p+q-k)}+\mathcal{O}(\tau^4)\Big) ~,
\eeqn
where the $\tau^5$ terms are contributed by $e^{i(p+q-k)\Delta \tau}$, while $ \mathcal{O}(\tau^4)$ are contributed by other terms. In the sub-horizon limit, only $\tau^5$ terms are most relevant, so in the original expression, we only need to keep  $e^{i(p+q-k)\Delta \tau}$ like terms. In particular, in such case, all the related modes are BD modes.

In Figure~\ref{loop_component_b}, suppose $\tau_U>\tau_1>\tau_2>\tau_L$, similarly we get
\beqn
I_b &\propto& \int \frac{d^3 q}{(2\pi)^3} k p q  \int_{\tau_L}^{\tau_U}\frac{d\tau_1 }{\tau_1} \int_{\tau_L}^{\tau_1}  \frac{d\tau_2 }{\tau_2}
u_{\bm p}\rq{}(\tau_1)u_{\bm p}\rq{}^*(\tau_2) u_{\bm q}\rq{}(\tau_1)u_{\bm q}\rq{}^*(\tau_2) u_{\bm k}\rq{}^*(\tau_1)u_{\bm k}\rq{}^*(\tau_2)
\nonumber\\
 &\propto&  \int \frac{d^3 q}{(2\pi)^3} k p q  \int_{\tau_L}^{\tau_U} d\tau_1   \int_{\tau_L}^{\tau_1} d\tau_2 \tau_1^2 \tau_2^2
e^{i(p+q)(\tau_2-\tau_1)}\Big(e^{ik\tau_1}+ c_{\bm k} e^{-i\theta_{\bm k}}e^{-ik\tau_1}\Big)\Big(e^{ ik\tau_2}+ c_{\bm k} e^{- i\theta_{\bm k}}e^{-ik\tau_2}\Big)
\nonumber\\
 &\propto&  \int \frac{d^3 q}{(2\pi)^3} k p q  \int_{\tau_L}^{\tau_U} d\tau_1   \int_{\tau_L}^{\tau_1} d\tau_2 \tau_1^2 \tau_2^2
 \Big( c_{\bm k} e^{-i\theta_{\bm k}}e^{i(p+q -k)(\tau_2-\tau_1) }+ c_{\bm k} e^{-i\theta_{\bm k}}e^{i(p+q+k)(\tau_2-\tau_1)}  +\cdots\Big)
\nonumber\\
 &\propto&  \int \frac{d^3 q}{(2\pi)^3} k p q  \Big(-i\frac{\tau_U^5-\tau_L^5}{5(p+q-k)}  c_{\bm k} e^{-i\theta_{\bm k}}
  -i\frac{\tau_U^5-\tau_L^5}{5(p+q+k)}  c_{\bm k} e^{-i\theta_{\bm k}}+\mathcal{O}(\tau^4)\Big)~,
\eeqn
Similarly $\tau^5$ terms are contributed by $e^{i(p+q\pm k)\Delta \tau}$. Now, we have two such terms. The first one corresponds to the folded limit case, while the second one vanishes when $k(\tau_U-\tau_L) \gtrsim \mathcal{O}(1)$ according to the previous one-loop analysis. Thus, we only need to keep $e^{i(p+q-k)\Delta \tau}$ terms and now, the late time $\tau_2$ external mode is non-BD mode.

For Figure~\ref{loop_component_c},  the analysis is nearly identical to that in Figure~\ref{loop_component_b}.

Besides the behavior of building components mentioned above, there are several other interesting properties for the whole diagram:
\begin{itemize}

\item The time sequence of these interacting vertices should be all time-ordered or anti-time-ordered. There is no mixing. In another word, in the in-in formalism, the relevant
contribution comes from $\zeta^2 H...H$ and $H...H\zeta^2$ which are related by some complex conjugation.
\beqn\label{2pt_in_in}
\EV{\zeta_{\bm k_1}(\tau)\zeta_{\bm k_2}(\tau)}
 &=&
\sum_{ m=0}^{\infty}    i^m
\int _{  \tau _0}^\tau  d\tilde{\tau}_m  \int_{\tau_0}^{\tilde{\tau}_m}  d\tilde{\tau}_{m-1} ...\int_{\tau_0}^{\tilde{\tau}_2}  d\tilde{\tau}_1
\EV{H_I(\tilde{\tau}_1)...H_I(\tilde{\tau}_m) \zeta_{\bm k_1}^I(\tau)\zeta_{\bm k_2}^I(\tau) }_0
\nonumber \\&&
+\sum_{n =0}^{\infty}    (-i)^n
 \int _{\tau_0}^{\tau}  d\tau_1 \int _{\tau_0}^{\tau_1}  d\tau_2...\int _{\tau_0}^{\tau_{n-1}}  d\tau_n
\EV{ \zeta_{\bm k_1}^I(\tau)\zeta_{\bm k_2}^I(\tau)  H_I( \tau_1 )...H_I( \tau_n) }_0
\nonumber\\ &=&
2\Real \sum_{n =0}^{\infty}    (-i)^n
 \int _{\tau_0}^{\tau}  d\tau_1 \int _{\tau_0}^{\tau_1}  d\tau_2...\int _{\tau_0}^{\tau_{n-1}}  d\tau_n
\EV{  \zeta_{\bm k_1}^I(\tau)\zeta_{\bm k_2}^I(\tau)  H_I( \tau_1 )...H_I( \tau_n) }_0
 ~.  \nonumber\\
\eeqn
The reason is that when we go along the loop chain, we start from time $\tau$ and finally go back to time $\tau$. So, there must be some turning and extremal loops. These loops should be non-BD loops if we require the $\tau^5$ terms. While we only keep terms up to first order in  non-BD coefficients $c_{\bm k}$, there can only be one non-BD loop and the only possible configurations are V-shaped with all time-ordered or anti-time-ordered interaction time sequence.

\item The two time of vertex at each loop should be consecutive. The reason is not to disturb the time integration and get as high power as possible.
\end{itemize}

All these conditions are verified by explicit calculations.

Remember that we only keep the lowest order term in $ c_{\bm k}$, which means that we can only have one non-BD mode in whole diagram. In order to get as high order terms
as possible, the diagram can only be composed of lots of  BD loops (as shown in Figure~\ref{loop_component_a}) and \emph{one} non-BD loop (as shown in Figure~\ref{loop_component_b}). The final dominant diagram is V-shaped loop chain diagram (as shown in Figure~\ref{Vshape_loop}) where the tip of
V is the non-BD loop (see Figure~\ref{loop_component_b}). The non-BD mode has the  earliest time and connects the non-BD loop with other BD loop.

So, the final contribution is of the form (The time ordering is $\tau_{2L-1}>\tau_{2L}>\tau_{2L-3}>...>\tau_3>\tau_4>\tau_1>\tau_2$)
\beqn
\mathcal I_L(k;\tau_0,\tau_{2L+2}) &=&\prod_{j=1}^L \int \frac{d^3 q_j}{(2\pi)^3} k p_j q_j \int_{\tau_0}^{\tau_{2j+2}} d\tau_{2j-1}\int_{\tau_0}^{\tau_{2j-1}} d\tau_{2j } \;  \tau_{2j -1}^2 \tau_{2j }^2\; e^{i(p_j+q_j-k)(\tau_{2j}-\tau_{2j-1})} \nonumber \\
&=& \int \frac{d^3 q_L}{(2\pi)^3} k p_L q_L \int_{\tau_0}^{\tau_{2L+2}} d\tau_{2L-1}\int_{\tau_0}^{\tau_{2L-1}} d\tau_{2L }
\; \tau_{2L -1}^2 \tau_{2L }^2\;
\nonumber \\ && \times
 e^{i(p_L+q_L-k)(\tau_{2L}-\tau_{2L-1})}\mathcal I_{L-1}(k;\tau_0,\tau_{2L }) ~.
\eeqn
Here, we make an assumption or approximation:
\beqn
\Real\mathcal  I_L(k;\tau_0,\tau_{2L+2})&=&\Real\Big[ \int \frac{d^3 q_L}{(2\pi)^3} k p_L q_L \int_{\tau_0}^{\tau_{2L+2}} d\tau_{2L-1}\int_{\tau_0}^{\tau_{2L-1}} d\tau_{2L } \;  \tau_{2L -1}^2 \tau_{2L }^2
\nonumber \\ && \times
  e^{i(p_L+q_L-k)(\tau_{2L}-\tau_{2L-1})} \Real\mathcal  I_{L-1}(k;\tau_0,\tau_{2L }) \Big]~.
\eeqn
Mathematically, it will give rise to very easy and interesting result. The other part starts from two-loop. We drop this two-loop contribution for the reason that shall be explained at the end of this subsection.

As we calculated before ($\mathcal I_0=1$),
\beqn
\mathcal I_1= \int \frac{d^3 q_1}{(2\pi)^3} k p_1 q_1 \int_{\tau_0}^{\tau_{4}} d\tau_{1}\int_{\tau_0}^{\tau_{1}} d\tau_{2  } \;  \tau_{1}^2 \tau_{2  }^2\; e^{i(p_1+q_1 -k)(\tau_{2}-\tau_{1})} \mathcal I_{0}  \rightarrow \Real\mathcal I_1=\frac{k^5(\tau_4^5-\tau_0^5)}{1200 \pi} ~.
\nonumber\\ \eeqn
By mathematical induction, we can show that under the previous approximation
\be
\Real\mathcal  I_L(k;\tau_0,\tau_{2L+2})=\frac{1}{L!} \Big( \frac{k^5(\tau_{2L+2}^5-\tau_0^5)}{1200 \pi} \Big)^L~.
\ee
\emph{Proof}: Suppose it holds for $\mathcal  I_{L-1}$, then
\beqn
\Real\mathcal  I_L&=&\Real \int \frac{d^3 q_L}{(2\pi)^3} k p_L q_L \int_{\tau_0}^{\tau_{2L+2}} d\tau_{2L-1}\int_{\tau_0}^{\tau_{2L-1}} d\tau_{2L } \;  \tau_{2L -1}^2 \tau_{2L }^2\; e^{i(p_L+q_L-k)(\tau_{2L}-\tau_{2L-1})}
\nonumber \\ && \times
\frac{1}{(L-1)!} \Big( \frac{k^5(\tau_{2L }^5-\tau_0^5)}{1200 \pi} \Big)^{L-1} \nonumber \\
&=&\Real\int \frac{d^3 q_L}{(2\pi)^3} k p_L q_L  \frac{1}{(L-1)!} \Big( \frac{k^5}{1200 \pi} \Big)^{L-1}
\cdot \frac{-i(\tau_{2L+2}^5-\tau_0^5)^L}{5L (p_L+q_L-k)}+\cdots \nonumber \\
&=&\frac{1}{L!} \Big( \frac{k^5(\tau_{2L+2}^5-\tau_0^5)}{1200 \pi} \Big)^L  ~,
\eeqn
where the momentum integral can be obtained from the following:
\beqn
\int  d^3 q \;  k p q \frac{1}{p+q-k}+\cdots
&=&\frac{\pi k^3}{4} \int_1^\infty d\mu \int_{-1}^1 d\nu\; (\mu^2-\nu^2) k^3 \frac{(\mu+\nu)(\mu-\nu)}{4}\frac{1}{(\mu-1)k}+\cdots \nonumber \\
&=&\frac{\pi k^5}{16} \int_0^\infty dz\;\frac{16}{15}\frac{1}{z}+\cdots
\eeqn
where we substitute $z=\mu-1$ after integrating $\nu$ and  $+\cdots$ represents all possible terms which can not be be written as $1/(p+q-k)$ or $1/(\mu-1)$ or $1/z$.

Using the basis function in Appendix~\ref{basis_function} and note that $b_1=A_1$,  we have
\beqn
&\rightarrow&\frac{\pi k^5}{16} \int_0^\infty dz \;   \frac{-16}{15}   T_1(z)+\cdots
\rightarrow \frac{\pi k^5}{16} \frac{-16}{15}\frac{-i\pi}{2}+\cdots= i\frac{\pi^2  k^5}{30}+\cdots  ~.
\eeqn

Next, we  need to do combinatorial counting for  V-shaped diagrams and restore the coefficient factors. For $L$ loop diagram, each loop can be on the left or the right leg of V except the loop at the tip of V, so there are $2^L/2$ diagrams.  In addition, in the series expansion of in-in formalism~(\ref{2pt_in_in}), we have a factor $2(- i )^{2L}$. And we should notice that there are three mode functions at each interacting vertex which are symmetric, implying a factor $(3\times3\times2)^L$. Finally, the coupling factor of interaction and the numeric factor of mode functions should also be included. After taking all of these into considerations, we are led to a factor
\be
2(- i)^{2L} \cdot (3\times3\times2)^L \cdot 2^L/2\cdot  \Big(\frac{H}{2 \sqrt{\epsilon}} \Big)^{6L}  \Big(\frac{-2\lambda}{H^4} \Big)^{2L} = \Big( - \frac{9\lambda^2}{4 \epsilon^3H^2} \Big)^L~.
\ee

So, the final result for the effective non-BD coefficient is
\beqn
\frac{ c_{\bm k}^{\text{eff}}(\tau)}{ c_{\bm k}}=\sum_{L=0}^\infty   \Big( - \frac{9\lambda^2}{4 \epsilon^3H^2} \Big)^L \frac{1}{L!} \Big( \frac{k^5(\tau^5-\tau_0^5)}{1200 \pi} \Big)^L
=\exp\Big(- \frac{3\lambda^2 k^5(\tau_{ }^5-\tau_0^5 )}{1600\pi  \epsilon^3H^2}\Big)~.
\eeqn

\noindent\emph{Remarks on the approximation}:
The real part of one loop result~(\ref{one_loop_real}) is exact. The calculations above show that higher loop corrections can be thought as the power of one loop result in some sense. But the even power of the imaginary part of one loop is real, which means that the imaginary part of one loop result is very important in higher loops. But anyway, we can regard them as the phase factor corrections  $e^{i\gamma}$ with $\gamma\sim \mathcal{O}(\tau_0^5,\tau^5)$. The rigorous treatment of this phase factor is beyond our capability due to the  \emph{log} term in Eq.~(\ref{one_loop_all}) which may require involved and subtle regularization and renormalization procedure~\cite{Senatore:2009cf,Weinberg:2005vy}. But we can justify the \emph{log} term from a physical perspective.

In one loop   calculation Eq.~\eqref{one_loop}, the cut-off divergence related part is $ \Ci(k \Lambda  (\tau - \tau_0))     -\log (k \Lambda  (\tau - \tau_0)) $. When $\tau-\tau_0\rightarrow 0$, it vanishes. Otherwise, the cosine integral function contribute nothing and we can just consider the \emph{log} part.  The divergence is expected to be canceled by the counter term which has the form $....\log(\Lambda_{phy} /\mu)$ where $\mu$ is the physical renormalization scale and the coefficient is exactly as   that of our loop result if the divergence is indeed canceled. Note that $\Lambda_{phy}=\Lambda k/a(\tau_{\text{mid}})$ with $\tau_0<\tau_{\text{mid}}<\tau$. So, the final result should be of the form $-\log(\Lambda_{phy}a(\tau_{\text{mid}}) (\tau - \tau_0))+\log(\Lambda_{phy} /\mu)=\log(\frac{H}{\mu}\frac{\tau-\tau_0}{-\tau_{\text{mid}}})$. In order for the one loop level perturbation result to be valid, $k^5(\tau^5-\tau_0^5)$ can not be too large and thus $\tau/\tau_0\sim \mathcal{O}(1)$ in the sub-horizon limit $|k\tau|,|k\tau_0|\gg 1$. After realizing this fact, the  \emph{log} term can only contribute finitely.  So, the imaginary part of Eq.~\eqref{one_loop} is $\Imag I_- \sim (\tau-\tau_0)\mathcal{O}(\tau_0^4,\tau_0^3\tau,...\tau^4)\sim \mathcal{O} (\tau^5-\tau_0^5)$. After applying the arguments to each loop of multi-loop calculation, we can conclude that the final contribution of one loop imaginary part is a phase factor correction   $e^{i\gamma}$ with $\gamma\sim \mathcal{O}(\tau_0^5,\tau^5)$. The exact form or coefficient is complicated, for simplicity, we ignore the phase factor correction and only consider the amplitude suppression.

\begin{figure}
\centering
 \includegraphics[width=0.4\textwidth]{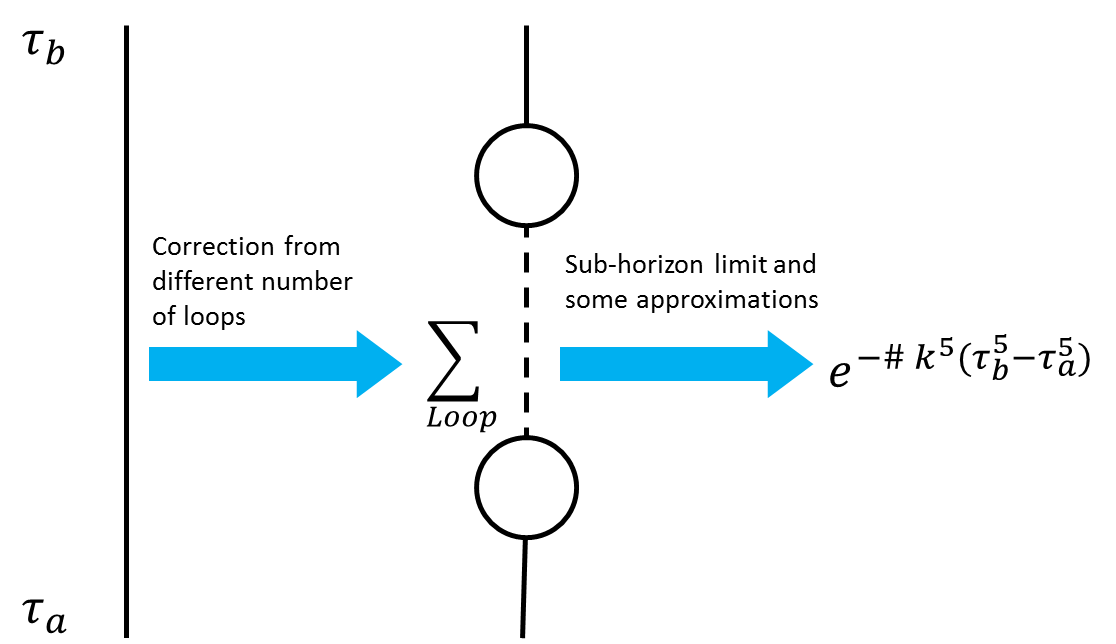}
\caption{After including loop correction, essentially, the effect is  renormalizing the tree level propagators by a exponential factor $\exp ({-\#k^5(\tau_b^5-\tau_a^5)} )$. This property has nothing to do with non-BD. Actually, it is the general result of loop corrections.
 }
\label{Loop_Exp}
\end{figure}

\section{Three-point function}\label{Three point function}

\begin{figure}

\subfigure[  $e^{i(k_1+k_2+k_3)\tau_V}+c_{\bm k_1}e^{i(-k_1+k_2+k_3)\tau_V}$ ]{\includegraphics[width=0.20\textwidth]{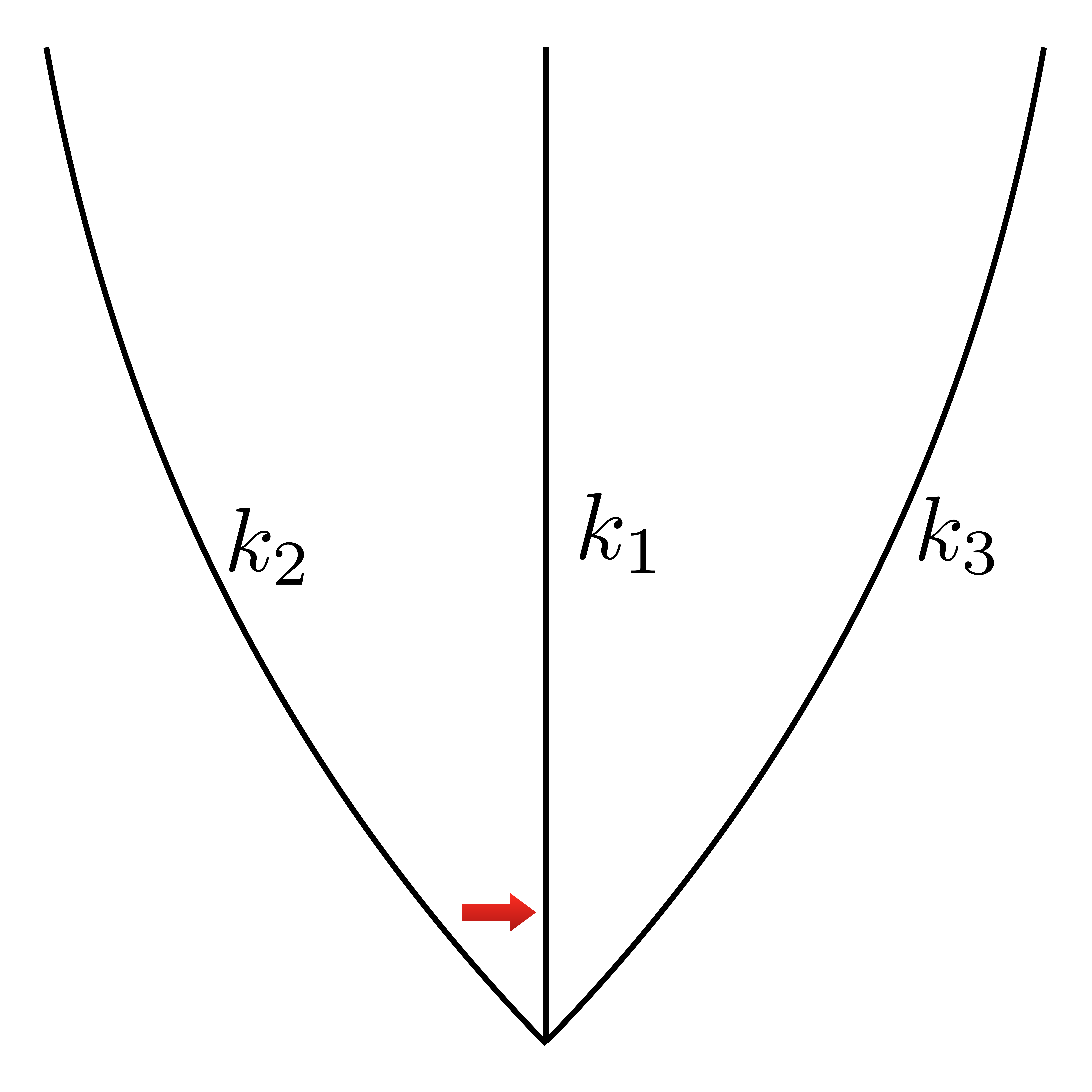}}
\quad
\subfigure[$J_{\text{BD}}(  L_1,L_2,L_3)$]
{\includegraphics[width=0.24\textwidth]{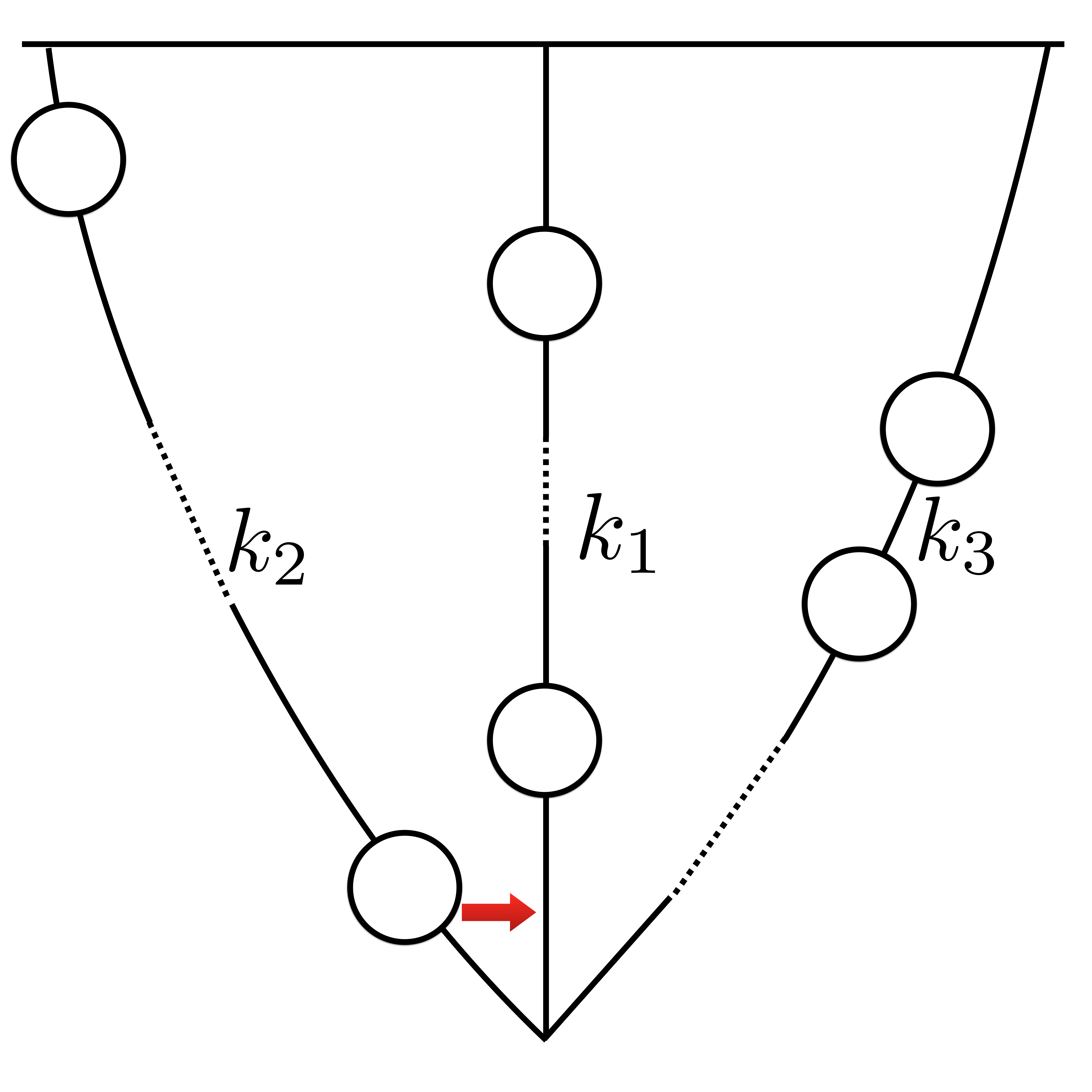}\label{3pt_Multi_Loop_BD}}
\quad
\subfigure[$e^{i(-k_1+k_2+k_3)\tau_V}+c_{\bm k_1}e^{i( k_1+k_2+k_3)\tau_V}$ ]{\includegraphics[width=0.20\textwidth]{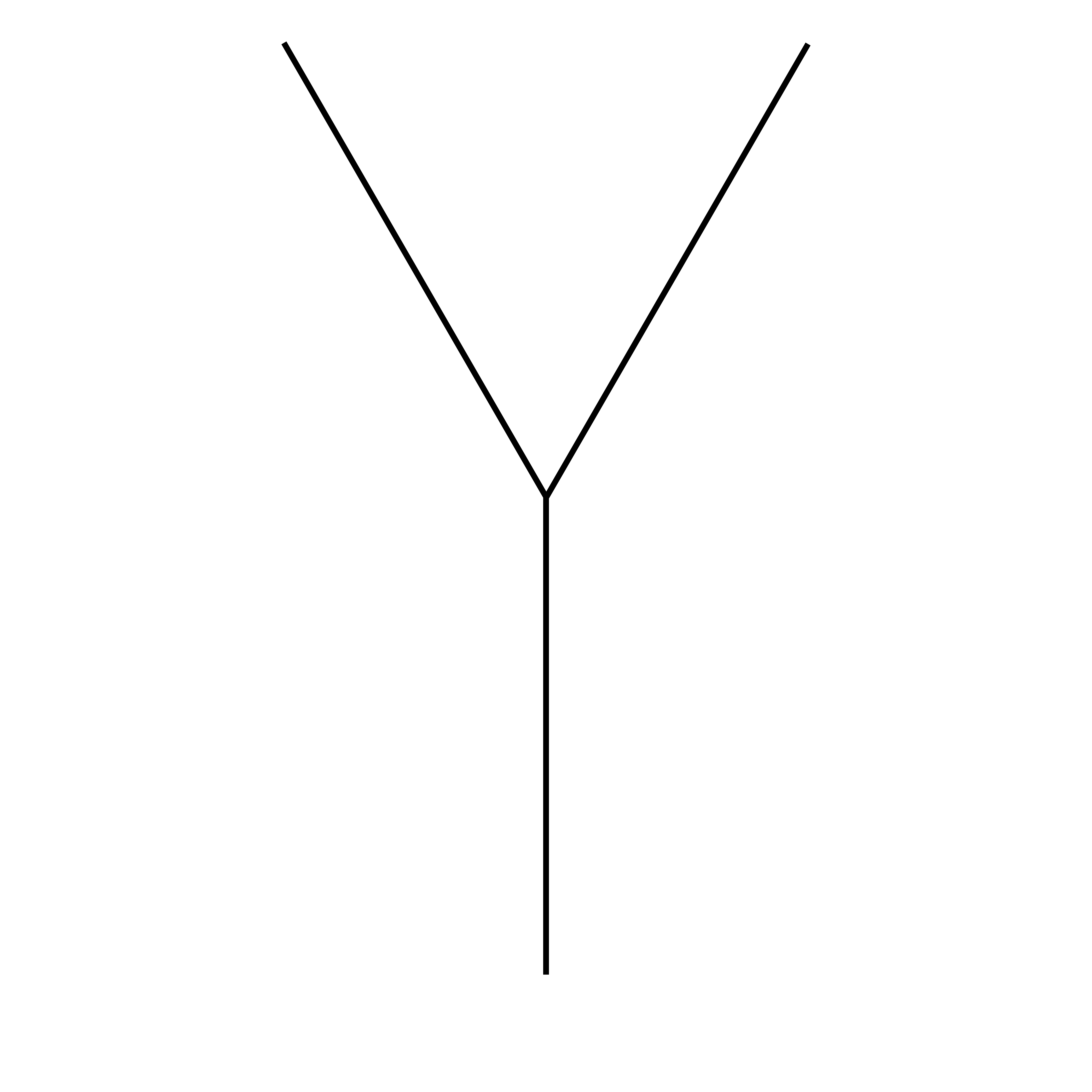}}
\quad
\subfigure[$J_{\text{Non-BD}}(\widetilde L_1,L_1,L_2,L_3)$]
{\includegraphics[width=0.24\textwidth]{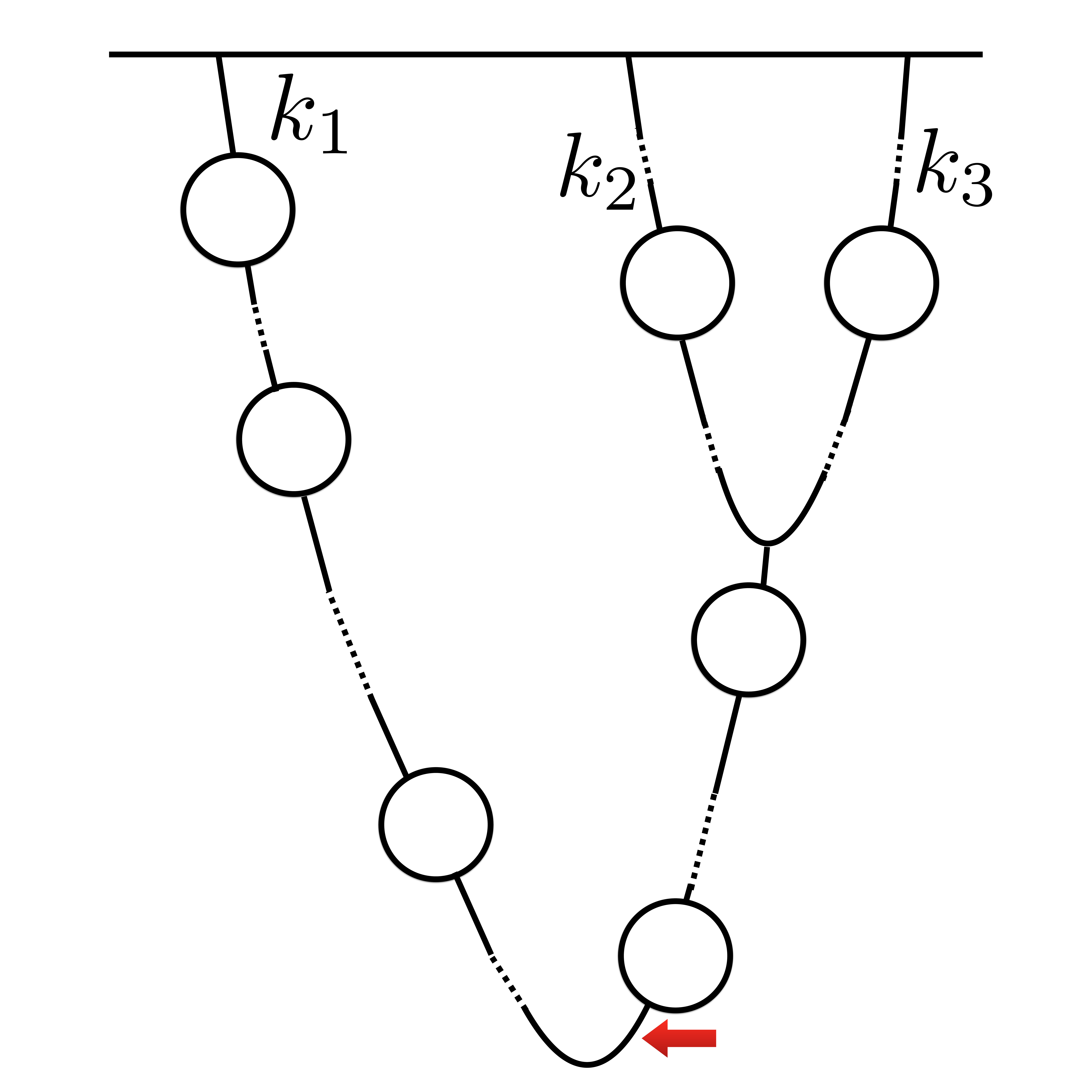}\label{3pt_Multi_Loop_NonBD_Extra}}
\caption{Three-point function. For BD three point function (3pt function unrelated with $c_{\bm k}$), (a) is the tree level diagram and (b) is the loop correction to 3pt function where each propagator in tree diagram is replaced by a chain of loops. For non-BD three point function (3pt function proportional to $c_{\bm k}$), both (a) and (c) can contribute a factor  $e^{i(k_2+k_3-k_1)\tau_V}$. Loop diagrams of non-BD 3pt functions (b) and (d) can be constructed from (a) and (c). The red arrow indicates the possible non-BD mode. It is interesting to note that when earlier time loop number in (d) is taken to be 0, its value is just (b), i.e. mathematically, $J_{\text{Non-BD}}(\widetilde L_1=0,L_1,L_2,L_3)=J_{\text{BD}}(  L_1,L_2,L_3)$.
}
\end{figure}

 \subsection{Tree-level result}
 Following the standard in-in formalism, we can easily obtain the tree level three-point function

 \beqn
  \EV{\zeta_{\bm k_1}(\tau)\zeta_{\bm k_2}(\tau)\zeta_{\bm k_3}(\tau)}^{\text{Tree}} &=&
 \EV{\zeta^I_{\bm k_1}(\tau)\zeta^I_{\bm k_2}(\tau)\zeta^I_{\bm k_3}(\tau)}_0+ 2\Imag \int_{\tau_0}^\tau d\tau_V
 \EV{\zeta^I_{\bm k_1}(\tau)\zeta^I_{\bm k_2}(\tau)\zeta^I_{\bm k_3}(\tau)H_I(\tau_V)}_0 \nonumber\\
 &=&(2\pi)^3 \delta(\bm k_1+\bm k_2+\bm k_3) 2 \Imag \int_{\tau_0}^\tau d\tau_V
   \frac{-2\lambda}{H^4\tau_V}\Big[3\times 2\Big]
   \prod_j u_{\bm k_j}(\tau)  u\rq{}_{\bm k_j}^*(\tau_V)
    \nonumber\\
 &=&(2\pi)^3 \delta(\bm k_1+\bm k_2+\bm k_3) 2 \Imag \Big[ u_{\bm k_1}(\tau) u_{\bm k_2}(\tau)u_{\bm k_3}(\tau)    \int_{\tau_0}^\tau  d\tau_V\;  \frac{-3\lambda}{2H   \epsilon^{3/2}}\tau_V^2
 \nonumber \\ &&  \times
 \sqrt{k_1 k_2 k_3}
   \Big(e^{i (k_1+k_2+k_3)\tau_V}  +c_{\bm k_1} e^{-i\theta_{\bm k_1}}e^{i(k_2+k_3-k_1)\tau_V} +2 \text{ perm.}\Big) \Big]
    \nonumber \\
 \eeqn


 \begin{figure}\label{3pt_one_loop}
\subfigure[Tree]{\includegraphics[width=0.32\textwidth]{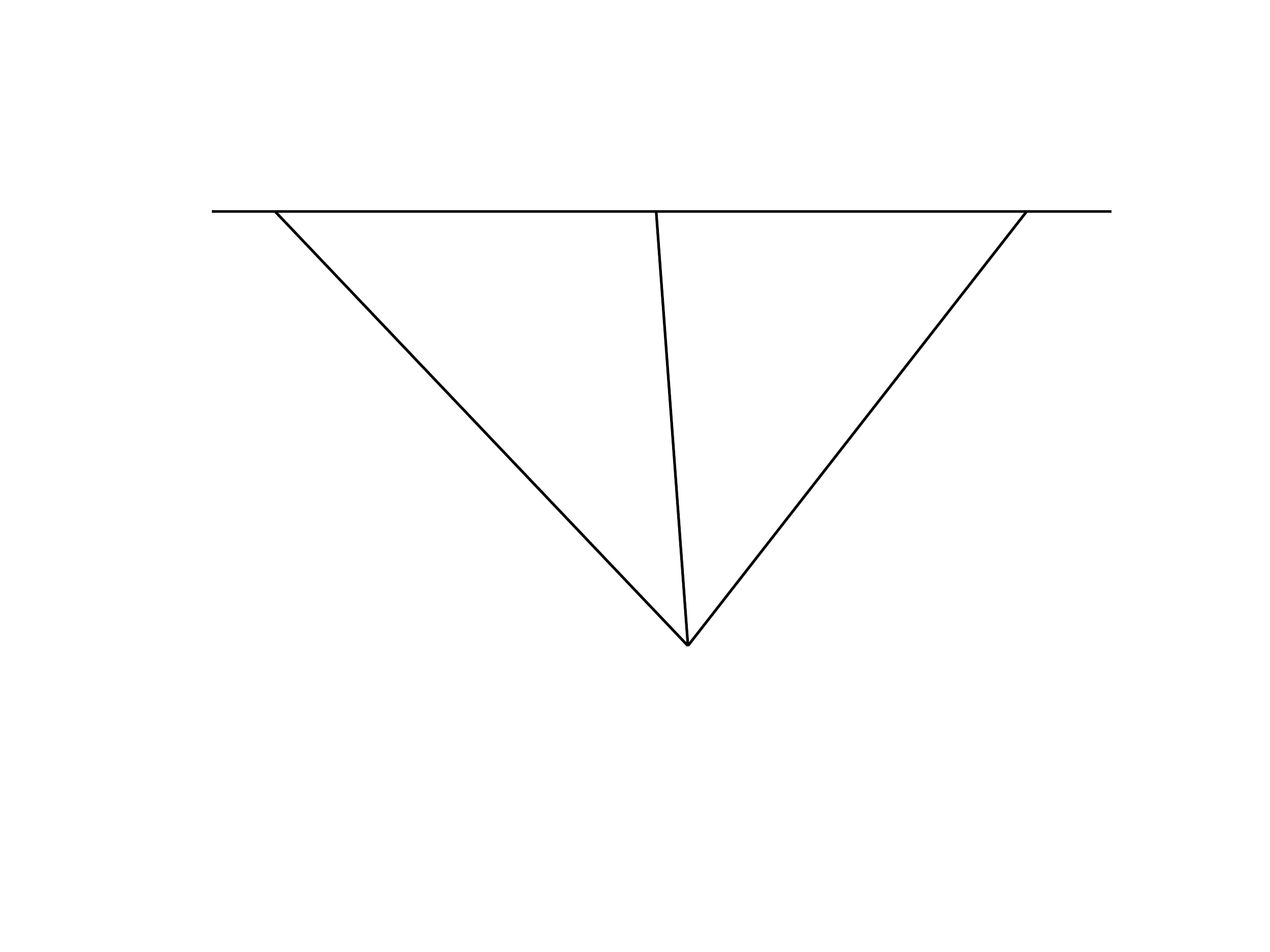}}
\quad
\subfigure[one-loop with $\tau_V<\tau_L$  ]{\includegraphics[width=0.3\textwidth]{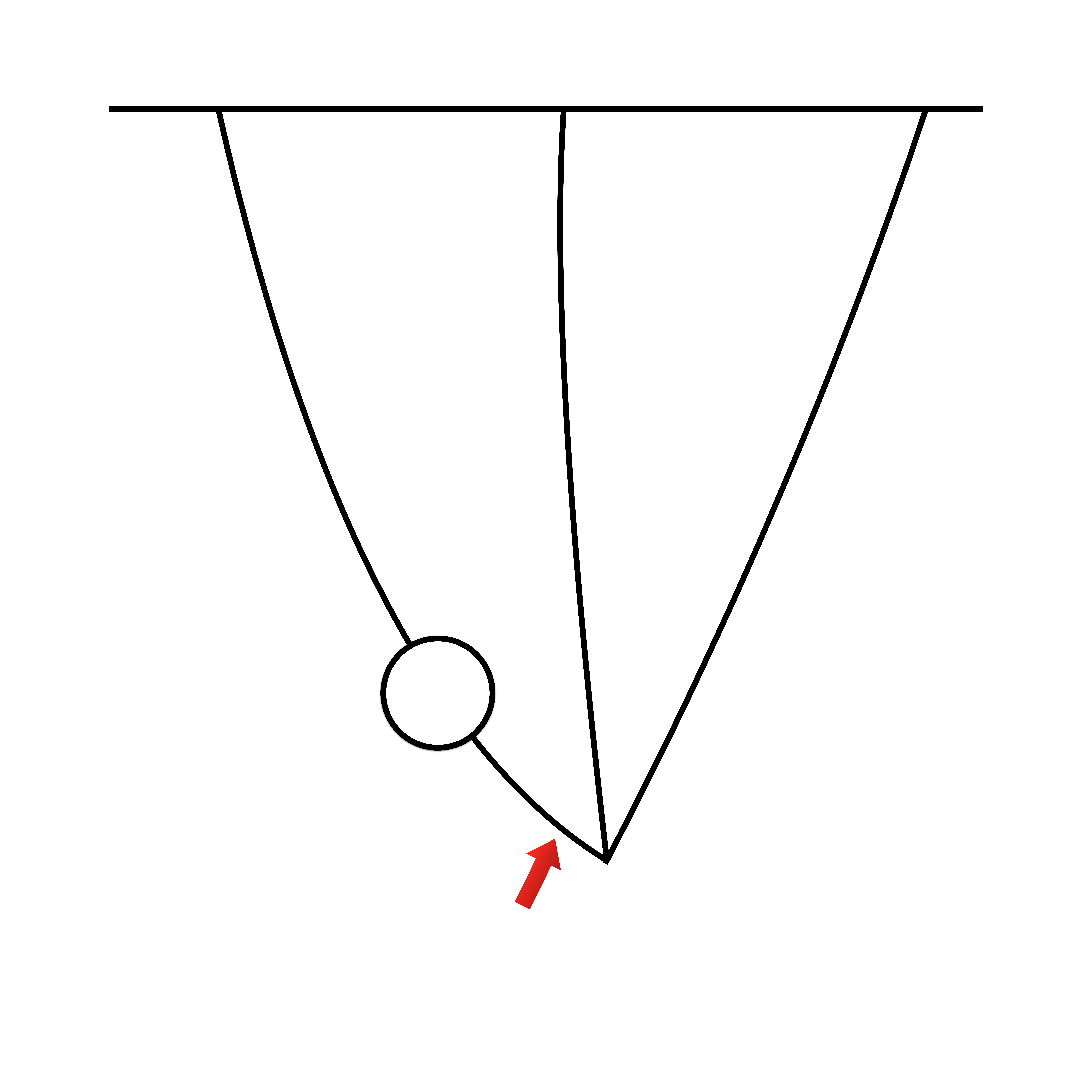}\label{3pt_one_loop_BDlike}}
\quad
\subfigure[one-loop with $\tau_V>\tau_L$  ]{\includegraphics[width=0.3 \textwidth]{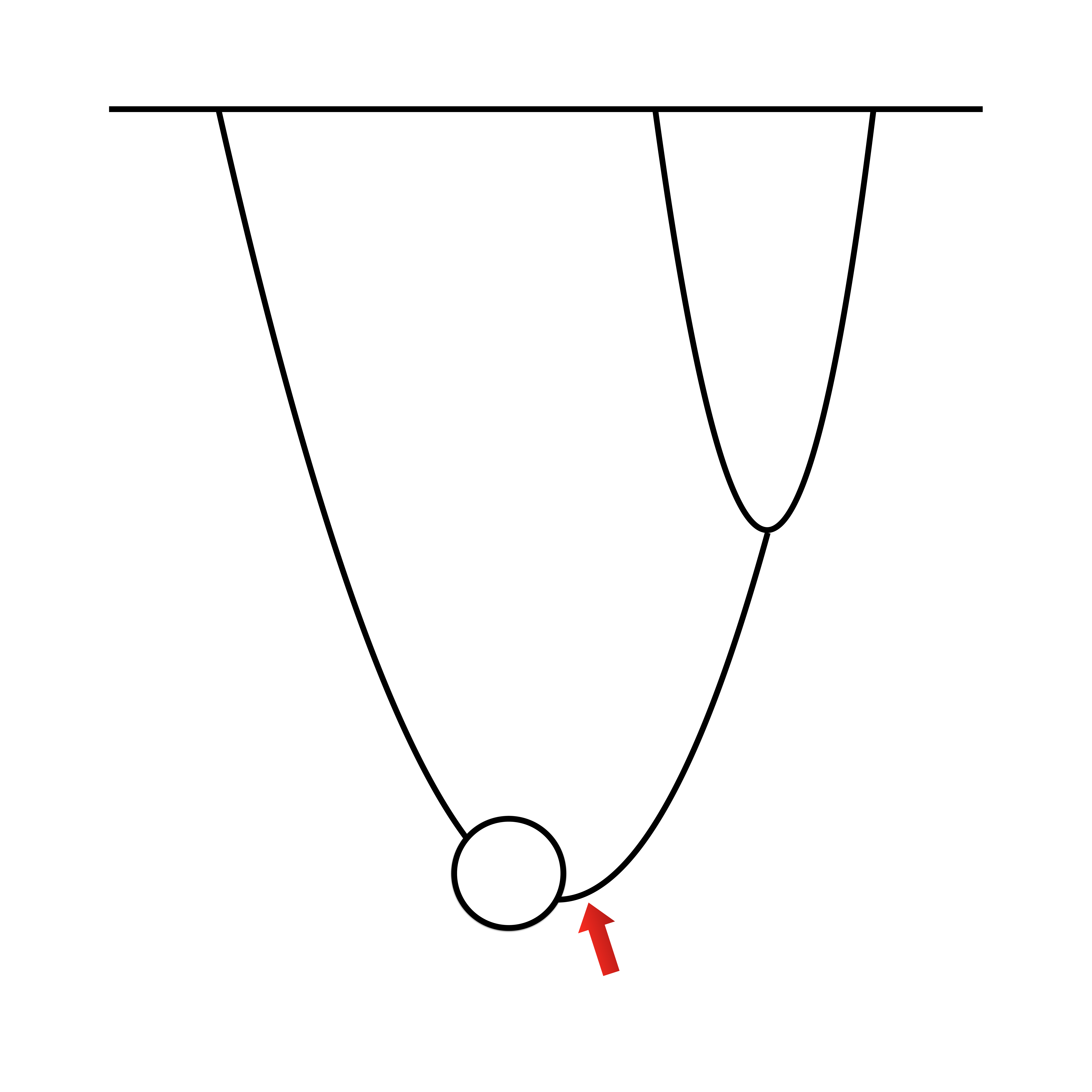}
\label{3pt_one_loop_Extra}}
\caption{Tree level and one-loop level non-BD 3pt function. There are two possible two one-loop diagrams corresponding to non-BD 3pt function (b)(c).}
\end{figure}


\subsection{Loop correction}

 It is well known that when  the non-BD initial condition is assumed, the folded limit non-Gaussianity will blow up.
 From the two point function calculation, we know that the effective non-BD coefficient will decay with time due to the loop correction. We expect that when we include loop corrections to 3pt function, the divergence will be cured.

For the three point functions, there are three external momenta. If we only consider the loop chain diagrams which modify three legs separately, namely those diagrams where three legs only meet at the original tree level interacting vertex with time $\tau_V$, things will become very easy. Other diagrams, either nested or connecting different legs, are thought to only result in  small corrections because their contributions are lower orders in $\tau$, which can be estimated from the uncertainly principle and number counting of the unconstrained time integrations.

For loop chain diagrams, we need to consider loop corrections on each leg. This means that even the BD three point functions will be corrected by loops. We first consider this case because it is easier to deal with due to the symmetry of three legs.

\subsubsection{Loop correction to BD three point function }

Explicit case studies show that the dominant diagrams have the following properties:
\begin{itemize}
\item Similar to the previous case, time sequence are all time-ordered or anti-time-ordered, i.e. dominant contributions are $H...H \zeta^3$ or $\zeta^3 H...H$ terms in in-in formalism. Because these two differ by some complex conjugation, we only need to analyze
time-ordered one $\zeta^3 H.....H$.
\beqn
\EV{\zeta^3}
 &=&
2\Real \sum_{n =0}^{\infty}    (-i)^n
 \int _{\tau_0}^{\tau}  d\tau_1 \int _{\tau_0}^{\tau_1}  d\tau_2...\int _{\tau_0}^{\tau_{n-1}}  d\tau_n
\EV{ \zeta^3 H_I( \tau_1 )...H_I( \tau_n) }_0
  \nonumber\\ &=&
  2\Real \sum_{n =1}^{\infty}   \sum_{m =1}^n
   (-i)^n   \Big(\int _{\tau_0}^{\tau}   d\tau_m  \Big)
   \Big(  \int _{\tau_m}^{\tau}  d\tau_1 \int _{\tau_m}^{\tau_1}  d\tau_2...\int _{\tau_m}^{\tau_{m-2}}  d\tau_{m-1}  \Big)
     \nonumber\\ &&\cdot
  \Big(  \int _{\tau_0}^{\tau_m}  d\tau_{m+1}  ...\int _{\tau_0}^{\tau_{n-1}}  d\tau_n  \Big)
\;\EV{ \zeta^3 H_I( \tau_1 )...H_I( \tau_m)...H_I( \tau_n) }_0
\nonumber\\ &=&
  2\Real \sum_{n_> =0}^{\infty}  \sum_{n_<=0}^{\infty}
   (-i)^{n_>+n_<+1}   \Big(\int _{\tau_0}^{\tau}   d\tau_V  \Big)
   \Big(  \int _{\tau_V}^{\tau}  d\tau_1 \int _{\tau_V}^{\tau_1}  d\tau_2...\int _{\tau_V}^{\tau_{n_>-1} }  d\tau_{n_>}  \Big)
     \nonumber\\ &&\cdot
  \Big(  \int _{\tau_0}^{\tau_V}  d\bar\tau_{ 1}\int _{\tau_0}^{\bar\tau_1}  d\bar\tau_2  ...\int _{\bar\tau_0}^{\bar\tau_{n_<-1}}  d\bar\tau_{n_<}  \Big)
\;\EV{ \zeta^3 H_I( \tau_1 )...H_I( \tau_{n>} )H_I( \tau_V)H_I(\bar\tau_1 )...H_I(\bar\tau_{n_<}) }_0
  \nonumber\\ &=&
   2\Imag  \int _{\tau_0}^{\tau}   d\tau_V  \Bigg\langle  \zeta^3\! \cdot
   \Big[ \Big( \sum_{n_> =0}^{\infty}    (-i)^{n_>  } \int _{\tau_V}^{\tau}  d\tau_1 \int _{\tau_V}^{\tau_1}  d\tau_2...\int _{\tau_V}^{\tau_{n_>-1} }  d\tau_{n_>} \Big)H_I( \tau_1 )...H_I( \tau_{n_>} ) \Big]
     \nonumber\\ &&\cdot H_I( \tau_V)\cdot
  \Big[\Big(\sum_{n_<=0}^{\infty}    (-i)^{ n_< } \int _{\tau_0}^{\tau_V}  d\bar\tau_{ 1}\int _{\tau_0}^{\bar\tau_1}  d\bar\tau_2  ...\int _{\bar\tau_0}^{\bar\tau_{n_<-1}}  d\bar\tau_{n_<}  \Big)
\; H_I(\bar\tau_1 )...H_I(\bar\tau_{n_<})\Big]  \Bigg\rangle_{\Large 0}
 \nonumber\\
\eeqn

where we use the tricks presented in the Appendix~\ref{in-in_formalism} and $\tau_m=\tau_V$ is the 3pt-tree level interacting vertex.
\item All loop vertices time are later than three point function interacting vertex $\tau_V<\tau_L$. This implies $n_<=0$.

\item For each leg, the loop chain has similar properties as those stated before including consecutive loop time.

\end{itemize}
So, the loop corrected three point function can be written as
\beqn
 \EV{\zeta_{\bm k_1}(\tau)\zeta_{\bm k_2}(\tau)\zeta_{\bm k_3}(\tau)}_{\text{BD}}^{\text{Loop}}
&=&(2\pi)^3 \delta(\bm k_1+\bm k_2+\bm k_3) 2 \Imag \Big[ u_{\bm k_1}(\tau) u_{\bm k_2}(\tau)u_{\bm k_3}(\tau)
\nonumber \\&& \times
 \int_{\tau_0}^\tau d\tau_V  \frac{-3\lambda}{2H   \epsilon^{3/2}}
\tau_V^2  \sqrt{k_1 k_2 k_3}  e^{i (k_1+k_2+k_3)\tau_V}Z_{\text{BD}}^{\text{Loop}}   \Big]~,
\eeqn
where $Z_{\text{BD}}^{\text{Loop}}\sim\Big[ \Big( \sum_{n_> =0}^{\infty}    (-i)^{n_>  } \int _{\tau_V}^{\tau}  d\tau_1 \int _{\tau_V}^{\tau_1}  d\tau_2...\int _{\tau_V}^{\tau_{n_>-1} }  d\tau_{n_>} \Big)H_I( \tau_1 )...H_I( \tau_{n_>} ) \Big] $ is the  loop contribution, essentially the product of mode functions of each loop and momentum integral. For diagram with $L_1, L_2, L_3$  loops at each leg respectively, the result contains the following term
\be
J_{\text{BD}}(L_1,L_2,L_3)=\mathcal I_{L_1}(k_1; \tau_V,\tau) \mathcal I_{L_2}(k_2; \tau_V,\tau)\mathcal I_{L_3}(k_3; \tau_V,\tau)~.
\ee
Recall $ \mathcal I_L(k;\tau_a,\tau_b)=\frac{1}{L!}\Big(\frac{k^5 (\tau_b^5-\tau_a^5) }{1200\pi}\Big)^L$. Next, we need to include coupling constants and do combinatoric counting. For each leg, the result is
\be
(\pm i)^{2L_j} (3\times 3\times2)^{L_j}\Big( \frac{-2\lambda}{H^4}\Big)^{2L_j} \Big( \frac{H}{2 \sqrt{\epsilon}}\Big)^{6L_j}
=\Big(-\frac{9}{8}\frac{\lambda^2}{H^2 \epsilon^3}\Big)^{L_j}~.
\ee
So,  the loop contribution factor with specific number of loops is
\beqn
(Z_{\text{BD}}^{\text{Loop}})_{L_1,L_2,L_3}&=&\Big(-\frac{9}{8}\frac{\lambda^2}{H^2 \epsilon^3}\Big)^{L_1+L_2+L_3}J_{\text{BD}}(L_1,L_2,L_3)
  \nonumber\\ &=&
  \prod_{j=1}^3  \Big(-\frac{9}{8}\frac{\lambda^2}{H^2 \epsilon^3}\Big)^{L_j}  \frac{1}{L_j!}\Big(\frac{k_j^5 (\tau ^5-\tau_V^5) }{1200\pi}\Big)^{L_j}
 =\prod_{j=1}^3  \frac{1}{L_j!}  \Big(-\frac{3\lambda^2k_j^5 (\tau ^5-\tau_V^5) }{3200\pi H^2 \epsilon^3}\Big)^{L_j}~.\nonumber\\
\eeqn

And finally, we need to sum over all possible  loops, yielding
\be
Z_{\text{BD}}^{\text{Loop}}  =\sum_{L_1,L_2,L_3=0}^\infty (Z_{\text{BD}}^{\text{Loop}})_{L_1,L_2,L_3}
 =\exp\Big(-\frac{3\lambda^2 }{3200\pi H^2 \epsilon^3} (k_1^5 +k_2^5+k_3^5)(\tau ^5-\tau_V^5)\Big)~.
\ee

This is just the philosophy presented in Figure~\ref{Loop_Exp}, i.e. replacing ``1'' with some exponential suppression factor and then you get the loop corrected results.

\subsubsection{Loop correction to non-BD three point function }

If there is one non-BD mode in the diagram, things are very similar to BD one but a little more complicated due to different behaviour of non-BD mode.


For illustration, we consider the one-loop case first. This loop is in the non-BD leg.   If we denote the loop time as  $\tau_1, \tau_2$ ($\tau_1 > \tau_2$), there are two diagrams with $\tau_{1,2}>\tau_V$ or $\tau_{1,2}<\tau_V$. Diagrams with $\tau_1>\tau_V>\tau_2$ means that the loop time are not consecutive and can not contribute highest power terms from previous analysis.

For three point function, there are two types of diagrams:
\begin{itemize}
\item
Type 1 (Figure~\ref{3pt_one_loop_BDlike}): If $\tau_V<\tau_2<\tau_1<\tau$, the non-BD mode is at $\tau_V$ and this case is similar to BD one.
\beqn
J^{(1)} &=&\int_{\tau_0}^\tau d\tau_V\; c_{\bm k_1}e^{-i\theta_{\bm k_1}} \tau_V^2 \sqrt{k_1 k_2 k_3} e^{i(k_2+k_3-k_1)\tau_V}
\int_{\tau_V}^\tau d\tau_1   \int_{\tau_V}^{\tau_1} d\tau_2\; \tau_1^2\tau_2^2 e^{i(p+q-k_1)(\tau_2-\tau_1)}~.\nonumber\\
 \eeqn

 \item
Type 2 (Figure~\ref{3pt_one_loop_Extra}): If $\tau_2<\tau_1<\tau_V<\tau$, the non-BD mode is at $\tau_2$ which is the earliest time.
 \beqn
J^{(2)} &=&\int_{\tau_0}^\tau d\tau_V\;   \tau_V^2 \sqrt{k_1  k_2 k_3} e^{i( k_2+k_3-k_1)\tau_V}  \int_{\tau_0}^{\tau_V} d\tau_1
 \int_{\tau_0}^{\tau_1} d\tau_2\; \tau_1^2\tau_2^2 e^{i(p+q )(\tau_2-\tau_1)}c_{\bm k_1}e^{-i\theta_{\bm k_1}} e^{i  k_1 (\tau_1-\tau_2)}~.  \nonumber\\
 \eeqn
\end{itemize}
We can see that the structure of these equations are the same as the previous ones. The only difference is the upper and lower limit of time integration.  It is not hard to generalize to the case where there are $L_1$ loops later than $\tau_V$ and $\widetilde L_1$  loops earlier than $\tau_V$ as shown in Figure~\ref{3pt_Multi_Loop_NonBD_Extra}.

Although there are two types of diagrams, actually they are unified with the same structure. The diagram in Figure~\ref{3pt_Multi_Loop_BD} is  a special case of Figure~\ref{3pt_Multi_Loop_NonBD_Extra} by setting $\widetilde L_1=0$. And we have $J_{\text{Non-BD}}(\widetilde L_1=0,L_1,L_2,L_3)=J_{\text{BD}}(  L_1,L_2,L_3)$. Thus, the loop contribution in general is
\be
J_{\text{Non-BD}}(\widetilde L_1, L_1,L_2,L_3)=\mathcal I_{\widetilde L_1}(k_1; \tau_0,\tau_V) \mathcal I_{L_1}(k_1; \tau_V,\tau)\mathcal  I_{L_2}(k_2; \tau_V,\tau)\mathcal I_{L_3}(k_3; \tau_V,\tau)~.
\ee

The coefficient is the product of the previous BD one and those contributed by $\widetilde L_1$  loops between time $\tau_0$ and $\tau_V$,
\beqn
&& \Big(-\frac{9}{8}\frac{\lambda^2}{H^2 \epsilon^3}\Big)^{L_1+L_2+L_3} \times
 (\pm i)^{2\widetilde L_1}2^{\widetilde L_1} (3\times 3\times 2)^{\widetilde L_1}\Big( \frac{-2\lambda}{H^4}\Big)^{2\widetilde L_1} \Big( \frac{H}{2 \sqrt{\epsilon}}\Big)^{6\widetilde L_1}
 \nonumber \\& =&
 \Big(-\frac{9}{8}\frac{\lambda^2}{H^2 \epsilon^3}\Big)^{L_1+L_2+L_3}\Big(-\frac{9}{4}\frac{\lambda^2}{H^2 \epsilon^3}\Big)^{\widetilde L_1}  ~.
 \eeqn

So, the final loop corrected non-BD three-point correlation function is
\beqn
 \EV{\zeta_{\bm k_1}(\tau)\zeta_{\bm k_2}(\tau)\zeta_{\bm k_3}(\tau)}_{\text{Non-BD}}^{\text{Loop}}
&=&(2\pi)^3 \delta(\bm k_1+\bm k_2+\bm k_3) 2 \Imag \Big[ u_{\bm k_1}(\tau) u_{\bm k_2}(\tau)u_{\bm k_3}(\tau)
  \int_{\tau_0}^\tau d\tau_V    \frac{-3\lambda}{2H  \epsilon^{3/2}}
  \nonumber\\&& \times
\tau_V^2  \sqrt{k_1 k_2 k_3}
   \Big( c_{\bm k_1} e^{-i\theta_{\bm k_1}}e^{i(k_2+k_3-k_1)\tau_V}Z_{\text{Non-BD}}^{\text{Loop}} +2 \text{ perm.} \Big) \Big] ~,  \nonumber\\
\eeqn
with the  loop correction    given by
\beqn
Z_{\text{Non-BD}}^{\text{Loop}}&=&\sum_{\widetilde L_1, L_1,L_2,L_3=0}^\infty \Big(-\frac{9}{8}\frac{\lambda^2}{H^2 \epsilon^3}\Big)^{L_1+L_2+L_3}
\Big(-\frac{9}{4}\frac{\lambda^2}{H^2 \epsilon^3}\Big)^{\widetilde L_1}   J_{\text{Non-BD}}(\widetilde L_1, L_1,L_2,L_3)
\nonumber\\
&=&\exp\Big[-\frac{3\lambda^2 }{3200\pi H^2 \epsilon^3}\Big(2k_1^5(\tau_V^5-\tau_0^5)+ (k_1^5 +k_2^5+k_3^5)(\tau ^5-\tau_V^5)\Big) \Big]~.
\eeqn

It is very interesting to notice that
\be
\frac{Z_{\text{Non-BD}}^{\text{Loop}}(\tau_V)}{Z_{\text{BD}}^{\text{Loop}}(\tau_V)}=\frac{c_{\bm k}^{\text{eff}}(\tau_V)}{1}=\frac{c_{\text{Non-BD}}^{\text{eff}}( \tau_V)}{c_{\text{BD}}^{\text{eff}}( \tau_V)}~,
\ee
where ``1'' essentially is just the effective BD coefficients to the lowest order~(\ref{consistency_quantization}). So, the ratio between non-BD and BD effects in different sectors are the same.

The relevant integrals of 3pt function are:
 \be
  \int_{\tau_0}^\tau d\tau_V\; \tau_V^2  e^{-B\Big( (k_1^5+k_2^5+k_3^5)  (\tau^5-\tau_V^5 )\Big) }   e^{i(k_1+k_2+k_3 )\tau_V}~,
 \ee
for BD one and
 \be
  \int_{\tau_0}^\tau d\tau_V\; \tau_V^2  e^{-B\Big(2k_1^5  (\tau_V^5-\tau_0^5) +(k_1^5+k_2^5+k_3^5)  (\tau^5-\tau_V^5 )\Big) }   e^{i(k_2+k_3-k_1)\tau_V}~,
 \ee
for non-BD one,  where $ B\equiv \frac{3\lambda^2 }{3200\pi  \epsilon^3H^2}  $.

There are  exponential oscillation terms in the integrals. In order to find a characteristic scale of initial time, we switch to consider the simpler case by neglecting the exponential oscillations and study the following two integrals
 \be
R(k_1,k_2,k_3,\tau_0,\tau) =\int_{\tau_0}^\tau d\tau_V \; \tau_V^2  e^{-B\Big( (k_1^5+k_2^5+k_3^5)  (\tau^5-\tau_V^5 )\Big) }~,
 \ee
  \be
Q(k_1,k_2,k_3,\tau_0,\tau) =\int_{\tau_0}^\tau d\tau_V \; \tau_V^2  e^{-B\Big(2k_1^5  (\tau_V^5-\tau_0^5) +(k_1^5+k_2^5+k_3^5)  (\tau^5-\tau_V^5 )\Big) }~.
 \ee

 As for function $Q$, when $\tau_0 \rightarrow -\infty$ the function in the integral is highly suppressed by the exponential factor. It makes no sense to choose the infinitely past initial time because $Q$ will vanish in that case. Instead, we try to find the conditions for maximal 3pt function. We expect that there exists one initial time $\tau_{0m}$ which can maximize the integral $Q$.

\begin{figure}
\centering
 \includegraphics[width=0.8\textwidth]{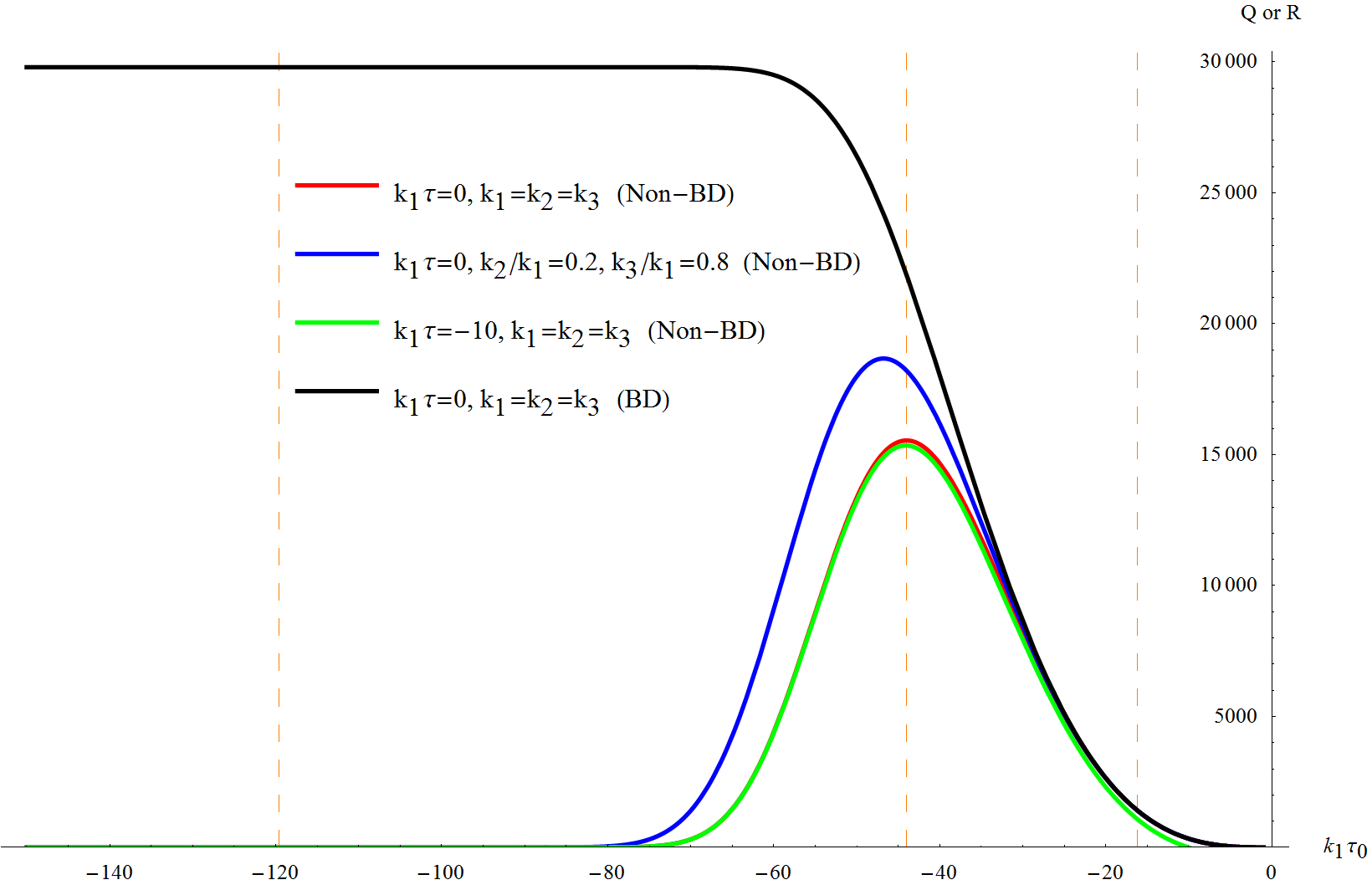}
 \caption{ The integral value $Q$ or $R$ (representing non-Gaussianity) as a function of initial time. Function $Q$ (corresponding to non-BD non-Gaussianity)  peaks at one specific time $\tau_{0m}$  and is very insensitive to the final time $\tau$ as well as the ratio of $k_{1,2,3}$. Function $R$ (corresponding to  BD non-Gaussianity) approaches a constant value when the initial time is early enough. Three vertical lines mark the initial time $e\tau_{0m},\tau_{0m},\tau_{0m}/e$. (Here $B=1.5459\times 10^{-9}$ corresponding to $f_{\text{NL}}=1$, $k_1\tau_{0m}=-44.0038$)}
\label{BD_Integral_time}
\end{figure}

So far our discussions are based on sub-horizon limit approximations $|k\tau|,|k\tau_0| \gg 1$, but mathematically, it is still meaningful to set $\tau=0$. This  can be justified from the fact that the function $Q$ is very insensitive to the final time $\tau$ if $|\tau_0|\gg |\tau|$ as well as from Figure~\ref{BD_Integral_time} where two curves with different initial time are compared.  For concreteness, we choose $k_1=k_2=k_3=k$ and $\tau=0$, then we get the function of $\tau_0$
 \be
 Q(k,k,k,\tau_0,0)=\frac{e^{2 B k^5\tau_0^5} \Big(\Gamma (\frac{3}{5} )
 -\Gamma (\frac{3}{5},- B k^5
   \tau_0^5 )\Big)}{5 ( B k^5 )^{3/5}}~.
 \ee
 We need to find $\frac{dQ}{d\tau_0}\Big|_{\tau_{0}=\tau_{0m}}=0$.
  Numerically, we can find that the integral is maximized when $-k^5 \tau_{0m}^5  B  \approx 0.255055$ with maximum value

 \be\label{Maximum_Value}
  Q(k,k,k, \tau_{0m},0)=\frac{0.0803601}{B^{3/5}k}~.
 \ee
This roughly calibrates the initial time for maximal non-Gaussianity and its  corresponding amplitude.  The weak dependence on the final time $\tau$ and different shape $(k_2/k_1,k_3/k_1)$ is shown in the Figure~\ref{BD_Integral_time}.

While for function $R$, it will saturate for early enough initial time (see Figure~\ref{BD_Integral_time}).

\subsection{Observational non-Gaussianity}

\subsubsection{Standard result on non-Gaussianity}
Before showing the non-Gaussianity under loop corrections, we first review the standard non-Gaussianity, i.e. the tree level result. In the standard procedure, the initial time is chosen to be past infinity. This will cause  divergence of the integral. In order to regulate the divergence, we need to adopt the so-called $i\varepsilon$  prescription which is well understood in standard quantum field theory, but a little problematic in cosmology.

In standard QFT,  $i\varepsilon$  prescription is valid and vital from both mathematics and physics. It not only cures the divergent problems in mathematics, but also ensures that physically the quantum system can evolve from the non-interacting vacuum state in the infinitely past to the true vacuum state with interaction at present.

However, in cosmology, our universe may start from a finite initial time. What's more, de-Sitter inflation has some different non-trivial  properties compared to flat space case. The most well-known and  serious problem for this $i\varepsilon$ prescription is the folded limit divergence of non-BD non-Gaussianity. Even for the BD case, $i\varepsilon$ prescription comes from the scattering problem where particles are initially far away from each other. This is not the case for inflation because the inflation originates from the sub-horizon scales.

Let us first present the result of standard non-Gaussianity. The BD non-Gaussianity can be obtained from the tree level 3pt function calculations by setting $\tau=0$:
\beqn
\EV{\zeta_{\bm k_1} \zeta_{\bm k_2} \zeta_{\bm k_3} }_{\text{BD}}^{\text{Tree}} \,\rq{}
&=&  2 \Imag \Big[ u_{\bm k_1}(0) u_{\bm k_2}(0)u_{\bm k_3}(0)
 \int_{-\infty}^{\tau=0} d\tau_V     \frac{-3\lambda}{2H  \epsilon^{3/2}}\tau_V^2
 \sqrt{k_1 k_2 k_3}  e^{i (k_1+k_2+k_3)\tau_V}      \Big]
 \nonumber  \\
&=& 2 \Imag \Big[  \Big(\frac{H}{2 \sqrt{\epsilon}} \Big)^3 ( k_1 k_2 k_3)^{-3/2}
 \int_{-\infty(1-i0^+)}^0 d\tau_V
  \frac{-3\lambda\tau_V^2  \sqrt{k_1 k_2 k_3}  }{2H   \epsilon^{3/2}}
 e^{i (k_1+k_2+k_3)\tau_V}      \Big]
 \nonumber  \\
&=&  \Big(\frac{H}{2 \sqrt{\epsilon}} \Big)^6 \frac{-24\lambda}{H^4}\frac{1}{k_1 k_2 k_3}
\Imag \Big[\frac{ 2i}{(k_1+k_2+k_3)^3}  \Big]
\nonumber  \\
&=&   \frac{-3H^2\lambda}{4 \epsilon^3} \frac{1}{(k_1+k_2+k_3)^3 k_1 k_2 k_3}
\eeqn
where prime denotes that $(2\pi)^3\delta(\bm k_1+\bm k_2+\bm k_3)$ is stripped, while the non-BD non-Gaussianity is given by
\beqn
\EV{\zeta_{\bm k_1} \zeta_{\bm k_2} \zeta_{\bm k_3} }_{\text{Non-BD}}^{\text{Tree}}  \,\rq{}
&=&  2 \Imag \Big[  \Big(\frac{H}{2 \sqrt{\epsilon}} \Big)^3 ( k_1 k_2 k_3)^{-3/2}
 \int_{-\infty(1-i0^+)}^0 d\tau_V\;  \frac{-3\lambda}{2H   \epsilon^{3/2}}
  \nonumber\\&&\times
\tau_V^2\sqrt{k_1 k_2 k_3}
  c_{\bm k_1}e^{-i\theta_{\bm k_1}}e^{i (-k_1+k_2+k_3)\tau_V} +2\text{ perm.}    \Big]  \nonumber\\
&=&  \frac{-3H^2\lambda}{4 \epsilon^3} \frac{1}{ k_1 k_2 k_3}
\Big( \frac{c_{\bm k_1}\cos(\theta_{\bm k_1})}{(-k_1+k_2+k_3)^3}+2\text{ perm.}\Big)
\eeqn
Note that for external line, we only consider the BD modes. Non-BD modes in the external lines essentially are just effectively renormalizing the BD part by contributing term
$\sum_i  \Real C_-(\bm k_i)  \EV{\zeta_{\bm k_1} \zeta_{\bm k_2} \zeta_{\bm k_3} }_{\text{BD}}$. So, we are not going to consider them anymore and alway assume the BD external
modes.

The non-Gaussianity can be characterized by the non-Gaussianity shape function $\mathcal{F}$ which is defined as
 \be
 \EV{\zeta_{\bm k_1} \zeta_{\bm k_2} \zeta_{\bm k_3} }=(2\pi)^7 \delta(\bm k_1+\bm k_2+\bm k_3)
  \frac{P_\zeta^2}{k_1^2 k_2^2 k_3^2} \mathcal{F}(k_2/k_1,k_3/k_1)~.
 \ee
where the scale invariance of the correlation functions has been used to show that $\mathcal{F}$ only depends on the ratios of different momenta.

So, for the standard non-Guassianty, the shape functions for BD and non-BD are given by
\beqn
\mathcal{F}(k_2/k_1,k_3/k_1)_{\text{BD} }^{\text{STD} }&=& \frac{-3\lambda}{H^2 \epsilon} \frac{k_1 k_2 k_3}{(k_1+k_2+k_3)^3}~,
 \nonumber\\
\mathcal{F}(k_2/k_1,k_3/k_1)_{\text{Non-BD}}^{\text{STD} }&=& \frac{-3\lambda}{H^2 \epsilon}  k_1 k_2 k_3  \Big( \frac{c_{\bm k_1}\cos(\theta_{\bm k_1})}{(-k_1+k_2+k_3)^3}+2\text{ perm.}\Big)~.
\eeqn

\begin{figure}
\subfigure[Standard BD non-Gaussianity ]{\includegraphics[width=0.48\textwidth]{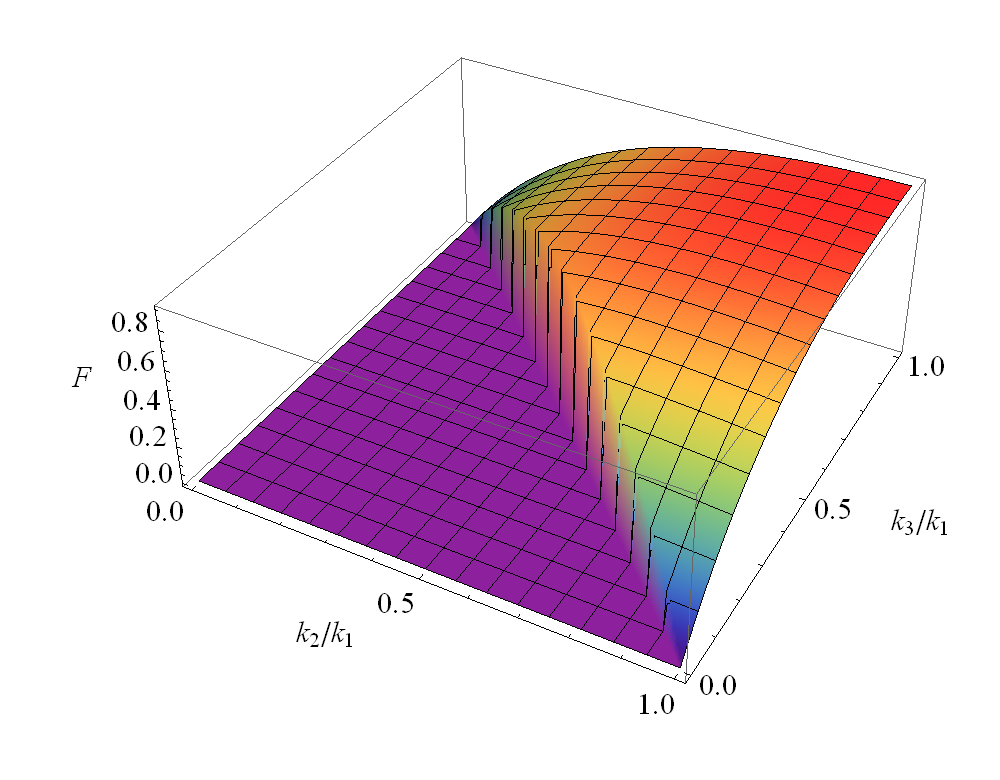}}
\quad
\subfigure[Standard non-BD non-Gaussianity ]{\includegraphics[width=0.48\textwidth]{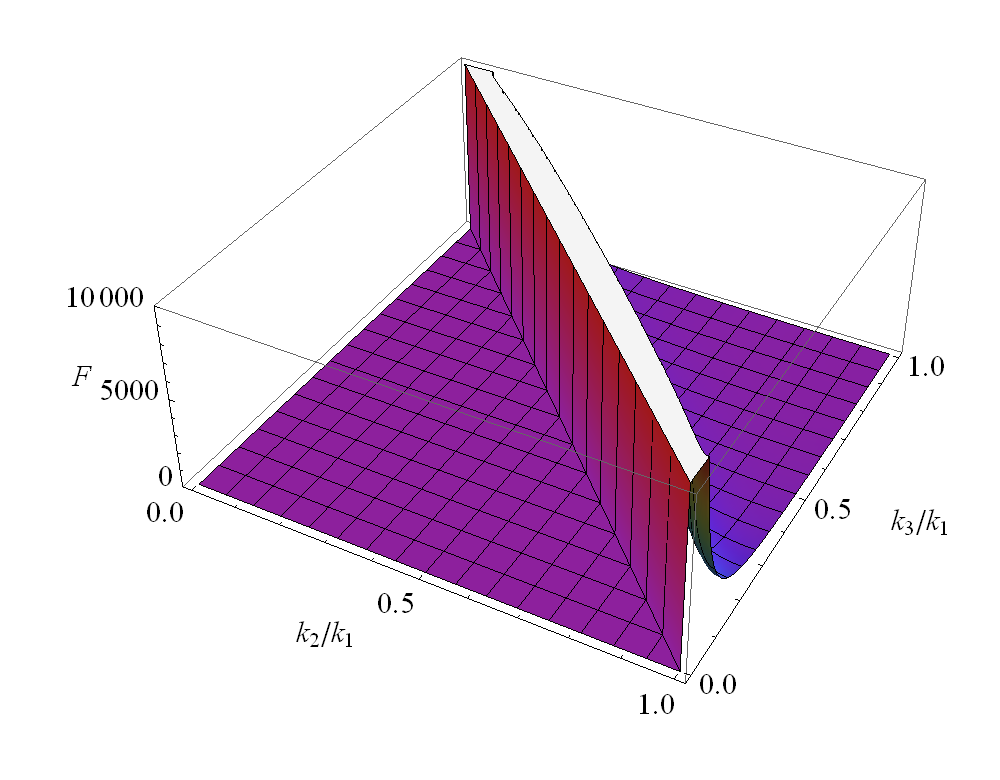}}
\quad
\caption{Standard non-Gaussianity in literature obtained through $i\varepsilon$ prescription. Note that the  non-BD non-Gaussianities in the folded limit are actually divergent (b). }
\label{StandardShape}
\end{figure}

The corresponding shapes can be seen from Figure~\ref{StandardShape}. Evidently,  non-Gaussianity diverges in the folded limit. But this divergence is unphysical. As we will show later, the decay of non-BD modes cures the divergence.

\begin{figure}
\subfigure[Tree level $\tau_{0 }=e\tau_{0m} $ ]{\includegraphics[width=0.49\textwidth]{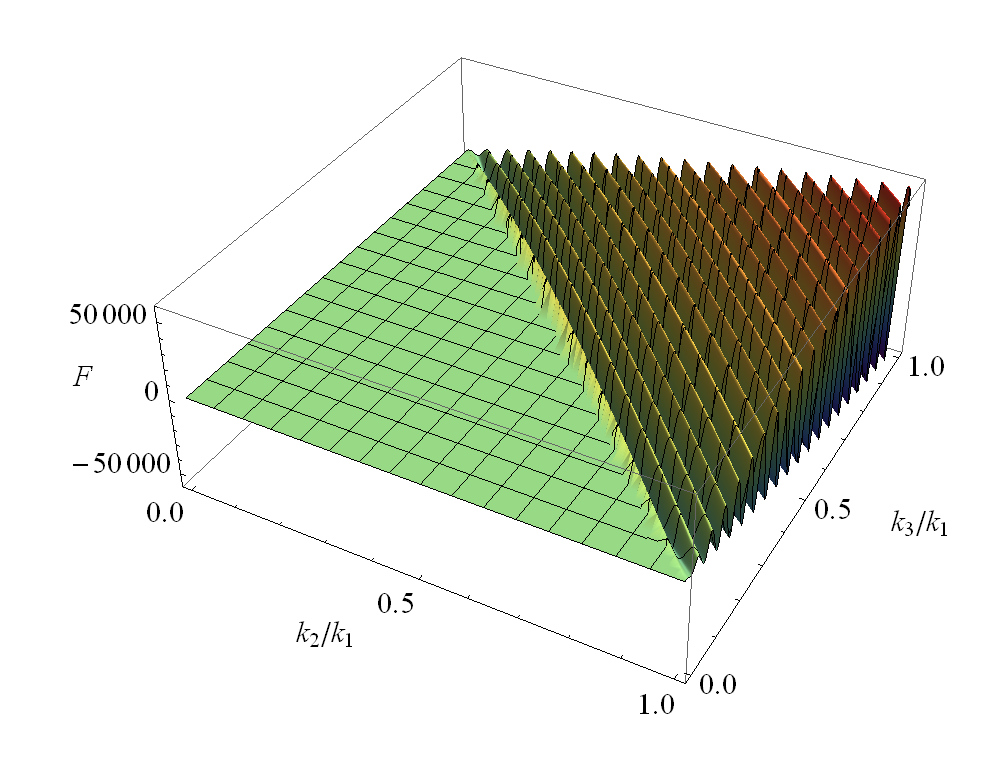}}
\quad
\subfigure[Loop corrected $\tau_{0 }=e\tau_{0m} $ ]{\includegraphics[width=0.49\textwidth]{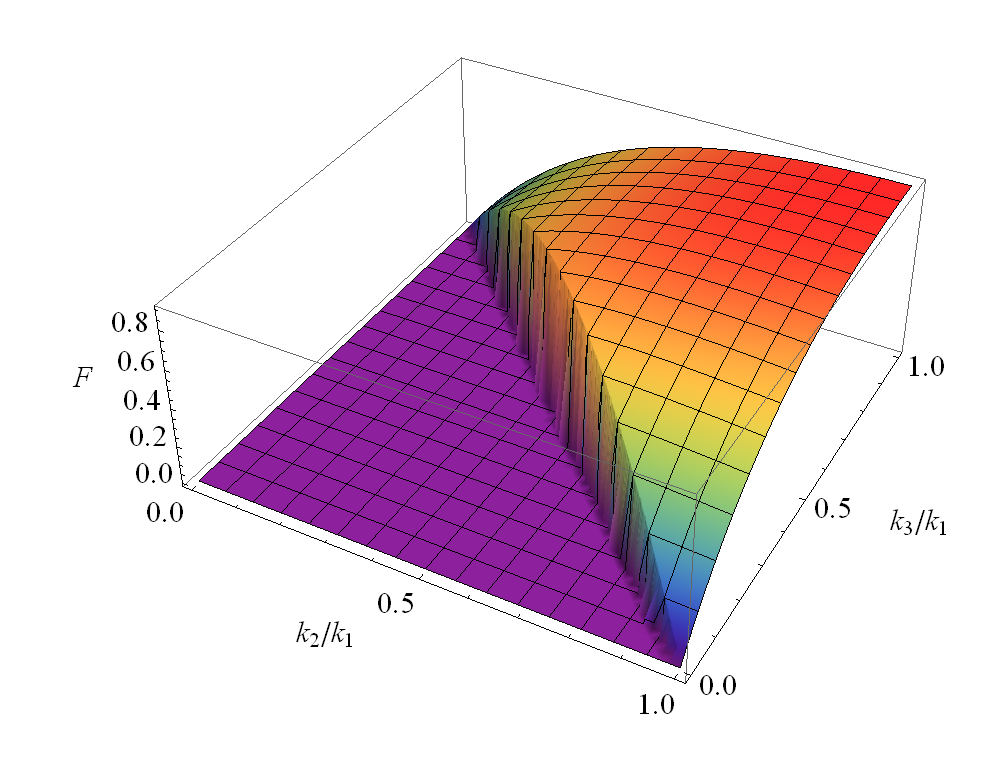}}

\caption{BD non-Gaussianity shape function at tree level (left) and loop level (right) for initial time $\tau_0=e\tau_{0m}$. At tree level, the sharp  initial time cut-off will give rise to a fast oscillating non-Gaussianity shape. After including loop correction, the oscillating behavior is suppressed to nearly vanishing value and we nearly recover the usual BD non-Gaussianity shape. }
\label{BDShape}
\end{figure}

\begin{figure}
\subfigure[Tree level $\tau_{0 }=e\tau_{0m} $ ]{\includegraphics[width=0.49\textwidth]{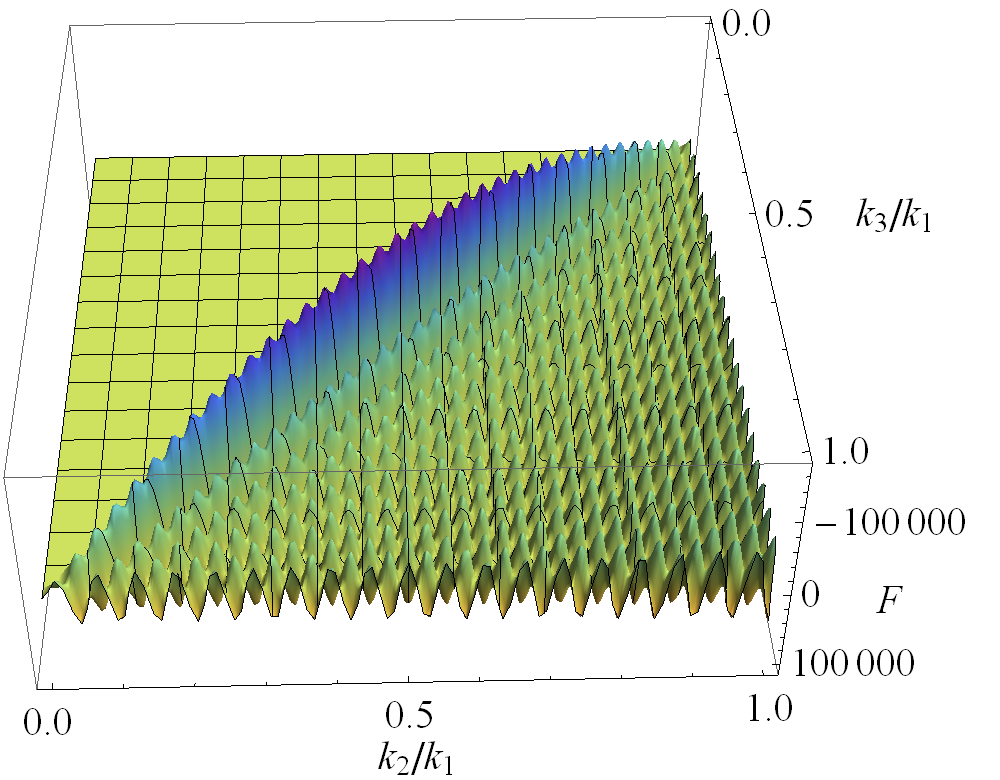}}
\quad
\subfigure[Loop corrected $\tau_{0 }=e\tau_{0m} $ ]{\includegraphics[width=0.49\textwidth]{LoopSharpNonBD.png}}

\caption{Non-BD non-Gaussianity shape function at tree level (left) and loop level (right) for initial time $\tau_0=e\tau_{0m}$ (without Guassian smearing initial time). By including the loop correction, the amplitude is suppressed to very small value. }
\label{SharpNonBDShape}
\end{figure}

\begin{figure}[!htb]
\subfigure[Tree level $\tau_{0c}=\tau_{0m}/e,\tau_{0w}=|\tau_{0c}|/3$]{\includegraphics[width=0.48 \textwidth]{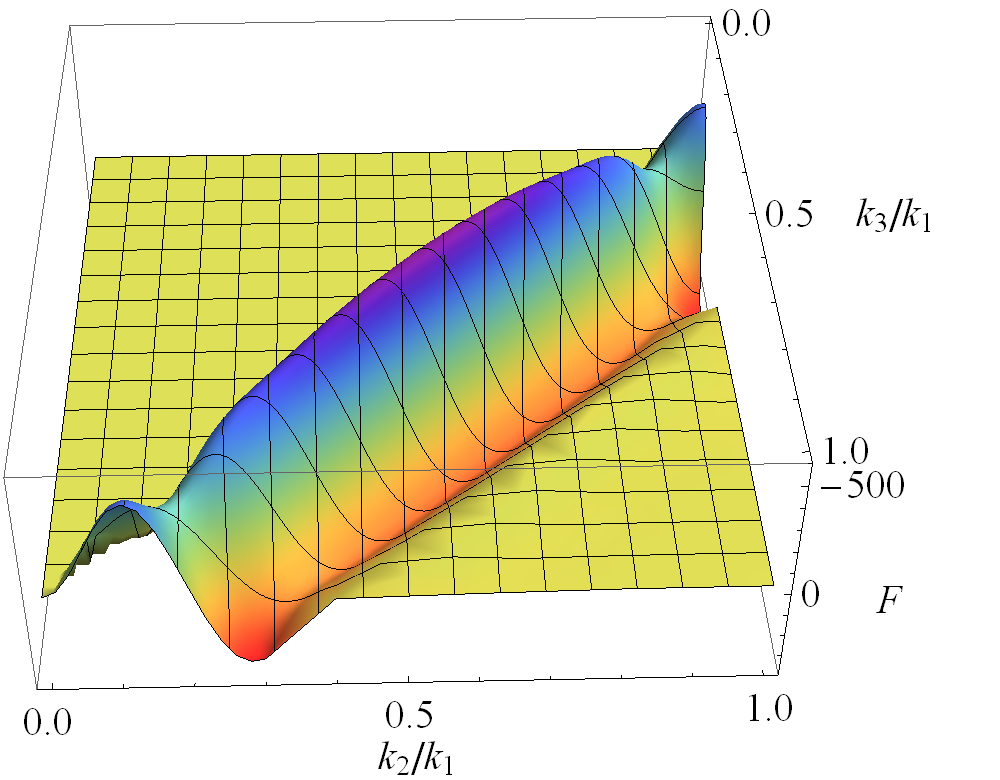}}
\quad
\subfigure[Loop corrected $\tau_{0c}=\tau_{0m}/e,\tau_{0w}=|\tau_{0c}|/3$]{\includegraphics[width=0.48\textwidth]{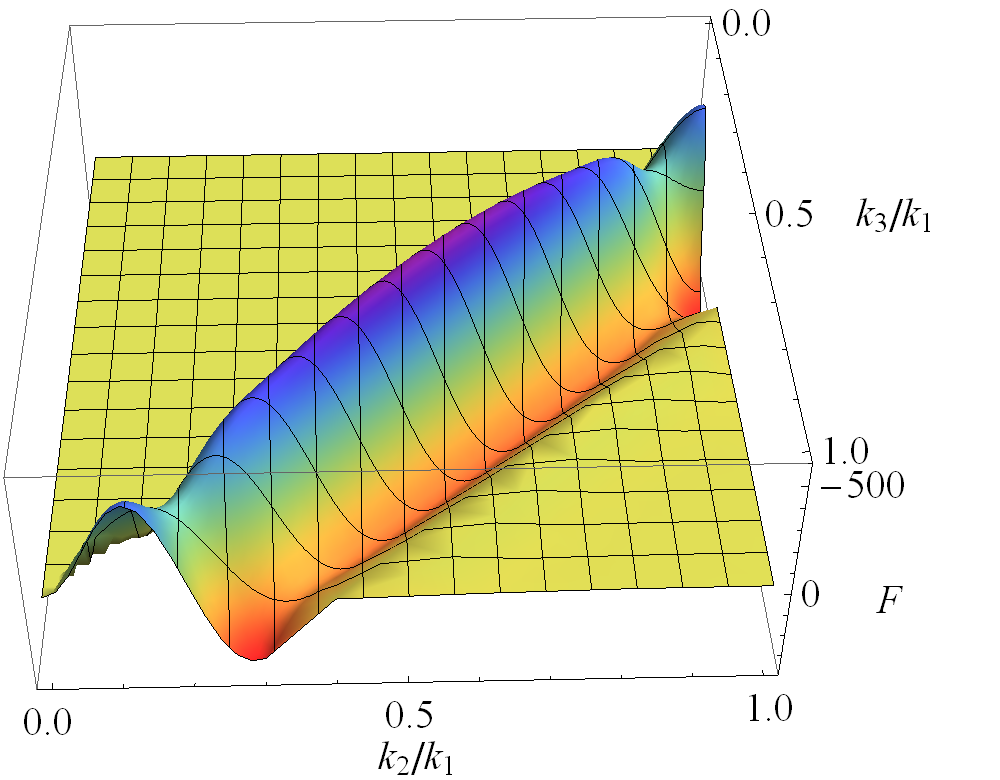}}

\subfigure[Tree level $\tau_{0c}=\tau_{0m},\tau_{0w}=|\tau_{0c}|/3$]{\includegraphics[width=0.48 \textwidth]{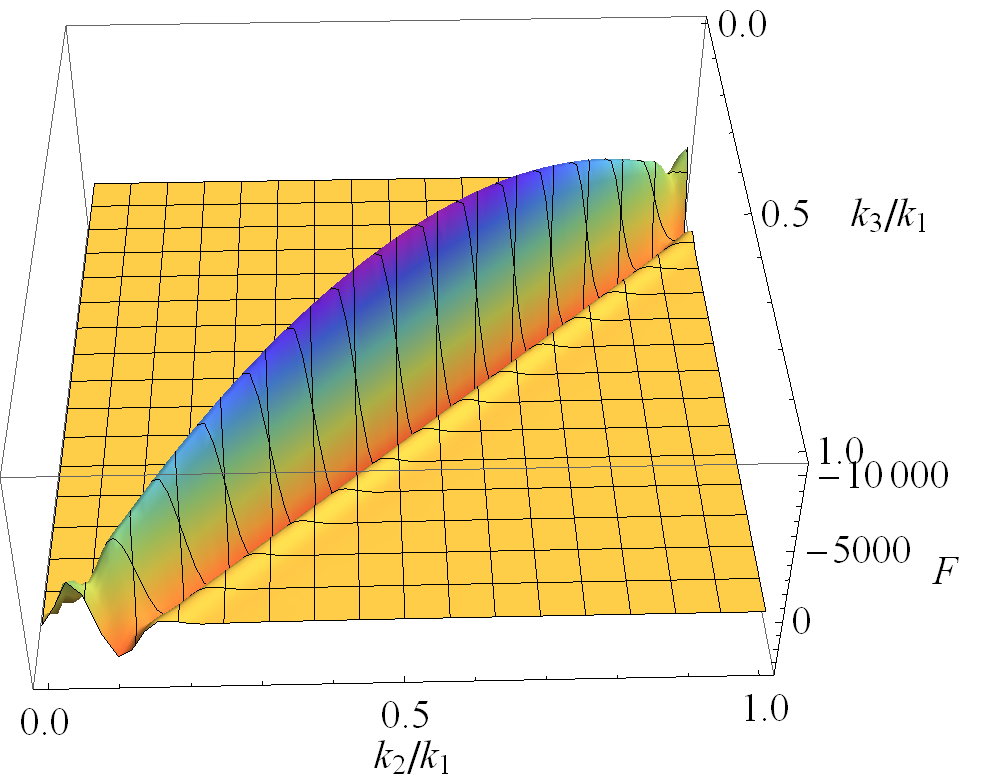}}
\quad
\subfigure[Loop corrected $\tau_{0c}=\tau_{0m},\tau_{0w}=|\tau_{0c}|/3$]{\includegraphics[width=0.48\textwidth]{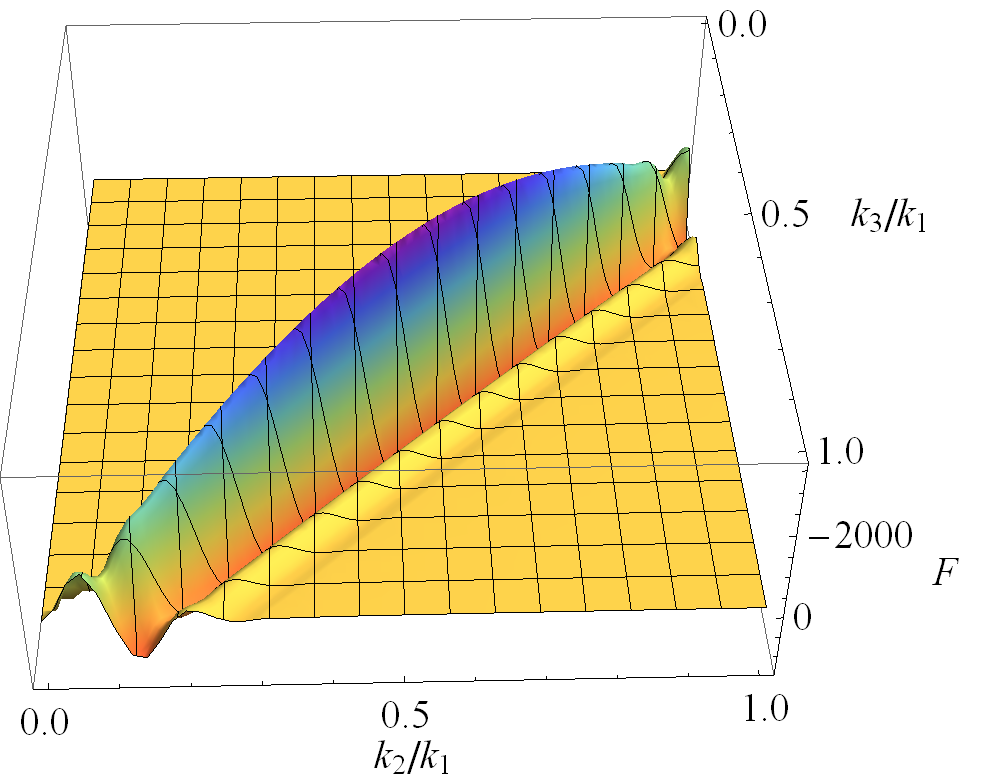}}

\subfigure[Tree level $\tau_{0c}=e\tau_{0m},\tau_{0w}=|\tau_{0c}|/3$ ]{\includegraphics[width=0.48\textwidth]{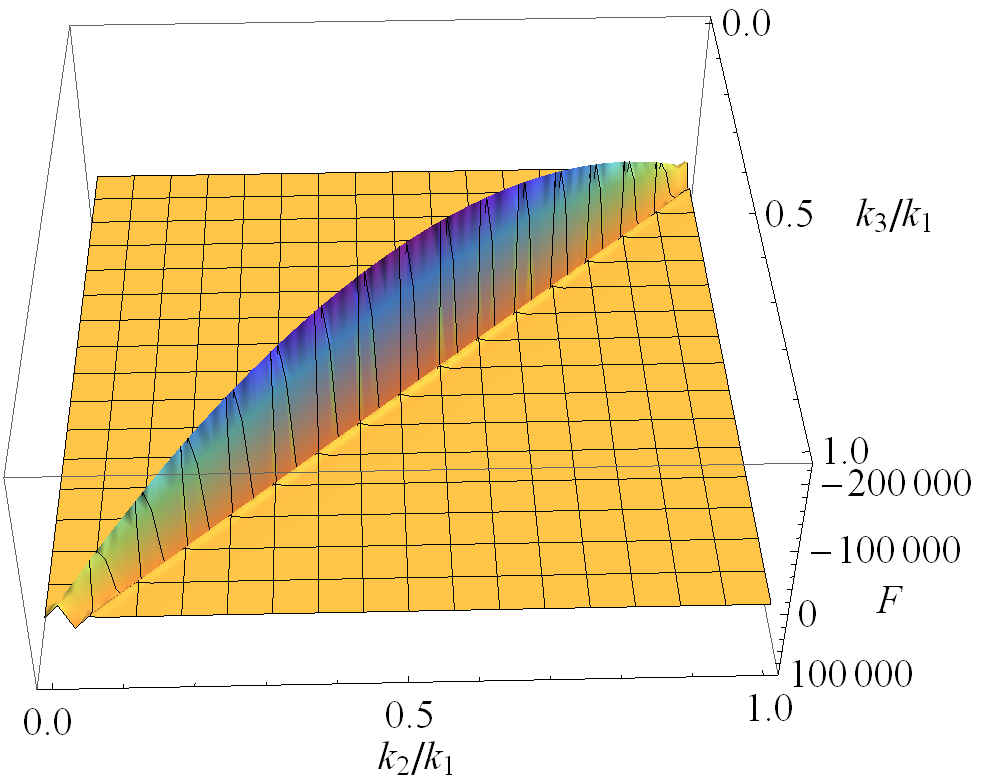}}
\quad
\subfigure[Loop corrected $\tau_{0c}=e\tau_{0m},\tau_{0w}=|\tau_{0c}|/3 $ ]{\includegraphics[width=0.48\textwidth]{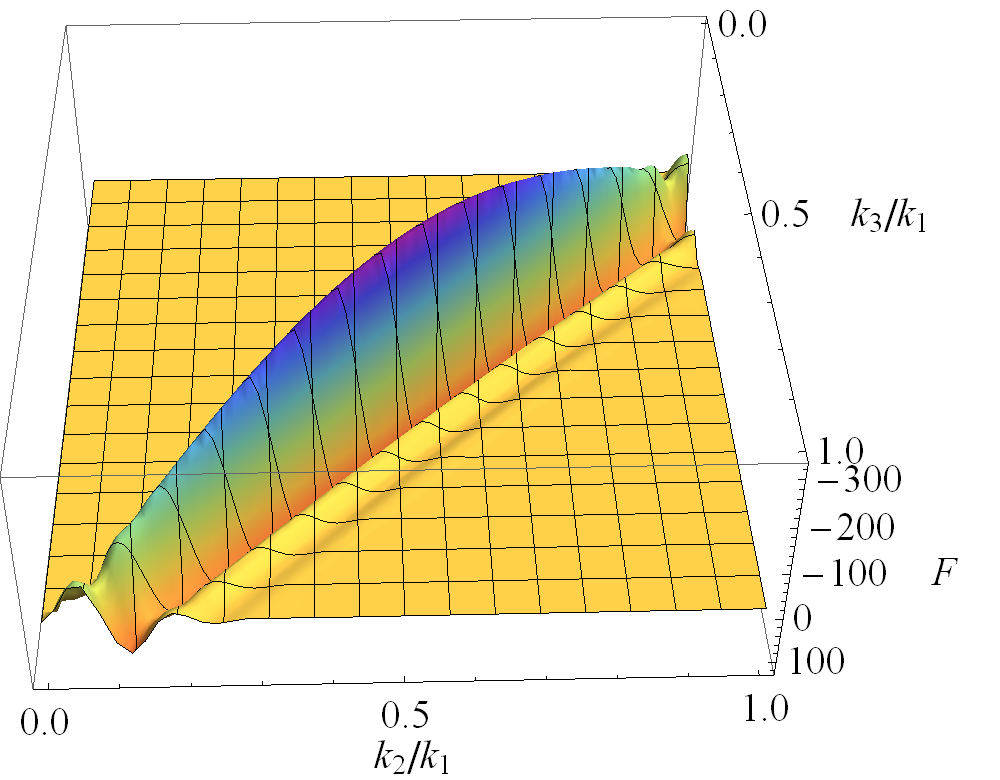}}
 \caption{Non-BD non-Gaussianity shape function at tree level (left) and loop level (right) for different initial time  $\tau_{0m}/e$ (upper), $\tau_{0m}$ (middle) and $e\tau_{0m}$ (down), filtered by a Gaussian weight function \eqref{gaussian-filter} for the initial time. Note two prominent features: folded shape peak at  $k_2+k_3\sim k_1$ and squeezed limit shape at $k_2\sim 0$ or $k_3\sim 0$. For folded shape,  if we only consider the tree level,  the peak value will blows up with earlier initial time and finally diverges for past infinity initial time. But, if we consider the loop corrections, the folded shape peak will be suppressed.  }
\label{nonBDShape}
\end{figure}

\subsubsection{Non-Gaussianity with loop correction}

Next, we need to consider the loop corrections. The  non-BD non-Gaussianity is maximized  roughly when non-BD modes are generated or excited at $\tau_{0m}$ instead of the infinitely past.

 Previously, we only handle the effective value under  the sub-horizon approximations, i.e. $|k\tau| \gg  1$. But physically we expect that in the super-horizon case, the decay is quite slow due to the frozen of modes. So, super-horizon and sub-horizon admit  completely different behaviors. Tentatively, we can find an  intermediate time $\tau_{\text{int}}$ ($|k\tau_{\text{int}}| \gtrsim 1 $)  to connect these two pieces. When $|\tau_V|>|\tau_{\text{int}}|$, previous sub-horizon approximated results are reliable. While for $|\tau_V|<|\tau_{\text{int}}|$, we can just simply ignore possible loop corrections (which are expected to be very small due to limited time integrations as well as nearly frozen super-horizon modes)  and only consider the tree level results with different initial conditions---the renormalized non-BD coefficients at $\tau_{\text{int}}$ or more explicitly, $c_{\bm k} \rightarrow c_{\bm k}^{\text{eff}}(\tau_{\text{int}})$.  So, the loop corrections can be evaluated in the following  way
 \be
 \int_{\tau_0}^0 d\tau_V\; c_{\bm k}\cdots \rightarrow
  \int_{\tau_0}^{\tau_{\text{int}}} d\tau_V\; c_{\bm k}Z_{\text{Non-BD}}^{\text{loop}}\cdots+ \int_{\tau_{\text{int}}}^0 d\tau_V\; c_{\bm k}^{\text{eff}}(\tau_{\text{int}})\cdots~,
 \ee
 where $\cdots$ denotes the tree level relevant terms.  It is very interesting to note that $c_{\bm k}^{\text{eff}}(\tau_V)\approx c_{\bm k} Z_{\text{Non-BD}}^{\text{loop}}(\tau_V)$ for $|\tau_0| \gg |\tau|,|\tau_V| $. This suggests that mathematically, the formula for $Z_{\text{Non-BD}}^{\text{loop}}$ can also be used in the super-horizon limit due to its similar behavior. Similar consideration also holds for the  BD non-Gaussianities.

Based on these arguments, we can still use our previous result derived in the sub-horizon limit to calculate the observable super-horizon non-Gaussianities simply by setting $\tau=0$:
\beqn
\EV{\zeta_{\bm k_1} \zeta_{\bm k_2 }\zeta_{\bm k_3} }_{\text{BD}}^{\text{Loop}} \,\rq{}
&=&  \Big(\frac{H}{2 \sqrt{\epsilon}} \Big)^6 \frac{-24\lambda}{H^4}\frac{1}{k_1 k_2 k_3}
 \Imag \Big[  \int_{\tau_0}^0 d\tau_V\; \tau_V ^2  e^{i ( k_1+k_2+k_3)\tau_V}
Z_{\text{BD}}^{\text{loop}}( \tau_V)  \Big]~,
 \nonumber \\
 \EV{\zeta_{\bm k_1} \zeta_{\bm k_2 }\zeta_{\bm k_3} }_{\text{Non-BD}}^{\text{Loop}} \,\rq{}
&=&  \Big(\frac{H}{2 \sqrt{\epsilon}} \Big)^6 \frac{-24\lambda}{H^4}\frac{1}{k_1 k_2 k_3}
 \Imag \Big[  \int_{\tau_0}^0 d\tau_V\; \tau_V ^2     c_{\bm k_1}  e^{-i\theta_{\bm k_1}}
 \nonumber\\&&\qquad\qquad\times
     e^{i (-k_1+k_2+k_3)\tau_V}Z_{\text{Non-BD}}^{\text{loop}}( \tau_V)
   +2\text{ perm.}   \Big]~.
\eeqn
where  $ Z_{\text{Non-BD}}^{\text{loop}}( \tau_V)  = \exp\Big[-B\Big(2k_1^5(\tau_V^5-\tau_0^5) +(k_1^5+k_2^5+k_3^5)  (0^5-\tau_V^5 )\Big)  \Big],Z_{\text{ BD}}^{\text{loop}}( \tau_V)  = \exp\Big[-B\Big( (k_1^5+k_2^5+k_3^5)  (0^5-\tau_V^5 )\Big)  \Big]  $.  They can be regarded as the loop corrections or the renormalization factors to the tree diagram. Especially, note that  they don't show decay behavior when $\tau_V$ is pretty small, consistent with our previous physical picture for super-horizon modes.

As we stated before, the loop corrections, in principle, also contain a phase factor  $\exp(i \gamma)$ with $\gamma\sim B k^5 \tau_V^5$. We don't consider them because the exact expression is unknown and may be very complicated. The above rough form can be understood from the DRG method. What we want to emphasize is that the amplitude decay is sufficient to suppress the divergence and the fast oscillating phase factor can only be more beneficial due to the dramatic cancellations between positive and negative parts.

Finally, we  obtain the  BD and non-BD   non-Gaussianity shape functions with loop corrections:
 \beqn
\mathcal{F}(k_2/k_1,k_3/k_1,\tau_0)_{\text{BD}}^{\text{Loop}}&=&  \frac{-3\lambda}{2H^2 \epsilon}  k_1 k_2 k_3    \Big[  \int_{\tau_0}^0 d\tau_V\; \tau_V ^2
e^{-B\Big(  (k_1^5+k_2^5+k_3^5)  (0^5-\tau_V^5 )\Big) }  \sin\Big((k_1+k_2+k_3 )\tau_V  \Big)      \Big]
 \nonumber \\
\mathcal{F}(k_2/k_1,k_3/k_1,\tau_0)_{\text{Non-BD}}^{\text{Loop}}&=& \frac{-3\lambda}{2H^2 \epsilon}  k_1 k_2 k_3    \Big[  \int_{\tau_0}^0 d\tau_V\; \tau_V ^2  c_{\bm k_1}
 e^{-B\Big(2k_1^5(\tau_V^5-\tau_0^5)+ (k_1^5+k_2^5+k_3^5)  (0^5-\tau_V^5 )\Big) }
  \nonumber\\&&\times
  \sin\Big((k_2+k_3-k_1)\tau_V-\theta_{\bm k_1} \Big) +2\text{ perm.}     \Big]
 \eeqn

Next, we give some plots for non-Gaussianity.  Recall that in general single field inflation, the power spectrum $P_\zeta=\frac{H^2}{8\pi^2   \epsilon}$ and $\Sigma=H^2 \epsilon$ ($c_s=1$ in our model).
 The non-Gaussianity estimator $f_{\text{NL}}=-\frac{10}{81}\frac{\lambda}{\Sigma}=-\frac{10}{81}\frac{\lambda}{H^2 \epsilon}$~\cite{Chen:2006nt}. So, we can express the exponential decay factor in terms of observable quantities as
 \be
 B =\frac{3\lambda^2 }{3200\pi  \epsilon^3H^2}
  =\frac{19683\pi}{40000} P_\zeta f_{\text{NL}}^2\approx    1.55 P_\zeta f_{\text{NL}}^2~.
 \ee

We choose parameters $P_\zeta=10^{-9}, f_{\text{NL}}=1$~\cite{Ade:2013ydc}. For non-BD parameters, we use $c_{\bm k}=0.1, \theta_{\bm k} =0$.    The shape for BD and non-BD non-Gaussianity are shown in Fig.~\ref{BDShape} and Fig.~\ref{SharpNonBDShape}, Fig.~\ref{nonBDShape}.

{\noindent\bf BD non-Gaussianity shape:  }
The non-Gaussianity shape for BD part is shown in Fig.~\ref{BDShape}. For BD non-Gaussianity,  if we choose one initial time sharp cut-off, at tree level  the non-Gaussianity shape function shows oscillating behavior due to the oscillating term in the integral. But at loop level, as long as the initial time is not too late which is always the case because the BD starts from very very early time and in principle from nearly past infinity, the oscillating behavior disappears and we nearly recover the usual BD non-Gaussianity shape.

{\noindent\bf Non-BD non-Gaussianity shape:  }
For  non-BD non-Gaussianity, it peaks at one specific initial time roughly. Earlier or later initial time can only generate smaller observational non-Gaussianity. What's more,  loop corrections cure the folded divergence behavior of  non-Gaussianity as we emphasize   before.   If the initial time is much earlier, which means substantial time for non-BD state to decay, it may be very difficult to observe the remnants of the non-BD information experimentally.

Due to the highly sensitive dependence on initial time, the final shape of non-Gaussianity may show some oscillating features which is not generic and depends on the details of $c_k,\theta_k,\tau_0(k)$ and so  on. We expect that in   reality, these highly sensitive dependence is fragile and will be averaged or smoothed due to complicated behavior of these functions. The important and generic part is the non-oscillating part with relatively weak dependence on initial time. Note that for BD part, we do not need to use this smoothing functions. Because, in principle, the BD  exists from the very early beginning, almost infinitely past. And our exponential correction term is enough to suppress the oscillating parts to get the standard BD non-Gaussianity shape. Nevertheless, for completeness, we provide a typical plot for  the non-BD non-Gaussianity shape without smearing initial time.

Therefore, we choose to filter   the slowly varying non-oscillating parts by averaging the initial time with a Gaussian distribution centered at $\tau_{0c}$ with width $\tau_{0w}$:
\be \label{gaussian-filter}
W(\tau_0,\tau_{0c},\tau_{0w})=\frac{1}{\sqrt{2\pi }\tau_{0w}}
\exp\Big({-\frac{(\tau_0-\tau_{0c})^2}{2\tau_{0w}^2}}\Big)
\ee
With this smoothing function, the observable non-Gaussianity shape function is
\be
\mathcal{F}_{\text{ave}}(k_2/k_1,k_3/k_1,\tau_{0c},\tau_{0w})_{\text{Non-BD}}^{\text{Loop }}
=\int_{- \infty}^0 d\tau_0 \; W(\tau_0,\tau_{0c},\tau_{0w})\mathcal{F}(k_2/k_1,k_3/k_1,\tau_{0c},\tau_{0w})_{\text{Non-BD}}^{\text{Loop }}
\ee

 Under Gaussian smoothing, the exponential function $e^{i (k-k_0) \tau_0}$ will be transformed into a  smooth and non-oscillating Guassian function of $k$ centering at $k_0$ with width $1/\tau_{0w}$.

The non-Gaussianity shape of non-BD part (see Fig.~\ref{nonBDShape}) includes two features: folded shape peak and squeezed limit shape. The folded shape is mainly contributed by the $c_{\bm k_1}$: when $k_2+k_3-k_1 \sim 0$ or $|(k_2+k_3-k_1)\tau_0|\lesssim  \pi$, the sine function in the integral will contribute coherently with oscillations. If we only   consider the tree level result, the folded shape value will blow up for very early initial time. But if we include the loop corrected exponential term, its value will be suppressed to nearly vanishing. While the off-diagonal corner ($k_2\sim 0$ part and $ k_3\sim 0$ part) shape  is contributed by the second and third term in the non-BD non-Gaussianity shape function, namely $c_{\bm k_2},c_{\bm k_3}$ term.  When $k_2\sim 0$ and thus $k_1+k_2-k_3\sim   k_2\sim 0$, the sine function $c_{\bm k_3}\sin((k_1+k_2-k_3)\tau_V)$ in the integral will  contribute  coherently as long as $|(k_1+k_2-k3)\tau_0|\lesssim  \pi $ and thus give rise to large value. The  arguments are similar to the folded shape one.  Similar arguments apply for $k_3\sim 0$.

At tree level, the earlier the initial time, the larger the folded shape peak value. While at loop level, the folded peak will be suppressed to very small value.

 \subsection{Non-interacting limit }
 In the following part, we are going to consider the non-interacting limit $\lambda\rightarrow 0$ and show that standard results for BD non-Gaussianity can be recovered.

For BD non-Gaussianity,
\beqn\label{BD_non_interacting_limit}
\mathcal{F} _{\text{BD}}&=&  \frac{-3\lambda}{2H^2 \epsilon}  k_1 k_2 k_3    \Big[  \int_{\tau_0}^0 d\tau_V\; \tau_V ^2
e^{-B\Big(  (k_1^5+k_2^5+k_3^5)  (0^5-\tau_V^5 )\Big) }  \sin\Big((k_1+k_2+k_3 )\tau_V  \Big)      \Big]
\nonumber\\
&=&  \frac{-3\lambda}{2H^2 \epsilon} \frac{- k_1 k_2 k_3 }{(k_1+k_2+k_3)^3}
  \Big[  \int_0^{x_0} dx\; x^2
e^{-D x^5}  \sin x      \Big]~,
\eeqn
 where  we define
 $x=-(k_1+k_2+k_3)\tau_V,x_0=-(k_1+k_2+k_3)\tau_0, D=B(k_1^5+k_2^5+k_3^5)/(k_1+k_2+k_3)^5$ .  The integral   can be evaluated in the following way through integration by parts
 \beqn
   \int_0^{x_0} x^2 e^{-D x^5}  \sin x \; dx
   &=&x^2 e^{-D x^5}(-\cos x)\Big|_0^{x_0}-   \int_0^{x_0}  e^{-D x^5}(-\cos x)(2x-5D x^6) \; dx
   \nonumber\\
    &=&-x^2 e^{-D x^5}\cos x\Big|_0^{x_0}+\int_0^{x_0} 2x e^{-D x^5} \cos x \; dx+D\times\cdots
      \nonumber\\
    &=&(-x^2 e^{-D x^5}\cos x+2x e^{-D x^5}\sin x) \Big|_0^{x_0}-\int_0^{x_0} 2  e^{-D x^5} \sin x \; dx+D\times\cdots
          \nonumber\\
    &=&(-x^2 e^{-D x^5}\cos x+2x e^{-D x^5}\sin x+2 e^{-D x^5}\cos x) \Big|_0^{x_0} +D\times\cdots~.
 \eeqn
In the non-interacting limit, $D\rightarrow 0, x_0\rightarrow \infty$, the above integral gives rise to a factor $-2$. Thus, we do recover the standard BD non-Gaussianity. Previously, the standard non-Gaussianity is obtained by $i\varepsilon$ prescription. Here the interaction coupling constant plays the role of $\varepsilon$ and regulates the divergence problem. In this sense, we give a natural explanation to the  problem of introducing $i\varepsilon$ in cosmology.

For non-BD non-Guassianity, things are a little more complicated due to the interplay of folded  limit and non-interacting limit. Physically, the correct order of taking limits should be like this: fix the coupling strength $B$ first and then examine the non-Gaussianity at folded limit, finally turn off the interactions gradually. After fixing the coupling strength, the initial time is roughly given by $\tau_{0m}$ instead of $-\infty$  in the $i\varepsilon$ prescription, which can only give a trivial vanishing result in our case.
Near the folded limit, we can perform Taylor expansion for $\delta k$, then this integral is a normal one and vanishes when $\delta k\rightarrow 0$.
 So, for a fix coupling strength, the exact folded limit non-Gaussianity vanishes. Then, we take the non-interacting limit, which by continuity also gives rise to a vanishing folded limit non-Guassianity. However for a fixed given coupling strength, globally the amplitude will increase if we let coupling strength goes to zero and choose  proper initial time. More specifically, if the shape is not too folded ($B(k/\delta k)^5$ is still very small with $\delta k=k_2+k_3-k_1$ or other permutations), the method for BD one can be applied here. Then, we recover the $1/\delta k^3$ factor in $\mathcal{F}_{\text{Non-BD}}$.   While the other integral factor (which is $-2$ for BD one) is highly sentive to the intial time.  The scale of amplitude is roughly given by Eq.~\eqref{Maximum_Value}, scaling like $B^{-3/5}$. This amplitude decreases very quickly once a different initial time is chosen and will  vanish if we start from past infinity.

In particular, if we tune the coupling constant $\lambda$ to be pretty small, for example equivalently let $f_{\text{NL}}=10^{-6}$ which is mathematically meaningful nevertheless, we can get non-BD non-Gaussianity which is nearly divergent at the folded limit  provided that  the proper initial time is chosen. Actually, the amplitude of the folded limit non-Gaussianity is proportional to $f_{\text{NL}}^{-6/5}$~\eqref{Maximum_Value}. However, once the initial time is slightly different, the amplitude   decreases dramatically.

 The conclusion is that the loop corrections is vital for non-BD three point functions. Even in the extremely weak interacting limit and we can fine tune the initial time delicately to get the ordinary divergent folded non-Guassianity, this  apparent divergence breaks down once   a little bit different initial time are considered. This is \emph{not} the feature of previous non-BD three point functions in literature. From this aspect, it is very challenging to observe the imprints of initial non-BD state at present, especially for large $k$ modes.


\section{Conclusion}\label{Conclusion}

In this paper, we develop the techniques of calculating one loop diagrams. And we discover a recursion relation which enables us to deal with infinite loop calculations and do resummations. By using these techniques, we show that the decay of non-BD coefficients are consistent with the previous cut-off result except an order one decay factor difference. Our method is enlightening and may shed light on the future loop  calculations and resummations in the general context.


Furthermore, we analyze the non-Gaussianity under loop corrections. As we expect from the decay of effective non-BD coefficients, the usual divergent non-BD non-Gaussianity at folded limit gets smoothed. What's more, the loop corrected non-BD non-Gaussianities peak at specific initial time and are very sensitive to these initial time. Once we deviate a little bit, these non-Gaussianities will decrease dramatically. So, we conclude that the non-BD  non-Gaussianities are very fragile to loop interactions and initial time. Thus, as long as the non-BD state is set up at early enough time, the imprints of these non-Gaussianties on observations may be difficult. These are very different from the previous results in literature where folded limit non-Gaussianities are dominated by the non-BD one due to the divergent behavior.

Besides, we also show that even for BD non-Gaussianity, the loop corrections can have significant influence, playing the role of infinitesimal regulator like $i\varepsilon$ prescription. The loop correction can not only regulate the divergence problem but also recover the usual result in literature. Thus, loop corrections provide a natural way of introducing $i\varepsilon$ in cosmology in a more natural and physical way.

Our results are derived based on sub-horizon limit approximations. It is well known that the sub-horizon modes do not feel the presence of gravity much and their behaviour resemble  the flat Minkowski  space case. So are the loop corrections. One may wonder whether the same properties that we have discussed already exist in flat space quantum field theory. The answer is yes or no. On the one hand, the UV limit of the cosmological perturbations indeed return to flat space quantum field theory. But on the other hand, in usual treatment of flat space quantum field theory, we are interested in the in-out amplitude. Interactions are shut off at the asymptotic past and future. However, in our case, the fluctuations keep on interacting in the asymptotic past. Also, the expansion of the universe exposes anything odd in the UV, if not diluted by the expansion of the universe, to observables at macroscopic scales. Finally, inflation needs a start and may have features, so asymptotic Lorentz symmetry or de Sitter symmetry may not help to determine the vacuum. Those reasons explain the difference between our work and a conventional treatment of flat space quantum field theory.

Although we only consider the general single field inflation  and rely on some approximations,   the conclusions are expected to apply in more general cases.

\acknowledgments

We thank Junyu Liu for initial collaboration during his internship. We thank Xingang Chen,  Lam Hui  and Gary Shiu for very helpful discussions. This work iss supported by the CRF Grants of the Government of the Hong Kong SAR under HUKST4/CRF/13G.

\appendix
\section{Basics of in-in formalism }\label{in-in_formalism}

We can use the in-in formalism to compute the correlation function~\cite{Weinberg:2005vy, Chen:2010xka, Wang:2013eqj}. The basic formula is
\beqn
\EV{Q(\tau)}&=&\Big\langle 0 \Big|\Big[ \bar{\mathcal{T}}\exp\Big( i\int_{\tau_0}^{\tau}H_I(\tilde\tau\rq{})d\tilde\tau\rq{}\Big)\Big]  Q^I(\tau)\Big[ \mathcal{T}\exp\Big(-i\int_{\tau_0}^{\tau}H_I(\tau\rq{})d\tau\rq{}\Big)\Big]  \Big|0\Big\rangle \nonumber \\
&=&
\sum_{n,m=0}^{\infty}    i^m (-i)^n
\int _{  \tau _0}^\tau  d\tilde{\tau}_1  \int _{\tilde{\tau}_1}^\tau  d\tilde{\tau}_2 ...\int _{\tilde{\tau}_{m-1}}^\tau  d\tilde{\tau}_m \quad
 \int _{\tau_0}^{\tau}  d\tau_1 \int _{\tau_0}^{\tau_1}  d\tau_2...\int _{\tau_0}^{\tau_{n-1}}  d\tau_n
  \nonumber \\ &&\times
\Big\langle 0 \Big| H_I(\tilde{\tau}_1)...H_I(\tilde{\tau}_m) Q^I(\tau) H_I( \tau_1 )...H_I( \tau_n)\Big |0\Big\rangle \nonumber \\
&=&
\sum_{n,m=0}^{\infty}   i^m (-i)^n
\int _{  \tau _0}^\tau  d\tilde{\tau}_m  \int_{\tau_0}^{\tilde{\tau}_m}  d\tilde{\tau}_{m-1} ...\int_{\tau_0}^{\tilde{\tau}_2}  d\tilde{\tau}_1 \quad
 \int _{\tau_0}^{\tau}  d\tau_1 \int _{\tau_0}^{\tau_1}  d\tau_2...\int _{\tau_0}^{\tau_{n-1}}  d\tau_n
 \nonumber \\ &&\times
\EV{H_I(\tilde{\tau}_1)...H_I(\tilde{\tau}_m) Q^I(\tau) H_I( \tau_1 )...H_I( \tau_n) }_0   ~.
\eeqn

Note that the factor $n!m!$ has been canceled by the time-orderings.

At a specific order of expansion, there are lots of terms due to different ways to connect the diagrams and different time sequences. We can represent them in different Feynman diagrams, where the time sequences can be seen from the relative positions of interacting vertices along the time axis, while the connecting ways can be understood as usual Feynman diagrams in QFT.

The basic type of integral has the following equivalent form  ($ \tau>\tau_1>\tau_2>...>\tau_{n-1}>\tau_n>\tau_0 $)
\be
 \int _{\tau_0}^{\tau}  d\tau_1 \int _{\tau_0}^{\tau_1}  d\tau_2...\int _{\tau_0}^{\tau_{n-1}}  d\tau_n \quad   f(\tau_1,...,\tau_n)
 = \int _{\tau_0}^{\tau}  d\tau_n  \int _{\tau_n}^{\tau}  d\tau_{n-1}...\int _{\tau_2}^{\tau }  d\tau_1 \quad   f(\tau_1,...,\tau_n)  ~.
\ee

Another very useful formula is  to leave one time $\tau_m$ at last.  ($ \tau>\tau_1>\tau_2>...>\tau_m>...>\tau_{n-1}>\tau_n>\tau_0 $)
\beqn
 &&\int _{\tau_0}^{\tau}  d\tau_1 \int _{\tau_0}^{\tau_1}  d\tau_2...\int _{\tau_0}^{\tau_{n-1}}  d\tau_n \; f(\tau_1,...,\tau_n)
\nonumber  \\&=&
     \Big(\int _{\tau_0}^{\tau}   d\tau_m  \Big)
  \Big(  \int _{\tau_m}^{\tau}  d\tau_1 \int _{\tau_m}^{\tau_1}  d\tau_2...\int _{\tau_m}^{\tau_{m-2}}  d\tau_{m-1}  \Big)
  \Big(  \int _{\tau_0}^{\tau_m}  d\tau_{m+1}  ...\int _{\tau_0}^{\tau_{n-1}}  d\tau_n  \Big)
\; f(\tau_1,...,\tau_n) ~.   \nonumber\\
 \eeqn

 Usually, $f$ is the product of some mode functions and can be factorized into different parts: $f(\tau_1,...,\tau_n)=\prod_j f_j(\tau_j)$. So, this formula is so convenient that we can deal with different parts separately.
 For example, when calculating the loop diagrams of three point function,  the interacting vertex for three external leg is special and we can leave it to the last integration.

 The basic idea of the proof is that we swap $\tau_m$ and $\tau_1$ first, then swap   $\tau_m$ and $\tau_2$... Repeat this process until $\tau_m$ is between $\tau_{m-1}$ and $\tau_{m+1}$ which recover the standard form. The basic formula is
$
  \int _{\tau_0}^{\tau_p}   d\tau_m     \int _{\tau_m}^{\tau_p}   d\tau_j \cdots= \int _{\tau_0}^{\tau_p}   d\tau_j     \int _{\tau_0}^{\tau_j}   d\tau_m \cdots ~.
$

\section{Elliptical coordinate system}\label{elliptical_coordinate}

Consider two fixed points $A$, $B$ separated by distance $R$,  and $P$ is the moving point. The distance between $A, B, O$ (the origin) and $P$ is $r_A, r_B, r$. Also, assume the angle between $OA$ and $OP$ is $\theta$.

\begin{figure}
\centering
\includegraphics[width=0.5\textwidth]{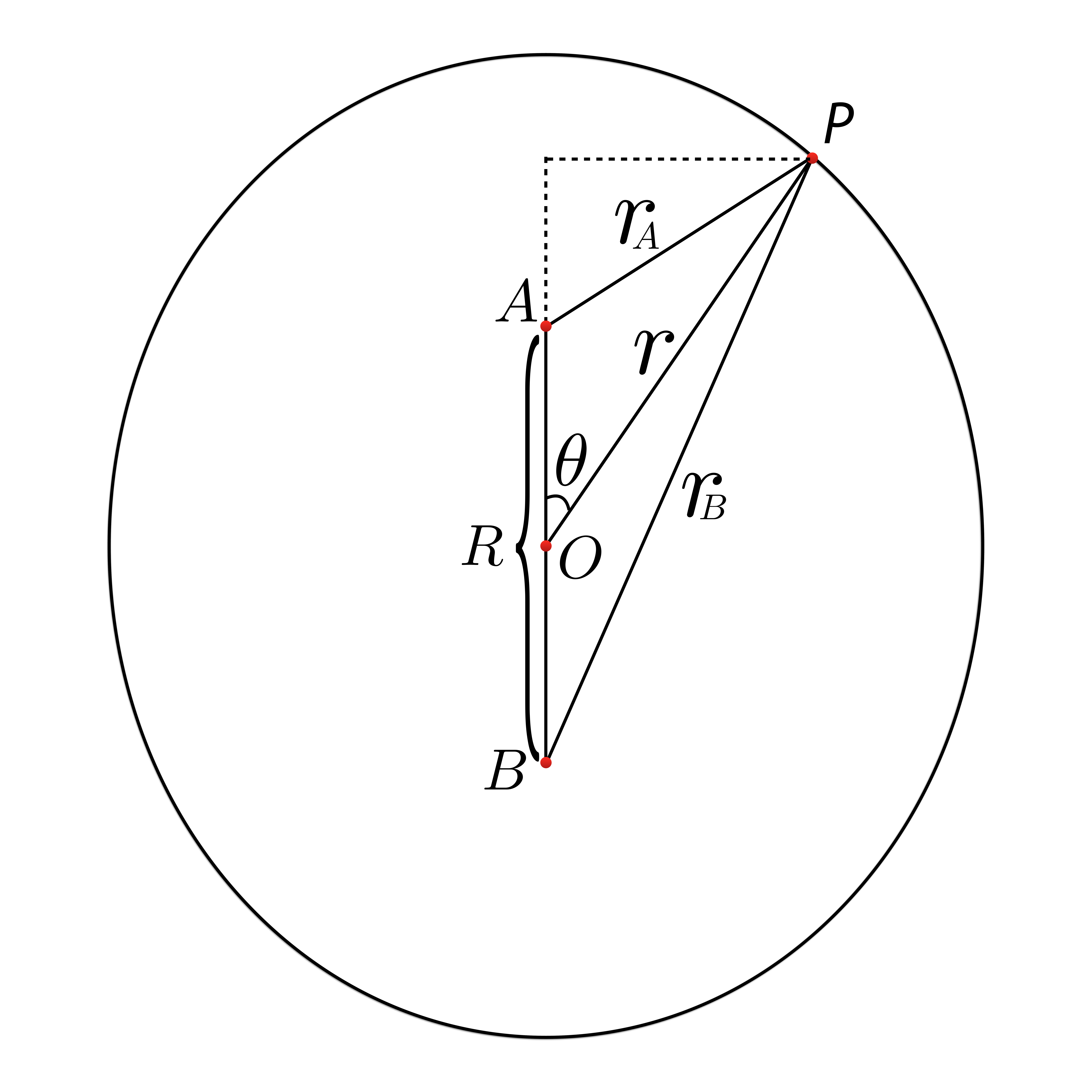}
\caption{Elliptical coordinate system}
\end{figure}

Define
\be
\mu=\frac{r_A+r_B}{ R}~, \qquad \nu=\frac{r_A-r_B}{ R}~,
\ee
then, we have
\beqn
r&=&\frac{R}{2}\sqrt{\mu^2+\nu^2-1}   ~, \\
z&=&r\cos\theta=-\frac{R}{2}\mu \nu       ~,   \\
\rho&=&r\sin\theta=\frac{R}{2}\sqrt{\mu^2+\nu^2-1-\mu^2\nu^2}~.
\eeqn
The Jacobi matrix is
\beqn
J=\frac{\partial(r,z)}{\partial(\mu,\nu)}=\PBK{
\frac{R}{2}\frac{\mu}{\sqrt{\mu^2+\nu^2-1}}
 &\frac{R}{2}\frac{\nu}{\sqrt{\mu^2+\nu^2-1}}\\
 -\frac{R}{2} \nu&-\frac{R}{2}\mu }  ~,
\eeqn

so,
\be
\det J=\frac{R^2}{4} \frac{-\mu^2+\nu^2}{\sqrt{\mu^2+\nu^2-1}}
\rightarrow  - \frac{R^3}{8} \frac{1}{r}(\mu^2-\nu^2)  ~.
\ee
In 3D, the volume element is
\beqn
dV&=&r^2 \sin \theta dr d\theta d\phi= -r dr d (r\cos\theta) d\phi  =-2\pi r dr dz
\nonumber\\
&\rightarrow&  2\pi r \frac{R^3}{8} \frac{1}{r}(\mu^2-\nu^2)  d\mu d\nu
=2\pi   \frac{R^3}{8} (\mu^2-\nu^2)  d\mu d\nu     ~.
\eeqn

In general, in $n$-dimensional space, the volume element (after integrating out the angular part) is
\be
dV=\frac{2\pi (\sqrt{\pi})^{n-3}}{\Gamma(\frac{n-1}{2})}\Big(\frac{R}{2} \Big)^n\Big( \mu^2+\nu^2-1-\mu^2\nu^2 \Big)^{\frac{n-3}{2}}(\mu^2-\nu^2)  d\mu d\nu  ~,
\ee

So, if we want to integrate the function $f(k,p,q)$, the calculation is transformed to
\beqn
I&=&\int d^n\bm q \int d^n \bm p\; \delta^{(n)}(\bm p+\bm q-\bm k)  f(k,p,q)  \nonumber \\
&=&\frac{2\pi (\sqrt{\pi})^{n-3}}{\Gamma(\frac{n-1}{2})}\Big(\frac{k}{2} \Big)^n
\int_1^\infty d\mu \int_{-1}^{1} d \nu
\Big( \mu^2+\nu^2-1-\mu^2\nu^2 \Big)^{\frac{n-3}{2}}(\mu^2-\nu^2) f(k,\frac{\mu+\nu}{2}k,\frac{\mu-\nu}{2}k)  ~.
 \nonumber \\
\eeqn

In three dimension, it simplifies as
\be
I=\frac{\pi k^3}{4}
\int_1^\infty d\mu \int_{-1}^{1} d \nu\; (\mu^2-\nu^2) f(k,\frac{\mu+\nu}{2}k,\frac{\mu-\nu}{2}k)  ~.
\ee

\section{New basis functions }\label{basis_function}

We introduce a set of basis function $T_n$:
\beqn
T_1(z)&=&\frac{e^{izu}-1}{z} ~,\\
T_n(z)&=&\frac{1}{z^n}\Big(1-\frac{i z u}{n-1} \Big) \Big(e^{i z u}-1-izu-...\frac{(izu)^{n-1}}{(n-1)!} \Big)
-\frac{(iu)^n}{(n-1)(n-1)!}\quad  n>1  ~.
\eeqn
There are some good properties for $T_n$:
\begin{enumerate}

\item They are regular near $z\sim0$.

\item For $n>1$, $T_n$ decreases very rapidly at $+\infty$ and is integrable without divergence
\be
\int_0^{+\infty} T_n(z) dz=\text{finite}~.
\ee

\item For $n=1$, the integration of $T_1$ from 0 to $+\infty$ will result in logarithmic divergence. We can introduce a cut-off on $z$ which yields
\be
\int_0^{\Lambda} T_1(z) dz= \Ci (|\Lambda u|)-\gamma_E
-\log(|\Lambda u|)+i\sgn(u)\cdot \Si(|\Lambda u|)~,
\ee
where $\Si, \Ci$ are Sine and Cosine integral function.
\end{enumerate}

Suppose we have a function with the following form
\be    K=   \sum_{n=1}^N  \frac{a_n e^{i z u}-b_n}{z^n}~, \ee
where $a_n, b_n$  are coefficients independent of $z$. Also, we require the above function is regular near $z\sim 0$ which means that $a_n, b_n$ are not totally independent. There must be some relations among them to ensure the regularity of the function. There are $2N$ coefficients $a_n, b_n$, but the regularity near $z=0$ will lead to $N$ constraint equations corresponding to each order of Taylor expansion of $   K$ in $1/z^n$. This implies that actually there are only $N$ free parameters.

So, we can decompose the above function in terms of the basis functions $T_n$
\be
\sum_n  \frac{a_n e^{i z u}-b_n}{z^n}=\sum_n  A_n T_n(z)~,
\ee
We can express $A_n$ in terms of $a_n, b_n$ by solving the above equation.
Also, note that in the basis functions $T_n$, $\frac{1}{z}$ only appears in $T_1$, so we can get a very important relation
\be A_1=b_1  ~.\ee

\bibliographystyle{JHEP}
\bibliography{NonBD_Bib}

\end{document}